# Engineering safe structures: recent advances in structural reliability modelling

**Adwait Sharma[1], Oindrila Kanjilal[2], C S Manohar[3]**

**Abstract**

This paper reviews advances over the last decade in the field of computational structural reliability modelling. The discussion is focused on problems of time-variant/time-invariant reliability at the component level. Three broad classes of problems are considered: (a) time-invariant reliability problems involving static problems, (b) problems of time-variant reliability analysis for deterministically parametered dynamical systems driven by random excitations, and (c) time-variant reliability analysis of randomly parametered dynamical systems subjected to random excitations. Four classes of approaches are discussed: (a) analytical methods based on first/second order reliability analyses (for time-invariant reliability analysis) and level crossing based approaches for time-variant reliability problems, (b) methods based on importance sampling strategies (including the Girsanov transformation based method for dynamical systems), (c) particle and trajectory splitting based methods, and (d) methods that employ machine learning based tools (primarily involving development of surrogate models and active learning strategies) in tackling reliability problems. The focus of the discussions is on methodological advances, and questions related to specific applications are not addressed. The review presents critical discussions on the relative merits of alternative approaches (in terms of computational efficiency, accuracy, scalability, treatment of rare events, ability to handle geometric complexities linked to failure surface, and suitability for high-dimensional problems) and identifies several directions for future research.

## 1.0 Motivations and scope of this paper

Civil engineering structures interact closely with nature, and this interaction forms the primary source of uncertainty in their behaviour. A principled treatment of these uncertainties is the prime requirement when engineering safe structures. At the outset, the sources of uncertainties can be conventionally classified into two broad categories: aleatory uncertainties (which arise from inherent natural variability in loads, materials, and environmental conditions), and epistemic uncertainties (which arise from incomplete knowledge, modelling idealizations, parameter estimation errors, and data limitations). This distinction is helpful in formulating of strategies for quantification, propagation and processing of the uncertainties. The interaction between structures and nature encompasses various aspects, and they can be summarized as follows:

- Loads: Most critical actions of design significance arise from extremes of natural processes. These processes themselves are highly nonlinear in nature. For instance, earthquakes arise from extremes of states of stress evolving due to tectonic activities, cyclonic winds and floods arise from extremes of fluid dynamic processes, and *mala fide* attacks arise from extremes of human mind. The observational records for many such events are sparse, and they contain short-range and long-range uncertainties. For instance, the number of earthquakes above a certain magnitude in a given region exhibits long-range uncertainties (spanning several decades/centuries) and the details of ground motion at a given location in the event of an earthquake exhibit short-range uncertainties (in terms of magnitude, direction, duration, and frequency content).
- Degradation mechanisms: Processes such as corrosion, fatigue, creep, alkali–silica reaction, and foundation settlement evolve slowly over decades, and their rates depend on environmental exposure conditions. If one considers the influence of climate change, involving altered temperature cycles, moisture patterns, and changes in atmospheric chemistry, the statistical properties of these degradation processes themselves become time varying. This motivates time-variant modelling of structural resistance and underscores the difficulty of predicting structural performance over long service periods.
- Material behaviour: Civil structures are made up of a combination of large volumes of naturally occurring materials (such as stone, sand, and soil) and manufactured materials (such as steel, cementitious materials and fabrics). Characterizing the material constitutive behaviour and bonding among these materials is beset with notable uncertainties.

[1] Formerly PhD Scholar, Department of Civil Engineering, IISc; Presently: Post-doctoral Researcher, Johns Hopkins University, Baltimore, Maryland, USA. Email: ashar125@jh.edu
[2] Formerly PhD Scholar, Department of Civil Engineering, IISc; Presently: Post-doctoral Researcher, LIMOS, CNRS (UMR 6158), Université Clermont-Auvergne, Clermont-Ferrand, France. Email: oindrila.kanjilal@uca.fr
[3] Author for Correspondence, Professor (Retired), Department of Civil Engineering, Indian Institute of Science. Email: manohar@iisc.ac.in

Besides the environmental loads, the structures also carry operational and accidental loads such as loads due to fire, presence of machinery, moving objects, and impacts. Notable uncertainties also arise due to various construction practices. Studies related to existing instrumented structures lead to inverse problems involving structural system identification wherein treating noisy sensor data poses notable challenges.

Extreme loads such as those due to strong earthquakes are rare, and including this aspect into the structural design framework has remained a challenge. A strategy that is commonly used is to permit structures to enter inelastic regimes when such extreme events indeed occur and aim to manage the resulting damage. In the context of earthquake engineering, this would mean that structures are designed to display controlled inelastic behaviour (with certain modes of failure being promoted while others being avoided by design). Treatment of uncertainties within such a framework, when the structure undergoes inelastic transient vibrations, and one aims to control the nature of resulting damage, poses significant challenges.

Within this setting, mathematical models serve as the primary tool for representing structural behaviour and assessing performance under uncertain loads and material properties. High-fidelity numerical models can incorporate geometric and material nonlinearities, anisotropy and inhomogeneities, multi-physics interactions, spatial variability, and both rapid and slowly evolving excitations. However, increasing the level of detail in a model inevitably increases the number of parameters, intricacies in specifying boundary conditions, and excitations. Many of these parameters are not easy to model due to limited experimental data, sparse extreme-event records, and incomplete characterization of degradation mechanisms. Thus, increasing model fidelity can paradoxically amplify epistemic uncertainty, rather than reducing it.

It is within this complex milieu of natural variability, sparse data, interacting uncertainties, and model complexity that the field of structural reliability modelling has emerged. The subject provides a systematic mathematical framework to model uncertainties, quantify their effects, and support reliability-based design and optimization. The framework of the subject also provides a systematic conduit for communication between structural engineers and society in terms of embedding societal expectations on resource allocation and levels of safety consistent with prevailing risks in the society due to other causes.

The present review focuses on methodological advances of the past 10-15 years, particularly in simulation-based reliability analysis, rare-event estimation, variance-reduction techniques, and incorporation of machine-learning-based tools into their folds. Some of the primary challenges here are associated with the treatment of high-dimensional representation of uncertainties, expensive finite element model evaluations, nonlinear failure domains, and the need to estimate very low probabilities of failure. While the primary focus of the review is on reliability modelling as a tool at the stage of design, we also briefly consider questions on laboratory qualification testing for reliability of highly reliable systems that allows for rare event simulations. The review covers problems of time-invariant and time-variant reliability modelling in which uncertainties can arise in system parameters and applied external actions.

This article is written on the occasion of the 75th anniversary of the Department of Civil Engineering at the Indian Institute of Science and follows the earlier review by Gupta and Manohar [2005], prepared when the Department had completed 50 years of functioning. The present paper thus also serves as a report on the progress achieved in the intervening two decades. During this period, several expository books and review articles have been published. Thus, the works of Haldar [2006] and Gardoni [2017] provide comprehensive snapshots of the prevailing state-of-the-art. The monograph by Hurtado [2004] is perhaps the earliest work to signal the emergence of machine learning-based methods in reliability modelling. The work by Faber et al., [2008] is representative of several documents published by the Joint Committee on Structural Safety (JCSS). The works of Lemaire et al., [2013], Melchers and Beck [2018], and Der Kiureghian [2022] remain authoritative textbooks on the subject of structural reliability. A few works focusing on specific topics have also been published: thus, the book by Yuen [2010] focuses on Bayesian methods for modelling instrumented civil structures, Au and Wang [2014] on the application of subset simulation methods in structural reliability modelling, Wang [2021] on lifetime reliability models, and Zhu and Keshtegar [2025] on optimization techniques in reliability-index based methods. The handbook by Ghanem et al., [2017] covers a wide spectrum of issues related to the treatment of uncertainty in computational models. The papers by Breitung [2015, 2019, 2021] highlight the insights that reliability-index-based methods provide vis-à-vis simulation-based methods, which often produce an estimate of failure probability with insights into the structural behaviour remaining hidden. The paper by Tabandeh et al., [2022] focuses on importance sampling (IS) methods and discusses two classes of approaches which approximate either the density function or the limit state function. The paper by Song and Kawai [2023c] provide a methodological review with emphasis on Monte Carlo simulation-based methods. Several papers present reviews on reliability estimation methods which incorporate machine learning tools into their fold [Chojaczyk et al., 2015, Teixeira et al., 2021, Xu and Saleh, 2021, Afshari et al., 2022, and Moustapha et al., 2022]. These reviews broadly address issues related to surrogate modelling of the limit state function using artificial neural networks, Gaussian process regression, and support vector machines, and controlling the number of calls to the parent computational model by using active learning strategies. Recently, Ellingwood et al., [2025] have presented a historical synthesis of the growth of the subject of structural reliability, highlighting the roles played by reliability index-based methods, Monte Carlo simulation-based methods, stochastic dynamics, and design code development. The role played by JCSS in the overall growth of the subject of structural reliability has been discussed by Vrouwenvelder et al., [2025].

As has been noted, the focus of the present review is on methodological aspects of computational modelling of reliability and several important topics, such as, application-specific considerations (as in the treatment of loads due to earthquakes, wind, fire, and guideway unevenness), specific details of degradation mechanisms, issues related to design code calibration and structural optimization, and the framework of performance based structural engineering, are not covered. We focus on probabilistic approaches to model uncertainties, and other alternative uncertainty modelling frameworks (such as convex function models and models based on fuzzy variables) are also not considered. Given the vastness of the literature in this area of research, we further restrict the review to methods for component reliability assessment, with questions on system reliability not covered.

## 2.0 Basic notions and tools

The way an engineering structure responds to the applied load depends on the type and magnitude of the loading, and the stiffness and strength characteristics of the structure. The structural response is considered to be satisfactory if it satisfies a set of prescribed performance requirements, which are termed limit states. The violation of the limit states signals failure of the structure. For the purpose of reliability assessment, the limit state is described mathematically in terms of a function (called the limit state function, or the performance function) whose value depends on the applied load and the relevant structural parameters. A negative value of this function implies the violation of the underlying limit state. Depending on the probabilistic models used to represent the uncertainties, the limit state function will be either time-dependent or independent of time. Accordingly, the problem of structural reliability analysis can be classified into two broad categories, namely, time-invariant and time-variant problems.

## 2.1 Time-invariant reliability

Here, the uncertainties pertaining to the applied load and the structural properties are modelled either as a set of random variables or a set of random fields, depending on whether the uncertainties are spatially invariant or not. When random field models are adopted, it is assumed that they would be discretized into an equivalent set of random variables using stochastic finite element techniques. The uncertainties, in this setting, are collectively represented by a random vector $\boldsymbol{X}$, with a prescribed joint probability density function (PDF) $p_{\boldsymbol{X}}(\boldsymbol{x})$. The components of $\boldsymbol{X}$, in general, could be mutually dependent and non-Gaussian. In most engineering applications, complete statistical information about the random vector $\boldsymbol{X}$ is not available, and hence a complete specification of $p_{\boldsymbol{X}}(\boldsymbol{x})$ is seldom possible. In cases where the marginal pdfs and partial information on the measures of dependence are available, a multi-variate joint pdf for the non-Gaussian random vector can be constructed based on the applications of copula modelling [Nelsen 2006, Melchers and Beck 2018]. To quantify the reliability, one identifies a suitable performance function $g(\boldsymbol{X})$, and seeks to evaluate the probability of failure, which is expressed as a multi-fold integral of the form

$$P_F = \int_F p_{\boldsymbol{X}}(\boldsymbol{x})d\boldsymbol{x} = \int_{g(\boldsymbol{x})\leq 0} p_{\boldsymbol{X}}(\boldsymbol{x})d\boldsymbol{x} = \int_{-\infty}^{\infty} I\{g(\boldsymbol{x}) \leq 0\}p_{\boldsymbol{X}}(\boldsymbol{x})d\boldsymbol{x} \quad (1)$$

Here $F = \{g(\boldsymbol{X}) \leq 0\}$ denotes the failure event, and $I\{\cdot\}$ is the indicator operator such that $I\{g(\boldsymbol{x}) \leq 0\} = 1$ if $g(\boldsymbol{x}) \leq 0$ and zero otherwise. The regions $g(\boldsymbol{X}) > 0$ and $g(\boldsymbol{X}) < 0$, respectively, denote the safe and unsafe domains of the space spanned by $\boldsymbol{X}$, and the boundary $g(\boldsymbol{X}) = 0$ is called the limit state surface.

## 2.2 Time-variant reliability

Here one considers problems where one or more of the sources of uncertainty are taken to evolve in time. In this paper, the focus is on linear/nonlinear vibrating systems where only the applied excitations are modelled as random processes, with time as the evolution parameter. The uncertainties in the structural system parameters are taken to be time-invariant and are treated as a vector of random variables. The governing equation for such systems can be expressed as

$$\mathbf{M}(\boldsymbol{\Theta})\ddot{\boldsymbol{V}}(t) + \boldsymbol{D}_e\left[\boldsymbol{\Theta}, \boldsymbol{V}(t), \dot{\boldsymbol{V}}(t), t\right] + \boldsymbol{D}_h\left[\boldsymbol{\Theta}, \boldsymbol{V}(\tau), \dot{\boldsymbol{V}}(\tau), t; 0 \leq \tau \leq t\right] = \boldsymbol{F}(t);$$
$$\boldsymbol{V}(0) = \boldsymbol{V}_0, \dot{\boldsymbol{V}}(0) = \dot{\boldsymbol{V}}_0;\ t \geq 0 \quad (2)$$

where, a dot represents the derivative with respect to time. This semi-discretized equation of motion can be obtained from spatial discretization of the continuum model of the structure using the finite element method. Here, $\boldsymbol{V}(t)$ is the $n_d \times 1$ nodal displacement vector, $\mathbf{M}$ is the $n_d \times n_d$ mass matrix, $\boldsymbol{F}(t)$ is the applied excitation modelled as a $n_d \times 1$ vector-valued random process, $\boldsymbol{D}_e\left[\boldsymbol{\Theta}, \boldsymbol{V}(t), \dot{\boldsymbol{V}}(t), t\right]$ and $\boldsymbol{D}_h\left[\boldsymbol{\Theta}, \boldsymbol{V}(\tau), \dot{\boldsymbol{V}}(\tau), t; 0 \leq \tau \leq t\right]$ are $n_d \times 1$ vectors which model, respectively, the geometric and material nonlinearity in the structural system, and $\boldsymbol{\Theta}$ is a $n_S \times 1$ vector of system parameters modelled as random variables with joint PDF $p_{\boldsymbol{\Theta}}(\boldsymbol{\theta})$. The term $\boldsymbol{D}_h\left[\boldsymbol{\Theta}, \boldsymbol{V}(\tau), \dot{\boldsymbol{V}}(\tau), t; 0 \leq \tau \leq t\right]$ captures rate-independent memory effects and could also encompass stiffness and (or) strength degradation characteristics, which one may expect to occur due to the effect of cyclic loads. An effective means to model these inelastic effects is to represent $\boldsymbol{D}_h\left[\boldsymbol{\Theta}, \boldsymbol{V}(\tau), \dot{\boldsymbol{V}}(\tau), t; 0 \leq \tau \leq t\right]$ through a set of additional internal state variables [Wen 1976, Baber and Noori 1985, Mostaghel and Byrd 2000, and Triantafyllou and Koumousis 2014]. The excitation vector $\boldsymbol{F}(t)$, in the above equation, can represent a wide range of loading conditions, such as multi-component and differential support motions, surface tractions, and (or) parametric excitations. The components of $\boldsymbol{F}(t)$, in general, could be non-stationary, non-white, and (or) non-Gaussian in nature. When non-white excitations are considered, the forcing function can be obtained as the output of additional linear/nonlinear filters driven by white noise excitations. Similarly, the components of $\boldsymbol{\Theta}$ could be dependent and non-Gaussian. In addition, when random field models for the system parameters are adopted, it is assumed that the vector $\boldsymbol{\Theta}$ is obtained after discretizing the random fields into an equivalent set of random variables using the stochastic finite element analysis.

Within this framework, one can define an extended state vector $\boldsymbol{X}(t)$ that includes the structural system states $\boldsymbol{V}(t)$ and $\dot{\boldsymbol{V}}(t)$, the additional internal state variables of the hysteresis model, and the states associated with the augmented filter equations for representing excitations. The resulting governing equation can be cast as an Ito's SDE of the form

$$d\boldsymbol{X}(t) = \boldsymbol{A}[\boldsymbol{\Theta}, \boldsymbol{X}(t), t]dt + \boldsymbol{\sigma}[\boldsymbol{\Theta}, \boldsymbol{X}(t), t]d\boldsymbol{B}(t);$$
$$\boldsymbol{X}(0) = \boldsymbol{X}_0;\ t \geq 0 \tag{3}$$

Here, $\boldsymbol{X}(t)$ is the $p \times 1$ vector of system states, $\boldsymbol{A}[\boldsymbol{\Theta}, \boldsymbol{X}(t), t]$ is the $p \times 1$ drift vector, $\boldsymbol{\sigma}[\boldsymbol{\Theta}, \boldsymbol{X}(t), t]$ is the $p \times q$ diffusion coefficient matrix, and $\boldsymbol{B}(t)$ is a $q \times 1$ vector of correlated Brownian motion processes with $E_{\mathrm{P}}[\Delta\boldsymbol{B}(t)] = E_{\mathrm{P}}[\boldsymbol{B}(t + \Delta t) - \boldsymbol{B}(t)] = \boldsymbol{0}$ and $E_{\mathrm{P}}\left[\Delta B_i(t_1)\Delta B_j(t_2)\right] = C_{ij}\Delta t\delta(t_1 - t_2)$, for $t_1, t_2 \geq 0$ and $i, j = 1, \ldots, q$. Here, $C_{ij} = \rho_{ij}\sqrt{\mathfrak{I}_i\mathfrak{I}_j}$

is the $i,j$-th element of the $q \times q$ covariance matrix $\mathbf{C}$ and $\delta(\cdot)$ is the Dirac delta function. The quantities $\mathcal{I}_j, j = 1, \dots, q$, denote the intensities of the underlying Gaussian white noise processes, and the correlation coefficients satisfy the constraint $|\rho_{ij}| \leq 1$. For the special case when the components of $\boldsymbol{B}(t)$ are uncorrelated, the correlation coefficients are given by $\rho_{ij} = 1$ if $i = j$ and zero otherwise. Let $(\Omega, \mathcal{F}, \mathbb{P})$ be the underlying probability space, and $E_{\mathbb{P}}[\cdot]$ denote the expectation operator with respect to the probability measure $\mathbb{P}$.

In such systems, the problem of time-variant reliability analysis consists of determining the probability that a performance metric $h[\boldsymbol{\Theta}, \boldsymbol{X}(t)]$ stays within a permissible safe limit over a specified time duration. Here, $\boldsymbol{X}(t)$ is the response vector of the dynamical system, and $\boldsymbol{\Theta}$ denotes the vector of random system parameters. This problem of determining the probability of out-crossing the permissible safe limit by the structural dynamic response, as depicted in Figure 1, is commonly known as the first-passage problem. The performance measure could be in terms of the displacement, velocity, acceleration, or stress at any specified point in the structural system. The reliability of the dynamical system is given by

$$P_R = \mathbb{P}[\{h[\boldsymbol{\Theta}, \boldsymbol{X}(t)] < h^* \; \forall t \in [0, T]\}] \tag{4}$$

where $h^*$ is the permissible threshold value and $0 \leq t \leq T$ is the duration of random excitation.

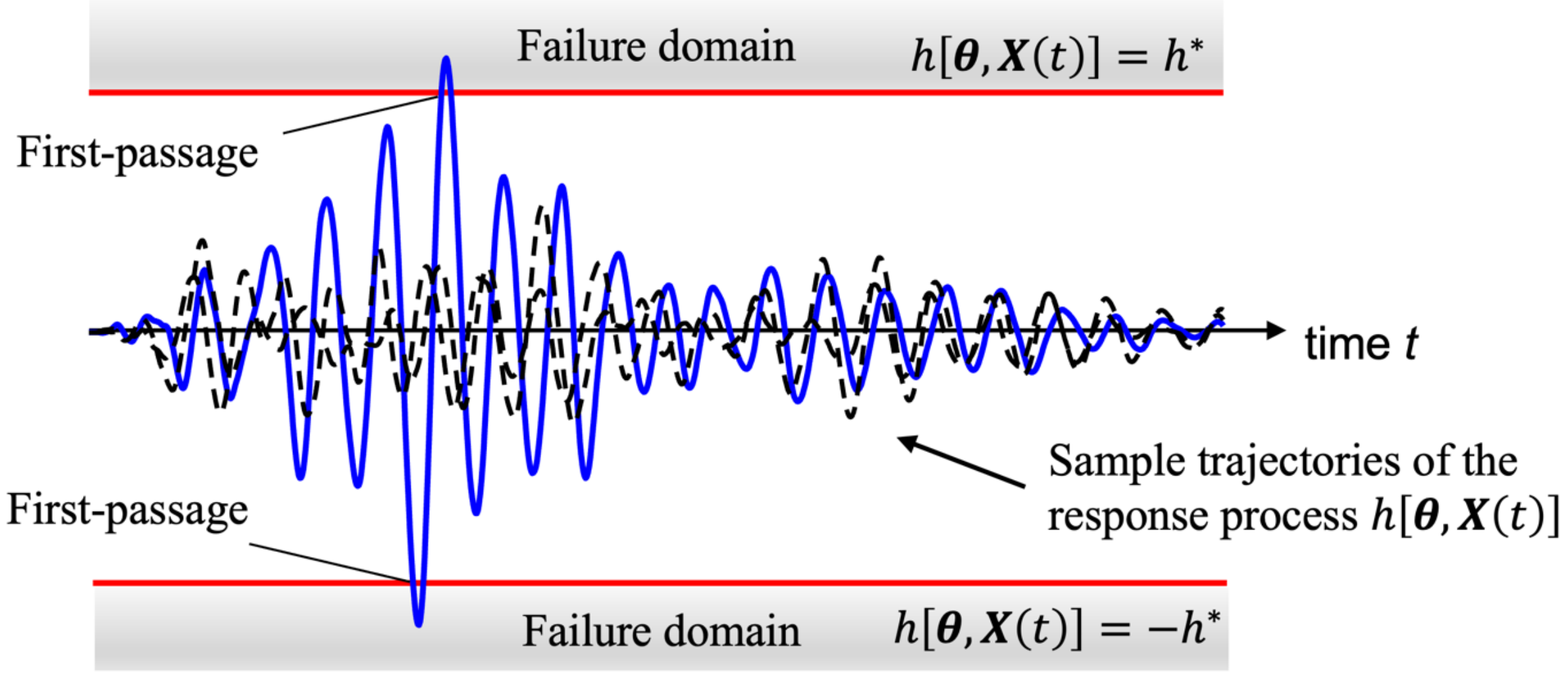

Figure 1: First-passage failure of a structural dynamic system.

The problem can be tackled by evaluating the probability that the maximum value of the response metric over the specified duration stays within the prescribed safe limits. Subsequently, Equation (4) can be re-expressed as

$$P_R = \mathbb{P}\left[\left\{\max_{0<t\leq T} h[\boldsymbol{\Theta}, \boldsymbol{X}(t)] < h^* \right\}\right] \quad (5)$$

The above re-formulation of the reliability problem in terms of the maximum value of the response converts the time dependent problem into a time-invariant format. The probability of failure, $P_F = 1 - P_R$, is then obtained as

$$P_F = \mathbb{P}[\{g[\boldsymbol{\Theta}, \boldsymbol{X}(t); 0 < t \leq T] \leq 0\}] = \mathbb{P}\left[\left\{h^* - \max_{0<t\leq T} h[\boldsymbol{\Theta}, \boldsymbol{X}(t)] < 0 \right\}\right], \quad (6)$$

where $g[\boldsymbol{\Theta}, \boldsymbol{X}(t); 0 < t \leq T] = h^* - \max_{0<t\leq T} h[\boldsymbol{\Theta}, \boldsymbol{X}(t)]$ is the associated limit-state function.

## 3.0 Analytical approximations

Exact analytical solution for the probability of failure is seldom possible due to several complicating features, such as presence of nonlinear and (or) implicitly defined performance functions, non-Gaussian and (or) non-stationary nature of the applied load, strong structural nonlinearity which makes the response non-Gaussian even for Gaussian inputs, presence of parametric excitations, and presence of multiple limit states as in system reliability problems among others. This section provides an overview of some of the commonly adopted approximate analytical methods for time-invariant and time-variant reliability problems.

### 3.1 First- and second- order reliability methods for time-invariant reliability analysis

The first-order reliability method (FORM) is formulated in the space of independent standard Gaussian random variables. The correlated non-Gaussian random vector $\boldsymbol{X}$ is transformed into a standard Gaussian vector $\boldsymbol{U}$ by applying the Nataf or the Rosenblatt transformations. Consequently, the limit-state function $g(\boldsymbol{X})$ also gets mapped into an equivalent function in the $\boldsymbol{U}$ space denoted by $G(\boldsymbol{U})$. An estimate of the probability of the failure event $F$ is obtained as [Madsen et al., 2006, Melchers and Beck 2018]

$$\hat{P}_F = 1 - \Phi(\beta_{HL}) \quad (7)$$

where the quantity $\beta_{HL} = \min_{G(\boldsymbol{u})=0} \sqrt{\boldsymbol{u}^{\mathrm{T}}\boldsymbol{u}}$, known as the Hasofer-Lind reliability index, represents the shortest distance from the origin to the limit state surface in the standard Gaussian space [Hasofer and Lind 1974]. The corresponding point on $G(\boldsymbol{U})$ closest to the origin is called the design point. Methods to solve the constrained optimization problem to locate the design point are widely developed [Haukaas and Der Kiureghian 2006, Breitung 2015, Wang et al., 2016, Keshtegar and Meng 2017]. It can be shown that evaluating $P_F$ according to the above equation is equivalent to replacing the nonlinear function $G(\boldsymbol{U})$ by its first-order Taylor series approximation about the design point, thereby linearizing the performance function.

When the limit state surface has significant curvature at the design point, an improvement to the FORM can be obtained by adopting a second-order Taylor series-based approximation of the limit state function about the design point. The resulting approach is known as the second-order reliability method (SORM) [Breitung 1984, Melchers and Beck 2018]. Several ways of constructing the quadratic surface have been studied in the literature [Der Kiureghian et al., 1987, Der Kiureghian and Stefano 1991, and Cai and Elishakoff 1994]. According to Breitung [1984], the probability of failure based on SORM can be obtained approximately as

$$\hat{P}_F = \Phi(-\beta_{HL}) \prod_{j=1}^{d} \left(1 + \beta_{HL}\kappa_j\right)^{-\frac{1}{2}} \tag{8}$$

where $\kappa_j; j = 1, \dots, d$ are the principal curvatures of the limit state surface about the design point, and $\beta_{HL}$ is the Hasofer-Lind reliability. It was shown by the author that the above estimate asymptotically approaches the true value of the failure probability as $\beta_{HL} \to \infty$. Further refinements to increase the accuracy of the SORM estimate for non-asymptotic conditions have been developed [Hohenbichler and Rackwitz 1988, Cai and Elishakoff 1994, Lee et al., 2012, Mansour and Olsson 2014, Hu and Du 2018].

### 3.2 Methods for time-variant reliability analysis

To evaluate the time-variant reliability by analytical approximation methods, it is convenient to first consider the probability of failure conditional on a specific realization $\boldsymbol{\Theta} = \boldsymbol{\theta}$ of the structural parameters. The unconditional probability of failure is then the expectation of the conditional failure probability with respect to the joint probability density function of $\boldsymbol{\Theta}$. This is mathematically expressed by rewriting Equation (6) as

$$P_F = \int_{\boldsymbol{\theta}} \mathbb{P}\left[\left\{h^* - \max_{0<t\leq T} h[\boldsymbol{\Theta}, \boldsymbol{X}(t)] \leq 0\right\} \middle| \boldsymbol{\Theta} = \boldsymbol{\theta}\right] p_{\boldsymbol{\Theta}}(\boldsymbol{\theta}) d\boldsymbol{\theta} = \int_{\boldsymbol{\theta}} P_{F|\boldsymbol{\Theta}}(\boldsymbol{\theta}) p_{\boldsymbol{\Theta}}(\boldsymbol{\theta}) d\boldsymbol{\theta} \quad (9)$$

The integral in Equation (9) can be evaluated by the fast integration technique, originally proposed by Wen and Chen [1987], which has been applied to determine the unconditional failure probability of linear and nonlinear structural dynamic systems [Madsen and Tvedt 1990, Chreng and Wen 1994, and Zayed et al., [2013]. The core idea here is to pose the evaluation of the expectation of the conditional failure probability function $P_{F|\boldsymbol{\Theta}}(\boldsymbol{\Theta})$ as an equivalent time-invariant reliability problem by introducing an auxiliary performance function, defined as

$$\tilde{G}(Z, \boldsymbol{V}) = Z - \Phi^{-1}\left[P_{F|\boldsymbol{\Theta}}\left(\Gamma^{-1}(\boldsymbol{V})\right)\right] \quad (10)$$

where $\boldsymbol{V} = \Gamma(\boldsymbol{\Theta})$ is the vector of standard Gaussian random variables and $\Gamma(\cdot)$ denotes the iso-probabilistic transformation that maps $\boldsymbol{\Theta}$ to the standard Gaussian space. $Z$ is an auxiliary standard Gaussian random variable that is independent of $\boldsymbol{V}$. The unconditional probability of failure, in terms of $\tilde{G}(Z, \boldsymbol{V})$, can thus be expressed as

$$P_F = \int_{\tilde{G}(z,\boldsymbol{v})\leq 0} p_{\boldsymbol{V}}(\boldsymbol{v}) p_Z(z) d\boldsymbol{v} dz, \quad (11)$$

Equation (11) is essentially a time-invariant reliability integral with limit-state function $\tilde{G}(Z, \boldsymbol{V}) = 0$, and can be evaluated using the methods in Section 4.1.

Another approach is the asymptotic approximation method proposed by Papadimitriou et al., [1997], where the idea is to rewrite the unconditional probability of failure as $P_F = \int_{\boldsymbol{\theta}} \exp[l(\boldsymbol{\theta})] d\boldsymbol{\theta}$, where $l(\boldsymbol{\theta}) = ln\, P_{F|\boldsymbol{\Theta}}(\boldsymbol{\theta}) + ln\, p_{\boldsymbol{\Theta}}(\boldsymbol{\theta})$. The resulting integral is then approximated by considering the Taylor series expansion of $l(\boldsymbol{\theta})$ about its maximizing point $\boldsymbol{\theta}^*$, and an estimate of the failure probability is obtained as

$$P_F \approx (2\pi)^{\frac{n_S}{2}} \frac{1}{\sqrt{det[\mathbf{L}(\boldsymbol{\theta}^*)]}} P_{F|\boldsymbol{\Theta}}(\boldsymbol{\theta}^*) \quad (12)$$

Here $\mathbf{L}(\boldsymbol{\theta}) = \{L_{jk}(\boldsymbol{\theta})\}$ is the Hessian matrix given by $L_{jk}(\boldsymbol{\theta}) = -\frac{\partial^2 l(\boldsymbol{\theta})}{\partial\theta_j \partial\theta_k}$ and $n_S$ is the dimension of the parameter vector $\boldsymbol{\Theta}$. A later study considers evaluating the unconditional failure probability in terms of the Taylor series expansion of the complementary conditional reliability function $P_{R|\boldsymbol{\Theta}}(\boldsymbol{\theta}) = 1 - P_{F|\boldsymbol{\Theta}}(\boldsymbol{\theta})$ about the expected value of $\boldsymbol{\Theta}$ (see Gupta and Manohar [2006])

$$P_F \approx 1 - P_{R|\boldsymbol{\Theta}}(\boldsymbol{\theta_m}) - \sum_{d=1}^{n_S} E_{p_{\boldsymbol{\Theta}}}\left[\Theta_d - \theta_{m,d}\right] \frac{dP_{R|\boldsymbol{\Theta}}(\boldsymbol{\theta_m})}{d\theta_d}$$

$$-\sum_{d_1=1}^{n_S} \sum_{d_2=1}^{n_S} E_{p_{\boldsymbol{\Theta}}}\left[\left(\Theta_{d_1} - \theta_{m,d_1}\right)\left(\Theta_{d_2} - \theta_{m,d_2}\right)\right] \frac{d^2 P_{R|\boldsymbol{\Theta}}(\boldsymbol{\theta_m})}{d\theta_{d_1} d\theta_{d_2}} \ldots, \quad (13)$$

where $\boldsymbol{\theta}_m = E_{p_{\boldsymbol{\Theta}}}[\boldsymbol{\Theta}]$ and, $\theta_d$ and $\theta_{m,d}$ denote the $d$-th dimension of $\boldsymbol{\Theta}$ and $\boldsymbol{\theta}_m$, respectively. Methods to evaluate the derivatives of $P_{R|\boldsymbol{\Theta}}(\boldsymbol{\theta})$, with respect to $\boldsymbol{\Theta}$, and the expectations in the above equation are discussed by Gupta and Manohar [2006].

### 3.3 Evaluation of the conditional time-variant failure probability

**Level-crossing approach:** In this approach, an estimate of the conditional failure probability is obtained by counting the number of times the response $h[\boldsymbol{\theta}, \boldsymbol{X}(t)]$ up-crosses the threshold $h^*$ in the time duration $[0, T]$. Let $\eta^+(h^*; 0, T, \boldsymbol{\theta})$ be the number of out-crossings from the safe domain. Under the assumption that $\eta^+(h^*; 0, T, \boldsymbol{\theta})$ has a Poisson distribution, an approximation to the failure probability can be obtained as [Soong and Grigoriu 1993, Lin and Cai 1995],

$$P_F = \mathbb{P}[\{\eta^+(h^*; 0, T, \boldsymbol{\theta}) > 0\}] = 1 - \mathbb{P}[\{\eta^+(h^*; 0, T, \boldsymbol{\theta}) = 0\}] \approx 1 - e^{-\int_0^T \nu^+(s; h^*, \boldsymbol{\theta}) ds} \quad (14)$$

The quantity $\nu^+(t; h^*, \boldsymbol{\theta})$ denotes the mean rate of up-crossing the threshold $h^*$ at time *t*.

The level-crossing rate can be determined by the well-known Rice's formula

$$\nu^+(h^*, t) = \int_0^{\infty} \dot{z} f(h^*, \dot{z}; t) d\dot{z}, \quad (15)$$

The general solution of Rice's formula has been derived for the case in which the response process $h[\boldsymbol{\theta}, \boldsymbol{X}(t)]$ is a Gaussian random process [Li and Melchers 1993, Firouzi et al., 2018]. When the response process is non-Gaussian, a solution of the Rice formula is possible only for specific cases, such as when $h[\boldsymbol{\theta}, \boldsymbol{X}(t)]$ is a lognormal process (see Li et al., [2016]) or a translation process obtained through a monotonic nonlinear transformation of a Gaussian process [Grigoriu 1984, Ferrante et al., 2005].

**Probability density evolution method:** A different approach to evaluate the time-variant reliability is to quantify the probability of failure based on the PDF of the maximum value of the response processes. The probability density evolution method considers the random excitation to be characterized in terms of a set of mutually independent random variables $\boldsymbol{\Theta}_E$ obtained through series representation methods, such as the Karhunen-Loeve expansion. The extreme value of the response, given by $\max_{0<t\leq T} h[\boldsymbol{\theta}, \boldsymbol{X}(t)] = \max_{0<t\leq T} Z[\boldsymbol{\Theta}_E, t] = M(\boldsymbol{\Theta}_E)$, will thus depend on $\boldsymbol{\Theta}_E$. In this approach, a virtual stochastic process given by $Y(\tau) = M(\boldsymbol{\Theta}_E)\tau$ is introduced such that $\dot{Y}(\tau) = M(\boldsymbol{\Theta}_E)$ [Chen and Li 2005]. Here $\tau$ is a virtual time parameter and $\dot{Y}(\tau)$ represents the derivative of $Y(\tau)$ with respect to $\tau$. The joint pdf of $(Y(\tau), \boldsymbol{\Theta}_E)$ is determined based on the probability density evolution function [Li and Chen 2009]

$$\frac{\partial p_{Y\boldsymbol{\Theta}_E}(y, \boldsymbol{\theta}_E; \tau)}{\partial \tau} + M(\boldsymbol{\Theta}_E)\frac{\partial p_{Y\boldsymbol{\Theta}_E}(y, \boldsymbol{\theta}_E; \tau)}{\partial y} = 0$$

$$p_{Y\boldsymbol{\Theta}_E}(y, \boldsymbol{\theta}_E; \tau)\Big|_{\tau=0} = \delta(y) p_{\boldsymbol{\Theta}_E}(\boldsymbol{\theta}_E) \quad (16)$$

from which the pdf of $Y(\tau)$ is obtained as $p_Y(y;\tau) = \int_{-\infty}^{\infty} p_{Y\boldsymbol{\Theta}_E}(y, \boldsymbol{\theta}_E; \tau) d\boldsymbol{\theta}_E$. It may be observed that the PDF of $M(\boldsymbol{\Theta}_E)$ is equal to $p_Y(y;1)$. The probability of the failure event $F = \{M(\boldsymbol{\Theta}_E) \geq h_i^*\}$ is subsequently estimated as $P_F = \int_{h^*}^{\infty} p_Y(y;\tau)|_{\tau=1} dy$. Enhancements of this method to capture response PDFs of high-dimensional nonlinear systems have been developed by Chen and Lyu [2022] and Lyu and Chen [2022].

### 3.4 Monte Carlo simulation

The essential idea in Monte Carlo simulation (MCS) methods is to digitally generate, on a computer, random samples of the uncertain load and structural parameters according to the prescribed probabilistic laws, and to solve the structural analysis problem for the ensemble using deterministic procedures. This leads to the generation of an ensemble of sample solutions to the problem, which are further processed statistically to obtain an estimate of the failure probability. In the most elementary form of MCS, called the direct Monte Carlo method, the random realizations of the basic uncertain variables are directly generated from the prescribed probabilistic model, and the probability of failure is estimated as a sample average of indicator functions by counting the number of times the limit state function gets violated. Thus, for time-invariant reliability problems, the direct Monte Carlo estimator for the reliability integral in Equation (1) is given by

$$\hat{P}_F^{DMCS} = \frac{1}{N}\sum_{k=1}^{N} I\{g(\boldsymbol{X}^k) \leq 0\} \quad (17)$$

where, $\boldsymbol{X}^k; k = 1, \dots, N$ are random samples of $\boldsymbol{X}$ that conform to the joint PDF $p_{\boldsymbol{X}}(\boldsymbol{x})$ and $N$ is the sample size. Similarly, for time-variant problems, the failure probability estimator is given by

$$\begin{aligned} \hat{P}_F^{DMCS} &= \frac{1}{N}\sum_{k=1}^{N} I\{g[\boldsymbol{\Theta}^k, \boldsymbol{X}^k(t); 0 < t \leq T] \leq 0\} \\ &= \frac{1}{N}\sum_{k=1}^{N} I\left\{h^* - \max_{0<t\leq T} h[\boldsymbol{\Theta}^k, \boldsymbol{X}^k(t)] \leq 0\right\} \end{aligned} \tag{18}$$

where $\boldsymbol{\Theta}^k; k = 1, \dots, N$ are sample realizations of the structural system parameters obtained as random draws from $p_{\boldsymbol{\Theta}}(\boldsymbol{\theta})$, and $\boldsymbol{X}^k(t); k = 1, \dots, N$ are random realizations of the dynamical system states obtained by solving the associated governing differential equations.

The estimators in the above equations are unbiased for a finite sample size $N$ and have a coefficient of variation given by

$$\delta_{\hat{P}_F^{DMCS}} = \sqrt{\frac{P_F(1-P_F)}{NP_F}} \tag{19}$$

Furthermore, by the central limit theorem, the sampling distribution of $\hat{P}_F^{DMCS}$ approaches a normal distribution with mean value $P_F$ and variance $\frac{P_F(1-P_F)}{NP_F}$ as $N$ increases. Based on this result, one can obtain confidence intervals associated with an estimate of $P_F$ [Ang and Tang 2007]. Thus, although the Monte Carlo estimate for the probability of failure is also approximate in nature for a finite sample size, it is possible to establish a measure of acceptability of the solution in terms of the statistical properties of the estimator, such as the confidence intervals or the coefficient of variation. There is no counterpart to this for the analytical methods discussed earlier.

Despite these advantages of direct Monte Carlo simulation, a major limitation of the method lies in efficient estimation of rare event probabilities, such as the failure probability in well-designed structures. To estimate $P_F$ of the order of $10^{-6}$, a sample size as large as $10^6$ might not always succeed in simulating a single realization in the failure domain, as only 1 out of $10^6$ samples is expected to fail. In addition, to achieve a target $\delta_{\hat{P}_F^{DMCS}}$ of 10%, at least $10^8$ samples would be needed. While the accuracy can indeed be gained by increasing $N$, the improved estimates come at the expense of additional computational time, which can be demanding. Hence, recent research on MCS-based methods for structural reliability analysis has focused on the development of the so-called sampling variance reduction techniques, which aim to devise alternative estimators for $P_F$ that control the sampling variance without considerably increasing the sample size.

# 4.0 Importance Sampling

### 4.1 Importance sampling methods for time-invariant reliability analysis

Importance sampling has been one of the most widely used approaches for variance reduction in simulation-based structural reliability analysis. The underlying concept here is to draw samples of the random vector $\boldsymbol{X}$ from an auxiliary sampling density $q_{\boldsymbol{X}}(\boldsymbol{x})$ that can generate more samples in the failure domain in comparison to the nominal density $p_{\boldsymbol{X}}(\boldsymbol{x})$. Accordingly, the reliability integral in Equation (1) is rewritten as [Rubinstein and Kroese 2016]

$$P_F = \int_{-\infty}^{\infty} I\{g(\boldsymbol{x}) \leq 0\} \frac{p_X(x)}{q_X(x)} q_{\boldsymbol{X}}(\boldsymbol{x}) d\boldsymbol{x} = E_{q_X}\left[I\{g(\boldsymbol{X}) \leq 0\} \frac{p_X(\boldsymbol{X})}{q_X(\boldsymbol{X})}\right], \tag{20}$$

where $q_{\boldsymbol{X}}(\boldsymbol{x})$ is termed as the importance sampling probability density function (ISPDF) and $E_{q_X}[\cdot]$ is the mathematical expectation with respect to $q_{\boldsymbol{X}}(\boldsymbol{x})$. The ISPDF should be such that $q_{\boldsymbol{X}}(\boldsymbol{x}) > 0 \; \forall \; \boldsymbol{x}$ for which $\boldsymbol{p}_{\boldsymbol{X}}(\boldsymbol{x}) > 0$. An estimator for $P_F$ based on the above equation is obtained as

$$\hat{P}_F = \frac{1}{N} \sum_{k=1}^{N} I\{g(\boldsymbol{x}^k) \leq 0\} \frac{p_X(x^k)}{q_X(x^k)}, \tag{21}$$

where $\boldsymbol{X}^k; k = 1, \dots, N$ are random samples distributed according to $q_{\boldsymbol{X}}(\boldsymbol{x})$. It can be shown that the above estimator is unbiased, i.e., $E_{q_X}[\hat{P}_F] = P_F$, and has a variance given by

$$\begin{aligned} \text{Var}[\hat{P}_F] &= \frac{1}{N} \text{Var}_{q_X}\left[I\{g(\boldsymbol{X}) \leq 0\} \frac{p_X(\boldsymbol{X})}{q_X(\boldsymbol{X})}\right] \\ &= \frac{1}{N} E_{q_X}\left[\left(I\{g(\boldsymbol{X}) \leq 0\} \frac{p_X(\boldsymbol{X})}{q_X(\boldsymbol{X})} - P_F\right)^2\right] \end{aligned} \tag{22}$$

It is straightforward to show that the sampling variance becomes zero when

$$q_{\boldsymbol{X}}(\boldsymbol{x}) = q_{\boldsymbol{X}}^{opt}(\boldsymbol{x}) = \frac{I\{g(x) \leq 0\} p_X(x)}{P_F}, \tag{23}$$

where $q_{\boldsymbol{X}}^{opt}(\boldsymbol{x})$ represents the ideal choice of the ISPDF. However, it is evident that this optimal ISPDF is ineffective in practice as its determination requires prior knowledge of $P_F$, the very quantity to be estimated. Hence, in practical applications of importance sampling methods, sub-optimal approximations of the ideal ISPDF are obtained and used. The existing importance sampling schemes [Engelund and Rackwitz 1993, Schueller et al., 2004, Tabandeh et al., 2022] can be broadly classified into two groups: (1) FORM-based importance sampling, and (2) adaptive importance sampling.

### 4.1.1 FORM-based importance sampling

In FORM-based importance sampling, the idea is to locate the ISPDF $q_{\boldsymbol{X}}(\boldsymbol{x})$ around the design point of the limit-state function $g(\boldsymbol{x}) = 0$, where the likelihood of failure is high. The early works suggested that $q_{\boldsymbol{X}}(\boldsymbol{x})$ be uniform [Shinozuka 1983] or have a Gaussian distribution [Schueller and Stix 1987, Fujita and Rackwitz 1988] with mean located at the design point. In addition to a global design point, if the limit-state function has multiple local design points, a (weighted) mixture of ISPDFs centred at the different design points can be considered [Fujita and Rackwitz 1988, Melchers 1989], in which case $q_{\boldsymbol{X}}(\boldsymbol{x})$ is mathematically given by

$$q_{\boldsymbol{X}}(\boldsymbol{x}) = \sum_{i=1}^{n_{dp}} w_i q_{\boldsymbol{X}}^i(\boldsymbol{x}); \quad w_i = \frac{\Phi(\beta^i)}{\sum_{j=1}^{n_{dp}} \Phi(\beta^j)} \tag{24}$$

In the above equation, the PDF $q_{\boldsymbol{X}}^i(\boldsymbol{x})$ denotes the mode of the ISPDF $q_{\boldsymbol{X}}(\boldsymbol{x})$ centred at the $i$-th design point with reliability index $\beta^i$, and $w_i$ is the weight of the $i$-th mode. The studies by Grooteman [2008, 2011] have considered a radial sampling scheme to obtain an effective ISPDF. Here, a hypersphere of radius $\beta$ is chosen, and the domain of the ISPDF is restricted to samples that lie outside this sphere. This idea was initially proposed by Harbitz [1986], where the author had taken $\beta$ to be equal to the FORM-based reliability index. The work by Grooteman [2008] has developed a robust scheme that iteratively determines the optimal value of $\beta$ using a line search algorithm, thus eliminating the need to perform a FORM analysis. A strategy to further improve the efficiency of the method by using a response surface model within the importance sampling framework was later suggested by the author [Grooteman 2011]. Here, the $\beta$-sphere and the response surface are constantly updated during the line search process until the optimum value of $\beta$ is reached. The study by Thedy and Liao [2021] proposes constructing multiple hyperspheres with different centres and radii in order to exclude even more samples that lie in the safe regions, which would not have been possible with a single hypersphere centred at the origin. Wei et al., [2023a] proposed an eccentric radial-based importance sampling, which is an improvement on the method proposed by Thedy and Liao [2021]. Meng et al., [2023] have proposed a strategy based on creating hyperspheres, called bubbles, in both safe and failure regions so that the performance function need not be called for samples lying within the same bubble.

### 4.1.2 Adaptive importance sampling

In adaptive importance sampling procedures, the philosophy is to iteratively update the sampling density function based on intermediate results obtained from a few rounds of pre-sampling. The early studies in this direction suggested an adaptive algorithm where the ISPDF is updated based on statistical moments estimated from pre-samples [Bucher 1988]. The idea here is to construct an ISPDF $q_{\boldsymbol{X}}(\boldsymbol{x})$ such that the mean and covariance of $q_{\boldsymbol{X}}(\boldsymbol{x})$ closely match those of the ideal ISPDF $q_{\boldsymbol{X}}^{opt}(\boldsymbol{x})$, i.e., $q_{\boldsymbol{X}}(\boldsymbol{x})$ should satisfy the condition

$$E_{q_X}[\boldsymbol{X}] = E_{p_X}[\boldsymbol{X}|g(\boldsymbol{X}) \leq 0]$$
$$E_{q_X}[\boldsymbol{X}^{\mathrm{T}}\boldsymbol{X}] = E_{p_X}[\boldsymbol{X}^{\mathrm{T}}\boldsymbol{X}|g(\boldsymbol{X}) \leq 0] \quad (25)$$

To start the iterations, a sample point $\bar{\boldsymbol{x}}_q$ lying in the failure domain is obtained and a Gaussian ISPDF centred at $\bar{\boldsymbol{x}}_q$, with an assumed covariance matrix $\mathbf{C}_q$, is constructed. A limited amount of simulation is then carried out, and the mean and covariance of the samples lying in the failure domain are estimated. These estimates are then used to update the mean and covariance matrix of $q_{\boldsymbol{X}}(\boldsymbol{x})$ for further sampling. Over the years, more refined adaptive approaches to approximate the optimal ISPDF have been developed, which are summarized below.

**Kernel-density approximation:** This approach uses kernel methods to approximate the optimal ISPDF. The methodology employs the generation of a set of samples of $\boldsymbol{X}$, $\widetilde{\boldsymbol{x}}^k; k = 1, \ldots, M$, belonging to the failure domain $F = \{g(\boldsymbol{X}) \leq 0\}$. The failure samples can be obtained either by standard rejection sampling from the original PDF $p_{\boldsymbol{X}}(\boldsymbol{x})$, or by the Markov chain Monte Carlo (MCMC) method as the intermediate states of a Markov chain whose stationary distribution is taken to be the ideal ISPDF. This idea was initially studied by Ang et al., [1992] and Au and Beck [1999], and later improved by Dai et al., [2012], where an approximation to the ideal ISPDF is obtained as

$$q_{\boldsymbol{X}}(\boldsymbol{x}) = \sum_{j=1}^{L} \sum_{k=1}^{M} \alpha_k^{(1)} k_1(\boldsymbol{x}, \widetilde{\boldsymbol{x}}^k) + \cdots + \alpha_k^{(l)} k_l(\boldsymbol{x}, \widetilde{\boldsymbol{x}}^k), \quad (26)$$

where $k_j(\boldsymbol{x}, \tilde{\boldsymbol{x}}^k)$ is a Gaussian kernel function given by

$$k_j(\boldsymbol{x}, \tilde{\boldsymbol{x}}^k) = \frac{1}{(\sqrt{2\pi}\sigma_j)^n} exp\left(\frac{1}{2\sigma_j^2}(\boldsymbol{x} - \tilde{\boldsymbol{x}}^k)^T(\boldsymbol{x} - \tilde{\boldsymbol{x}}^k)\right) \quad (27)$$

Here, $\sigma_j; j = 1, \dots, l$ are the bandwidth parameters and $\alpha_k^{(j)}; k = 1, \dots, M; j = 1, \dots, L$ are normalized weights satisfying $\sum_{k=1}^{M}\sum_{j=1}^{L}\alpha_k^{(j)} = 1$. These parameters are determined by solving a quadratic optimization problem which ensures that the cumulative distribution function of the fitted ISPDF matches the empirical CDF of the samples $\tilde{\boldsymbol{x}}^k; k = 1, \dots, M$. An alternative approach to construct the ISPDF is the wavelet density estimation technique [Dai et al., 2015], which involves construction of the importance sampling density by means of wavelet orthogonal bases using samples from the failure region generated through an MCMC simulation scheme.

**Principle of maximum entropy:** Selection of ISPDF based on the principle of maximum entropy has been proposed by Dai et al., [2016]. Here, a collection of samples, asymptotically distributed according to the ideal ISPDF, and denoted as $\tilde{\boldsymbol{x}}^k = \{\tilde{x}_1^k \quad \cdots \quad \tilde{x}_n^k\}^{\mathrm{T}}; k = 1, \dots, M$, is generated from the failure domain $F$ using MCMC simulation. The sub-optimal ISPDF $q_{\boldsymbol{X}}(\boldsymbol{x})$ is then taken to be the maximum entropy PDF that is consistent with the sampled dataset and also satisfies some specified moment constraints. Mathematically, this translates to finding $q_{\boldsymbol{X}}(\boldsymbol{x})$ that maximizes the Shannon entropy of $q_{\boldsymbol{X}}(\boldsymbol{x})$, given by

$$S[q_{\boldsymbol{X}}(\boldsymbol{x})] = -\int_{\mathbb{R}^n} q_{\boldsymbol{X}}(\boldsymbol{x}) \log[q_{\boldsymbol{X}}(\boldsymbol{x})]\, d\boldsymbol{x} \quad (28)$$

and satisfies the constraints

$$\int_{\mathbb{R}^n} \prod_{l=1}^{n} x_l^{\alpha_l}\, q_{\boldsymbol{X}}(\boldsymbol{x}) d\boldsymbol{x} = \hat{\mu}_\alpha \quad (29)$$

where $\hat{\mu}_\alpha = \frac{\sum_{k=1}^{M}\prod_{l=1}^{n}\left(\tilde{x}_l^k\right)^{\alpha^l}}{M}$ is an estimate of the joint moment computed from the failure samples of $\boldsymbol{X}$. In Equation (29), $\alpha^l; l = 1, \dots, n$ are non-negative integers describing the joint moment being computed and $\alpha = \sum_{l=1}^{n}\alpha^l$ denotes the total order of the moment. The method begins by first approximating the maximum entropy PDF by a multi-dimensional histogram constructed from the samples $\tilde{\boldsymbol{x}}^k = \{\tilde{x}_1^k \quad \cdots \quad \tilde{x}_n^k\}^{\mathrm{T}}; k = 1, \dots, M$. The conditional PDF of the histogram in each dimension is then represented as a series expansion of orthogonal Legendre polynomials. The coefficients of the series expansion in the different dimensions are selected to ensure that the approximate maximum entropy density $q_{\boldsymbol{X}}(\boldsymbol{x})$, reconstructed from the one-dimensional conditional PDFs, satisfies the empirical cumulative distribution of the samples $\tilde{\boldsymbol{x}}^k; k = 1, \dots, M$ and also the moment constraint in equation (29).

**The cross-entropy method:** The cross-entropy method (CE) for rare event estimation [Rubinstein and Kroese 2016] has been applied to select a near-optimal ISPDF for time-invariant structural reliability analysis [Kurtz and Song 2013, Wang and Song 2016]. These methods typically solve the reliability estimation problem in the space of standard Gaussian random variables $\boldsymbol{U} = \{U_1, \dots, U_n\}^{\mathrm{T}}$. The mapping from $\boldsymbol{X}$ to $\boldsymbol{U}$ is defined by an iso-probabilistic transformation $\boldsymbol{U} = \Gamma_X(\boldsymbol{X})$ and the limit-state function in the standard Gaussian space is given by $G(\boldsymbol{U}) = g\left(\Gamma_{\boldsymbol{X}}^{-1}(\boldsymbol{U})\right)$. The generic idea in cross-entropy-based importance sampling is to consider a parameterized family of ISPDFs $q_{\boldsymbol{U}}(\boldsymbol{u}; \boldsymbol{\lambda})$, which are indexed by the parameter vector $\boldsymbol{\lambda}$. The optimal parameter $\boldsymbol{\lambda}^*$ is determined such that $q_{\boldsymbol{U}}(\boldsymbol{u}; \boldsymbol{\lambda}^*)$ closely approximates the ideal ISPDF. This is achieved by minimizing a measure of distance between $q_{\boldsymbol{U}}(\boldsymbol{u}; \boldsymbol{\lambda})$ and $q_{\boldsymbol{U}}^{opt}(\boldsymbol{u})$, known as the Kullback-Leibler divergence, given by

$$D_{KL}\left(q_{\boldsymbol{U}}^{opt}(\boldsymbol{u}), q_{\boldsymbol{U}}(\boldsymbol{u}; \boldsymbol{\lambda})\right) = E_{q_{\boldsymbol{U}}^{opt}}\left[\ln \frac{q_{\boldsymbol{U}}^{opt}(\boldsymbol{U})}{q_{\boldsymbol{U}}(\boldsymbol{U}; \boldsymbol{\lambda})}\right], \tag{30}$$

where

$$q_{\boldsymbol{U}}^{opt}(u) = \frac{1}{P_F} I\{G(\boldsymbol{u}) \leq 0\} \phi_n(\boldsymbol{u}) \tag{31}$$

is the optimal IS density corresponding to the limit state function $G(\boldsymbol{U}) = 0$ and $\phi_n(\boldsymbol{u})$ is the $n$-dimensional standard Gaussian PDF. The optimal parameter $\boldsymbol{\lambda}^*$ is obtained by minimizing $D_{KL}\left(q_{\boldsymbol{U}}^{opt}(\boldsymbol{u}), q_{\boldsymbol{U}}(\boldsymbol{u}; \boldsymbol{\lambda})\right)$, which leads to the CE optimization problem

$$\boldsymbol{\lambda}^* = \underset{\boldsymbol{\lambda}}{\operatorname{argmax}}\, E_{\phi_n}\left[I\{G(\boldsymbol{U}) \leq 0\} \ln\left(q_{\boldsymbol{U}}(\boldsymbol{U}; \boldsymbol{\lambda})\right)\right]. \tag{32}$$

The expectation in Equation (32) is estimated using a set of samples from $\phi_n(\boldsymbol{u})$ to get an estimate $\hat{\boldsymbol{\lambda}}$ of $\boldsymbol{\lambda}^*$. If the failure event $F = \{\boldsymbol{u} \in \mathbb{R}^n : G(\boldsymbol{u}) \leq 0\}$ has a small probability of occurrence, a large number of samples are required to get a sufficient number of non-zero indicators for a good sample approximation.

The studies by Kurtz and Song [2013] and Wang and Song [2016] address this problem by defining a sequence of intermediate failure events $F_l = \{\boldsymbol{u} \in \mathbb{R}^n: G(\boldsymbol{u}) \leq \gamma_l\}; l = 1, \dots, L$. The parameters $\gamma_l$ satisfy $\gamma_1 > \gamma_2 > \cdots > \gamma_L = 0$, such that they define a sequence of nested failure events approaching the target failure event $F$. The CE optimization problem is then solved iteratively for the optimal ISPDF of the intermediate reliability problems, resulting in a sequence of parameter vectors such that the final parameter vector gives a good approximation of $\boldsymbol{\lambda}^*$. The optimal ISPDF corresponding to the $l$-th intermediate failure event is given by

$$q_U^l(\boldsymbol{u}) = \frac{1}{P_{F_l}} I\{G(\boldsymbol{u}) \leq \gamma_l\}\phi_n(\boldsymbol{u}), \tag{33}$$

where $P_{F_l}$ denotes the probability of the failure event $F_l$.

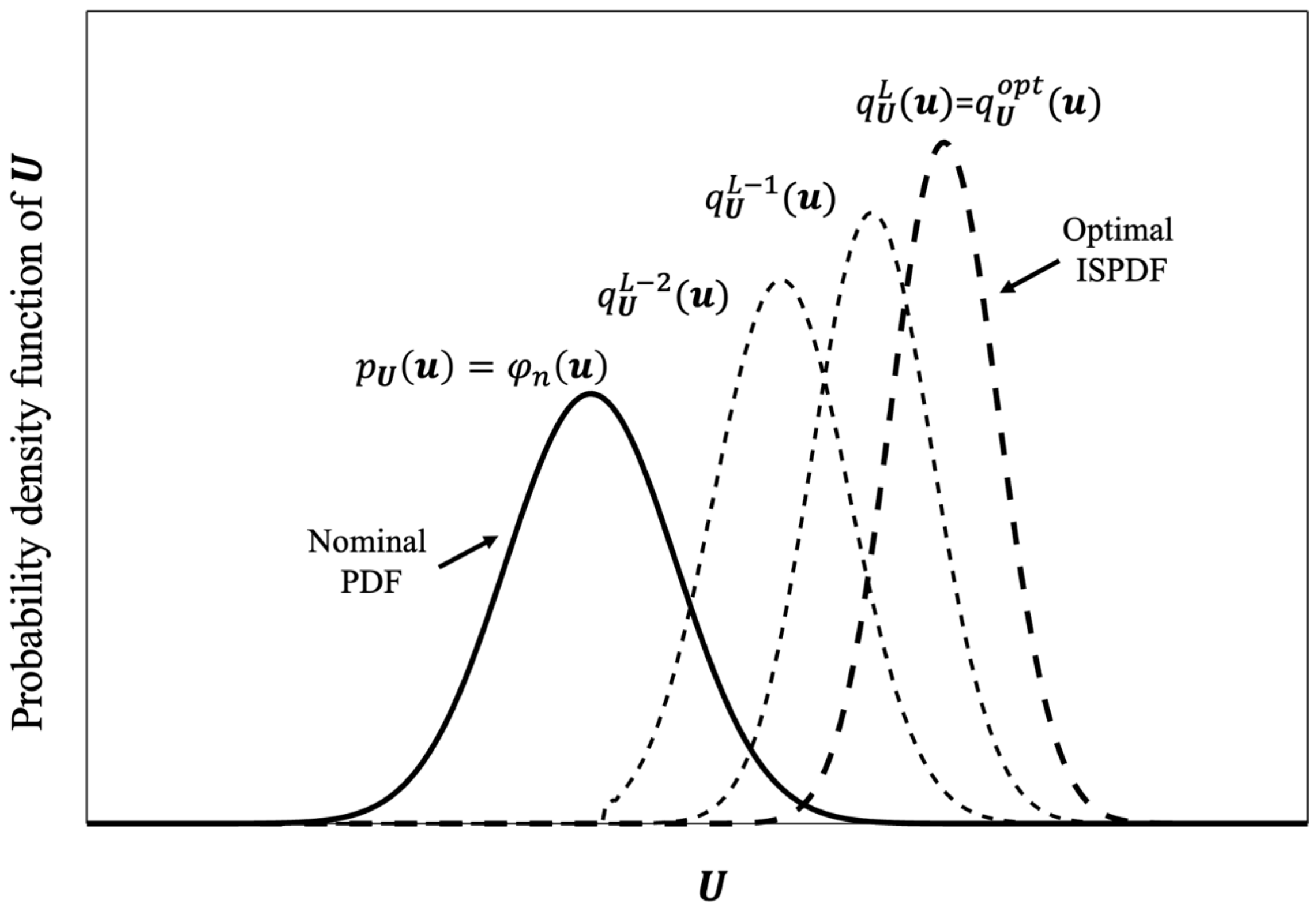

Figure 2: Sequence of intermediate densities in the cross-entropy method, defining a gradual transition between the nominal PDF (dominant in the safe domain) to the optimal ISPDF (dominant in the rare failure domain) (adapted from Kanjilal et al., (2021)].

An alternative approach is suggested by Papaioannou et al., [2019], which employs a gradual transition to the optimal ISPDF by substituting the indicator function of the failure event $F$ with a smooth approximation, given by

$$I\{G(\boldsymbol{u}) \leq 0\} = \lim_{\sigma \to 0} \Phi\left(-\frac{G(\boldsymbol{u})}{\sigma}\right), \tag{34}$$

where $\Phi(\cdot)$ is the standard Gaussian cumulative distribution function. Using this approximation, a sequence of target ISPDFs can be defined as

$$q_{\boldsymbol{U}}^{l}(\boldsymbol{u}) = \frac{1}{P_l}\Phi\left(-\frac{G(\boldsymbol{u})}{\sigma_l}\right)\phi_n(\boldsymbol{u}); l = 1, \dots, L, \tag{35}$$

where $P_l$ denotes the normalizing constant and $\infty > \sigma_1 > \sigma_2 > \cdots > \sigma_L > 0$. The PDF sequence, depicted in Figure 2 gradually approaches $q_{\boldsymbol{U}}^{opt}(\boldsymbol{u})$ as the parameter values $\sigma_l, l \geq 1$ decrease to 0.

Using the PDF sequence in Equations (33) or (35) within the CE method, one arrives at the following stochastic optimization problem to be solved at each intermediate level:

$$\boldsymbol{\lambda}_l = \underset{\boldsymbol{\lambda}}{\operatorname{argmax}}\, E_{\phi_n}\left[H_l(\boldsymbol{U}) \ln\left(q_{\boldsymbol{U}}(\boldsymbol{U}; \boldsymbol{\lambda})\right)\right], \tag{36}$$

where $H_l(\boldsymbol{U})$ is equal to either $I\{G(\boldsymbol{U}) \leq \gamma_l\}$ or $\Phi\left(-\frac{G(\boldsymbol{U})}{\sigma_l}\right)$. To obtain a good sample approximation of the objective function, the expectation in Equation (36) is estimated by importance sampling from $q_{\boldsymbol{U}}(\boldsymbol{u}; \hat{\boldsymbol{\lambda}}_{l-1})$, $\hat{\boldsymbol{\lambda}}_{l-1}$ being the estimate of $\boldsymbol{\lambda}_{l-1}$ in the previous step. Thus, the sample counterpart of the CE optimization problem at each level is given by

$$\hat{\boldsymbol{\lambda}}_l = \underset{\boldsymbol{\lambda}}{\operatorname{argmax}} \frac{1}{N_{CE}} \sum_{k=1}^{N_{CE}} \frac{\phi_n(\boldsymbol{u}^{l-1,k})}{q_{\boldsymbol{U}}(\boldsymbol{u}^{l-1,k}; \hat{\boldsymbol{\lambda}}_{l-1})} H_l(\boldsymbol{u}^{l-1,k}) \ln\left(q_{\boldsymbol{U}}(\boldsymbol{u}^{l-1,k}; \boldsymbol{\lambda})\right), \tag{37}$$

where $\boldsymbol{u}^{l-1,k}; k = 1, \dots, N_{CE}$ are independent samples drawn from $q_{\boldsymbol{U}}(\boldsymbol{u}; \hat{\boldsymbol{\lambda}}_{l-1})$. The probability of failure is estimated based on the last fitted density $q_{\boldsymbol{U}}(\boldsymbol{u}; \hat{\boldsymbol{\lambda}}_L)$ and the corresponding estimator is given by

$$\hat{P}_F = \frac{1}{N} \sum_{k=1}^{N} I\{G(\boldsymbol{u}^k) \leq 0\} \frac{\phi_n(\boldsymbol{u}^{L,k})}{q_{\boldsymbol{U}}(\boldsymbol{u}^{L,k}; \hat{\boldsymbol{\lambda}}_L)} \tag{38}$$

In order to get a good estimate of $\boldsymbol{\lambda}_l$ with a small sample size $N_{CE}$, the difference between the sampling PDF $q_{\boldsymbol{U}}(\boldsymbol{u}; \hat{\boldsymbol{\lambda}}_{l-1})$ and the target PDF $q_{\boldsymbol{U}}^{l}(\boldsymbol{u})$ should be small. This is ensured by selecting the parameters $\gamma_l$ or $\sigma_l$ through an adaptive mechanism [Kurtz and Song 2013, Papaioannou et al., 2019]. The objective function in Equation (37) is typically convex and the parameter vector $\hat{\boldsymbol{\lambda}}_l$ at each level can be obtained by equating the gradient of the objective function, with respect to $\boldsymbol{\lambda}$, to zero. Closed form update rules are derivable when $q_{\boldsymbol{U}}(\boldsymbol{u};\boldsymbol{\lambda})$ is selected as the unimodal Gaussian PDF [Kurtz and Song 2013] or the von-Mises Fisher Nakagami PDF [Wang and Song 2016, Papaioannou et al., 2019]. When a mixture distribution is considered as the parametric density family, the parameter update is accomplished by means of the expectation-maximization algorithm [Dempster et al., 1977, Geyer et al. 2019].

**Sequential importance sampling:** The sequential importance sampling method, developed by Papaioannou et al., [2016], is similar in spirit to the CE method, wherein a sequence of intermediate target PDFs are sampled adaptively, until the optimal ISPDF is reached. Specifically, the density sequence in Equation (35) is considered, and the parameters $\sigma_1, \dots, \sigma_L$ are selected adaptively such that the difference between the consecutive PDFs is small. One may observe that the normalizing constant $P_l$ can be expressed as

$$P_l = \int_{\boldsymbol{u}\in\mathbb{R}^n} \eta_l(\boldsymbol{u}) d\boldsymbol{u} = P_{l-1} \int_{\boldsymbol{u}\in\mathbb{R}^n} \frac{\eta_l(\boldsymbol{u})}{\eta_{l-1}(\boldsymbol{u})} q_{\boldsymbol{U}}^{l-1}(\boldsymbol{u}) d\boldsymbol{u}\,, \tag{39}$$

where $\eta_l(\boldsymbol{u}) = \Phi\left(-\frac{G(\boldsymbol{u})}{\sigma_l}\right)\phi_n(\boldsymbol{u})$, which implies that the ratio of the normalizing constant of two consecutive PDFs, denoted as

$$S_l = \frac{P_l}{P_{l-1}} = E_{q_{\boldsymbol{U}}^{l-1}}\left[\frac{\eta_l(\boldsymbol{U})}{\eta_{l-1}(\boldsymbol{U})}\right], \tag{40}$$

can be estimated with samples from the PDF $q_{\boldsymbol{U}}^{l-1}(\boldsymbol{u})$.

Noting that the probability of the failure event $F$ is the normalizing constant of the optimal ISPDF $q_{\boldsymbol{U}}^{opt}(\boldsymbol{u})$ in Equation (31), an estimator of the failure probability $P_F$ is obtained as

$$\hat{P}_F = \frac{\hat{P}_L}{N} \sum_{k=1}^{N} I\{G(\boldsymbol{u}^{L,k}) \leq 0\} \frac{\phi_n(\boldsymbol{u}^{L,k})}{\eta_L(\boldsymbol{u}^{L,k})}, \tag{41}$$

where $\boldsymbol{u}^{L,k}; k = 1, \dots, N_{SIS}$ are independent samples from $q_{\boldsymbol{U}}^{L}(\boldsymbol{u})$, the last PDF in the sequence, and $\hat{P}_L = \prod_{l=2}^{L} \hat{S}_l$.

The estimate $\hat{S}_l$ is obtained through sample approximation of the expectation in Equation (40),

$$\hat{S}_l = \frac{1}{N_{SIS}} \sum_{k=1}^{N_{SIS}} \frac{\eta_l(\boldsymbol{u}^{l-1,k})}{\eta_{l-1}(\boldsymbol{u}^{l-1,k})}, \tag{42}$$

where $\boldsymbol{u}^{l-1,k}; k = 1, \dots, N_{SIS}$ are independent samples from $q_{\boldsymbol{U}}^{l-1}(\boldsymbol{u})$. The sequence of PDFs $q_{\boldsymbol{U}}^{l}(\boldsymbol{u}); l \geq 1$ are sampled using the resample-move scheme, starting from samples from $q_{\boldsymbol{U}}^{1}(\boldsymbol{u}) = \phi_n(\boldsymbol{u})$. At any intermediate level, the samples from $q_{\boldsymbol{U}}^{l-1}(\boldsymbol{u})$ are weighted according to the next PDF in the sequence, using the weights $w_l(\boldsymbol{u}) = \frac{\eta_l(\boldsymbol{u})}{\eta_{l-1}(\boldsymbol{u})}$, and subsequently resampled to get unweighted samples distributed according to $q_{\boldsymbol{U}}^{l}(\boldsymbol{u})$. These samples are then propagated to the dominant regions in the support of $q_{\boldsymbol{U}}^{l}(\boldsymbol{u})$ using MCMC moves.

**Directional importance sampling:** In this category of methods, the probability of failure is estimated by importance sampling in the polar coordinates of the $\boldsymbol{U}$-space, given by $\boldsymbol{U} = R\boldsymbol{A}$. The sample pair $(r, \boldsymbol{a})$ represents the radius and direction of a corresponding sample $\boldsymbol{u}$ in the standard Gaussian space. The reliability integral in polar coordinates is expressed as

$$P_F = \int_{\mathcal{S}^{n-1}} \int_0^{\infty} I\{G(r\boldsymbol{a}) \leq 0\} p_R(r) p_{\boldsymbol{A}}(\boldsymbol{a}) dr d\boldsymbol{a}, \tag{43}$$

where $p_R(r)$ is the PDF of the radius vector such that $R^2 = \|\boldsymbol{U}\|_2^2$ follows the $\chi^2$- distribution with $n$ degrees of freedom, and $p_{\boldsymbol{A}}(\boldsymbol{a})$ is the PDF of the direction vector $\boldsymbol{A} = \boldsymbol{U}/\|\boldsymbol{U}\|_2$, given by the uniform distribution on the $n$-dimensional unit hypersphere $\mathcal{S}^{n-1}$. The integral over the support of $p_R(r)$ can be evaluated in closed form, which leads to

$$P_F = \int_{\mathcal{S}^{n-1}} \left[1 - P_{\chi_n^2}\left(r^2(\boldsymbol{a})\right)\right] p_{\boldsymbol{A}}(\boldsymbol{a}) d\boldsymbol{a}, \tag{44}$$

where $r(\boldsymbol{a})$ is the root of $G(r\boldsymbol{a}) = 0$ along the direction $\boldsymbol{a}$, and $P_{\chi_n^2}(\cdot)$ is the cumulative distribution function of a $\chi^2$ random variable with degrees of freedom $n$.

In directional importance sampling, the important directions that contribute most to the failure probability are efficiently sampled by introducing an ISPDF $q_{\boldsymbol{A}}(\boldsymbol{a})$ of $\boldsymbol{A}$, which re-formulates Equation (44) as

$$P_F = \int_{\mathcal{S}^{n-1}} \left[1 - P_{\chi_n^2}\left(r^2(\boldsymbol{a})\right)\right] \frac{p_{\boldsymbol{A}}(\boldsymbol{a})}{q_{\boldsymbol{A}}(\boldsymbol{a})} q_{\boldsymbol{A}}(\boldsymbol{a}) d\boldsymbol{a} = E_{q_A}\left[\left(1 - P_{\chi_n^2}\left(r^2(\boldsymbol{A})\right)\right) W(\boldsymbol{A})\right], \tag{45}$$

where $W(\boldsymbol{A}) = p_{\boldsymbol{A}}(\boldsymbol{A})/q_{\boldsymbol{A}}(\boldsymbol{A})$ is the importance weight function. The theoretically optimal ISPDF of $\boldsymbol{A}$ is given by

$$q_{\boldsymbol{A}}^{opt}(\boldsymbol{a}) = \frac{1}{P_F}\left(1 - P_{\chi_n^2}\left(r^2(\boldsymbol{a})\right)\right) p_{\boldsymbol{A}}(\boldsymbol{a}). \tag{46}$$

Studies on strategies to choose suitable ISPDF $q_{\boldsymbol{A}}(\boldsymbol{a})$ can be found in [Bjerager 1988, Ditlevsen et al., 1988, Ditlevsen and Bjerager 1989, Nie and Ellingwood 2000, 2004a, 2004b]. The CE method has been applied to select $q_{\boldsymbol{A}}(\boldsymbol{a})$ that closely approximates $q_{\boldsymbol{A}}^{opt}(\boldsymbol{a})$ [Shayanfar et al., 2018, Zhang et al., 2022]. In [Shayanfar et al., 2018], $q_{\boldsymbol{A}}(\boldsymbol{a})$ is selected such that it has a high probability density in the regions around the direction of the design point of the limit state function $G(\boldsymbol{u}) = 0$. In this approach, a sample direction $\boldsymbol{a}$ is generated through rotational transformation of a random vector sampled around the basis vector $e_n = (0,0,\dots,1)$. This vector is obtained through transformation of the polar angles of a $(n-1)$ dimensional vector sampled from a zero-mean Gaussian PDF, for which the optimal covariance matrix is determined using the CE method. The matrix defining the rotational transformation is selected such that it transforms $e_n = (0,0,\dots,1)$ into the unit vector pointing to the design point. A subsequent study by Zhang et al., [2022] proposed to approximate $q_{\boldsymbol{A}}^{opt}(\boldsymbol{a})$ by the von Mises-Fisher PDF, wherein the optimal parameters of the density family are determined by minimizing its Kullback-Leibler divergence from $q_{\boldsymbol{A}}^{opt}$. Another study by Jafari-Asl et al., [2022] incorporates the Harris Hawks optimization algorithm to conduct reliability estimation using directional simulation. Studies by Zhang et al., [2020a] and Song et al., [2011] have also used directional simulation to compute the sensitivity of the failure probability with respect to the input random variables.

The studies by Cheng et al., [2023] and Cheng et al., [2025] propose a sequential importance sampling method to estimate the failure probability in Equation (44). Herein, an auxiliary limit-state function $G(\alpha\boldsymbol{u})$ is considered, where $\alpha(>1)$ represents a magnification factor that amplifies the standard deviation of the random variables in $\boldsymbol{U}$. Consequently, the probability of the auxiliary failure event $F_\alpha = \{\boldsymbol{u} \in \mathbb{R}^n : G(\alpha\boldsymbol{u}) \le 0\}$ is larger than that of $F$ and can be efficiently estimated based on samples from $\phi_n(\boldsymbol{u})$. For a decreasing sequence of magnification factors $\alpha_1 > \alpha_2 > \cdots > \alpha_L = 1$, the auxiliary failure probabilities $P_{F_{\alpha_l}}$ gradually reduce to the target failure probability $P_F$ as the parameters $\alpha_l$ decrease to 1. An adaptive mechanism is typically adopted to select the sequence of values [Cheng et al., 2023]. A relation between the failure probabilities of two consecutive auxiliary limit state functions is obtained as

$$P_{F_{\alpha_l}} = \int_{\mathbb{R}^n} I\{G(\alpha_l\boldsymbol{u}) \le 0\}\phi_U(\boldsymbol{u})d\boldsymbol{u} = \int_{\mathcal{S}^{n-1}} \left[1 - P_{\chi_n^2}\left(r_l^2(\boldsymbol{a})\right)\right] p_{\boldsymbol{A}}(\boldsymbol{a})d\boldsymbol{a}$$

$$= P_{F_{\alpha_{l-1}}} \int_{\mathcal{S}^{n-1}} \frac{1-P_{\chi_n^2}\left(r_l^2(\boldsymbol{a})\right)}{1-P_{\chi_n^2}\left(r_{l-1}^2(\boldsymbol{a})\right)} \frac{1-P_{\chi_n^2}\left(r_{l-1}^2(\boldsymbol{a})\right)}{P_{F_{\alpha_{l-1}}}} p_{\boldsymbol{A}}(\boldsymbol{a}) d\boldsymbol{a} = P_{F_{\alpha_{l-1}}} E_{q_{\boldsymbol{A}}^{l-1}}[W_l(\boldsymbol{A})], \tag{47}$$

where $r_l(\boldsymbol{a})$ is the root of $G(\alpha_l r\boldsymbol{a}) = 0$ along the direction $\boldsymbol{a}$,

$$q_{\boldsymbol{A}}^{l-1}(\boldsymbol{a}) = \frac{1}{P_{F_{\alpha_{l-1}}}} \left(1 - P_{\chi_n^2}\left(r_{l-1}^2(\boldsymbol{a})\right)\right) p_{\boldsymbol{A}}(\boldsymbol{a}) \tag{48}$$

is the ISPDF of $\boldsymbol{A}$ and $W_l(\boldsymbol{a}) = \frac{1-P_{\chi_n^2}\left(r_l^2(\boldsymbol{a})\right)}{1-P_{\chi_n^2}\left(r_{l-1}^2(\boldsymbol{a})\right)}$ is the importance weight function. The roots $r_l(\boldsymbol{a})$ and $r_{l-1}(\boldsymbol{a})$ of $G(\alpha_l r\boldsymbol{a}) = 0$ and $G(\alpha_{l-1} r\boldsymbol{a}) = 0$, respectively, satisfy the condition $r_l(\boldsymbol{a}) = r_{l-1}(\boldsymbol{a})\, \alpha_{l-1}/\alpha_l$.

It follows that the probability of the target failure event $F$ can be expressed as

$$P_F = P_{F_{\alpha_1}} \prod_{l=2}^{L} S_l, \tag{49}$$

where

$$S_l = E_{q_{\boldsymbol{A}}^{l-1}}[W_l(\boldsymbol{A})]. \tag{50}$$

The estimate $\hat{P}_{F_{\alpha_1}}$ is obtained by direct MCS based on samples from $p_{\boldsymbol{A}}(\boldsymbol{a})$. The samples that fall in the domain of $F_{\alpha_1}$ are distributed according to $q_{\boldsymbol{A}}^1(\boldsymbol{a})$. At any intermediate level, a set of samples from the PDF $q_{\boldsymbol{A}}^{l-1}(\boldsymbol{a})$ are used to evaluate the expectation in Equation (50) to estimate $\hat{S}_l$. These samples are then resampled according to the weights $W_l(\boldsymbol{a})$ to get equivalent samples from the subsequent PDF $q_{\boldsymbol{A}}^l(\boldsymbol{a})$, which are propagated to the high probability density region of $q_{\boldsymbol{A}}^l(\boldsymbol{a})$ by MCMC moves.

**Inverse importance sampling:** Papakonstantinou et al., [2023] proposed a method termed approximate sampling target with post-processing adjustment (ASTPA). Here, the indicator function is smoothed using a logistic function defined as follows.

$$l_{g_u}(\boldsymbol{u}) = \left(1 + \exp\left(\frac{\frac{G(u)}{g_c} + \mu_g}{\left(\frac{\sqrt{3}}{\pi}\right)\sigma}\right)\right)^{-1} \tag{51}$$

Here, $g_c$ is a scaling parameter to be chosen according to the magnitude of the performance function. The suggested value of $\sigma$ = 0.1-0.8, and $\mu_g = -\left(\frac{\sqrt{3}}{\pi}\sigma\right)\ln\left(\frac{p}{1-p}\right)$ with a percentile value $p = 0.1$. This is used to define the following target pdf with an unknown normalizing constant $C_h$.

$$h(\boldsymbol{u}) = l_{g_u}(\boldsymbol{u}) p_U(\boldsymbol{u})/C_h \tag{52}$$

The authors propose a modified version of a quasi-Newton mass-preconditioned Hamiltonian MCMC to draw samples from the target pdf. The sampler estimates the Hessian inverse according to the following BFGS update.

$$\mathbf{W}_{k+1} = \left(\mathbf{I} - \frac{\boldsymbol{s}_k \boldsymbol{y}_k^T}{\boldsymbol{y}_k^T \boldsymbol{s}_k}\right) \mathbf{W}_k \left(\mathbf{I} - \frac{\boldsymbol{y}_k \boldsymbol{s}_k^T}{\boldsymbol{y}_k^T \boldsymbol{s}_k}\right) + \frac{\boldsymbol{s}_k \boldsymbol{s}_k^T}{\boldsymbol{y}_k^T \boldsymbol{s}_k} \tag{53}$$

Here, $\boldsymbol{s}_k = \boldsymbol{u}_{k+1} - \boldsymbol{u}_k$ and $\boldsymbol{y}_k = -\nabla\mathcal{L}(\boldsymbol{u}_{k+1}) + \nabla\mathcal{L}(\boldsymbol{u}_k)$, where $\mathcal{L}$ denotes the log target pdf. The estimated Hessian is used as a preconditioning mass in the Hamiltonian dynamics. Finally, since the samples are not drawn from the true conditional pdf, the failure probability is obtained using the importance sampling approach.

$$P_F = \int \frac{I(\boldsymbol{u} \in F) p_U(\boldsymbol{u})}{h(\boldsymbol{u})} h(\boldsymbol{u})\, d\boldsymbol{u} = C_h \int I(\boldsymbol{u} \in F) \frac{1}{l_{g_{\boldsymbol{u}}}(\boldsymbol{u})} h(\boldsymbol{u})\, d\boldsymbol{u} \tag{54}$$

This yields the following estimator for the probability of failure.

$$\hat{P}_F = C_h \left[\frac{1}{N} \sum_{i=1}^{N} \frac{I(\boldsymbol{u}_i \in F)}{l_{g_{\boldsymbol{u}}}(\boldsymbol{u})}\right] \tag{55}$$

The normalizing constant is estimated by fitting a suitable Gaussian mixture model $Q(\boldsymbol{u})$ using the following importance sampling estimator.

$$C_h = \frac{1}{M} \sum_{i=1}^{M} \frac{l_{g_{\boldsymbol{u}}}(\boldsymbol{u}_i') p_U(\boldsymbol{u}_i')}{Q(\boldsymbol{u}'_i)} \tag{56}$$

In a subsequent work, Eshra et al., [2025] extended the ASTPA approach to non-Gaussian space of random variables. Eshra and Papakonstantinou [2025] have also proposed a method where the quasi-Newton Hamiltonian MCMC, which necessitates gradient computations for sampling, is replaced by a gradient-free preconditioned Crank-Nicolson sampler assisted by a failure domain discovery stage in which samples are drawn from a scaled standard normal distribution.

### 4.2 Importance sampling methods for time-variant reliability analysis

For structural dynamic systems with excitations modeled as random processes, one can, in principle, discretize the excitation into an equivalent set of random variables, using series representations such as the Karhunen-Loeve expansion [Ghanem and Spanos 2003], and continue to employ the importance sampling (IS) methods for time-invariant reliability analysis to estimate the first-passage failure probability. However, the number of random variables entering the formulation is often very large, ranging from several hundred to thousands, which can severely limit the performance of importance sampling due to the curse of dimensionality and weight degeneracy. To this end, tailored IS methods have been developed for evaluating the failure probability of specific structural dynamic systems, which are outlined in Section 4.2.1. A more generic approach to circumvent the challenges of applying importance sampling for time-variant reliability estimation is to design a new probability measure that increases the likelihood of the failure event by directly manipulating the trajectory of the structural response process. This importance sampling probability measure can be designed using the Girsanov transformation framework, as described in Section 4.2.2. Extension of the IS methods to tackle randomness in structural parameters is described in Section 4.2.3.

#### 4.2.1 IS methods based on an equivalent time-invariant reliability problem

The idea here is to convert the time-variant reliability problem into a time-invariant form by discretizing the random excitation process into an equivalent set of random variables. Let the random vector $\boldsymbol{Z} = \{Z_1, \dots, Z_{n_z}\}$ comprise the random variables arising from the discretization of the random process. For the particular case of Gaussian excitation, $\boldsymbol{Z}$ comprises a set of mutually independent standard Gaussian random variables. Let the time interval $[0, T]$ be discretized into $n_T$ sub-intervals given by $[t_{j-1}, t_j]; j = 1, \dots, n_T$, $t_0 = 0$ and $t_{n_T} = T$, of step-size $\Delta t = t_j - t_{j-1} = \frac{T}{n_T}$. Based on this, the first-passage failure probability can be expressed as an equivalent series system reliability problem, wherein the failure event that occurs over the time duration $[0, T]$ is approximated by a sequence of instantaneous failure events, given by

$$F(\boldsymbol{\theta}) \approx \cup_{j=1}^{n_T} F_j(\boldsymbol{\theta}) = \cup_{j=1}^{n_T}\{g_j(\boldsymbol{\theta}, \boldsymbol{Z}) \leq 0\}. \tag{57}$$

In Equation (57), the failure event $F_j$ represents structural failure at the time instant $t = t_j$, with a limit-state function given by

$$g_j(\boldsymbol{\theta}, \boldsymbol{Z}) = h^* - h[\boldsymbol{\theta}, \widehat{\boldsymbol{X}}(t_j)] = h^* - h[\boldsymbol{\theta}, V_j(\boldsymbol{Z})], \tag{58}$$

where $\widehat{\boldsymbol{X}}(t)$ denotes the approximation of $\boldsymbol{X}(t)$ arising from the random process discretization and $V_j\colon \mathbb{R}^{n_z} \to \mathbb{R}^{n_p}$ is the mapping from $\boldsymbol{Z}$ to $\widehat{\boldsymbol{X}}(t)$ at the time instant $t = t_j$.

When the excitation is a Gaussian random process and the structural response $h(\boldsymbol{\theta}, \boldsymbol{X}(t))$ at any time instant is a linear function of $\boldsymbol{Z}$, a very efficient sampling method to evaluate the first-passage failure probability is proposed by Au and Beck [2001b]. The study suggests an ISPDF of the random vector $\boldsymbol{Z}$ defined as a weighted sum of Gaussian PDFs truncated on the domain of the instantaneous failure events, given by

$$q_{\boldsymbol{Z}}(\mathbf{z}; \boldsymbol{\theta}) = \sum_{j=1}^{n_T} w_j(\boldsymbol{\theta}) p_{\boldsymbol{Z}}\left(\boldsymbol{z} | \boldsymbol{z} \in F_j(\boldsymbol{\theta})\right) = \sum_{j=1}^{n_T} w_j(\boldsymbol{\theta}) \frac{p_{\boldsymbol{Z}}(\boldsymbol{z}) \mathrm{I}\{\boldsymbol{z} \in F_j(\boldsymbol{\theta})\}}{P[F_j(\boldsymbol{\theta})]} \tag{59}$$

The normalized weights of the ISPDF are given by $w_j(\boldsymbol{\theta}) = \mathbb{P}[F_j(\boldsymbol{\theta})]\left(\sum_{l=1}^{n_T} \mathbb{P}[F_l(\boldsymbol{\theta})]\right)^{-1}$, where $\mathbb{P}[F_j(\boldsymbol{\theta})]$ denotes the probability of occurrence of the instantaneous failure event $F_j(\boldsymbol{\theta})$. An estimator for the first-passage failure probability is subsequently obtained as

$$\hat{P}_F(\boldsymbol{\theta}) = \left(\sum_{l=1}^{n_T} \mathbb{P}[F_l(\boldsymbol{\theta})]\right) \times \frac{1}{N} \sum_{k=1}^{N} \left(\sum_{j=1}^{n_T} \mathrm{I}\{\boldsymbol{Z}^k \in F_j(\boldsymbol{\theta})\}\right)^{-1} \tag{60}$$

where $\boldsymbol{Z}^k; k = 1, \dots, N$ are independent draws from $q_{\boldsymbol{Z}}(\boldsymbol{z}; \boldsymbol{\theta})$.

For both linear and nonlinear dynamical systems, Au and Beck [2003b] proposed an importance sampling scheme in which the ISPDF is a weighted sum of Gaussian density functions with unit covariance matrices and centred at the design points of the instantaneous failure domains. When the system is linear, the solution of the design point, in terms of the impulse response function, is available in closed form [Der Kiureghian 2000]. For nonlinear systems, gradient-based optimization methods are typically used to find the design point. Such techniques require several response analyses and response sensitivity analyses, which can pose computational challenges. In this context, it has been noted that the realization of the input excitation corresponding to the design point is analogous to the critical excitation introduced by Drenick [1970, 1977]. For specific systems, determining the critical excitation provides a means to locate the design points for the ISPDF without resorting to expensive gradient-based algorithms (see, for example, [Koo et al., 2005, Salari and Safi 2017, Au 2006a, 2006b, Au et al., 2007a, 2007b]).

### 4.2.2 IS methods based on the Girsanov transformation

In Markovian systems governed by Ito's SDE of the form given in Equation (3), the IS probability measure can be designed by introducing additional control forces into the dynamical system [Grigoriu 2002, Oksendal 2003]. This process of replacing the original SDE by a modified equation with artificial controls is called the Girsanov transformation [Girsanov 1960]. The control forces are designed to drive the dynamical system's response to failure while keeping the failure probability estimator unbiased and reducing the associated sampling variance. The modified dynamical system is thus governed by an SDE of the form

$$\mathrm{d}\widetilde{\boldsymbol{X}}(t) = \boldsymbol{A}\big[\boldsymbol{\theta}, \widetilde{\boldsymbol{X}}(t), t\big]\mathrm{d}t + \boldsymbol{\sigma}\big[\boldsymbol{\theta}, \widetilde{\boldsymbol{X}}(t), t\big]\boldsymbol{u}\big[\boldsymbol{\theta}, \widetilde{\boldsymbol{X}}(t), t\big]\mathrm{d}t + \boldsymbol{\sigma}\big[\boldsymbol{\theta}, \widetilde{\boldsymbol{X}}(t), t\big]\mathrm{d}\widetilde{\boldsymbol{B}}(t);$$

$$\widetilde{\boldsymbol{X}}(0) = \boldsymbol{X}_0, t \geq 0 \tag{61}$$

where, $\boldsymbol{u}\big[\boldsymbol{\theta}, \widetilde{\boldsymbol{X}}(t), t\big]$ is the $q \times 1$ vector of control forces, and $\widetilde{\boldsymbol{B}}(t)$ is an Ito's process given by

$$\mathrm{d}\widetilde{\boldsymbol{B}}(t) = -\boldsymbol{u}\big[\boldsymbol{\theta}, \widetilde{\boldsymbol{X}}(t), t\big]\mathrm{d}t + \mathrm{d}\boldsymbol{B}(t);\ \widetilde{\boldsymbol{B}}(0) = 0;\ t \geq 0 \tag{62}$$

The schematic in Figure 3 illustrates the generic principle of importance sampling based on the Girsanov transformation.

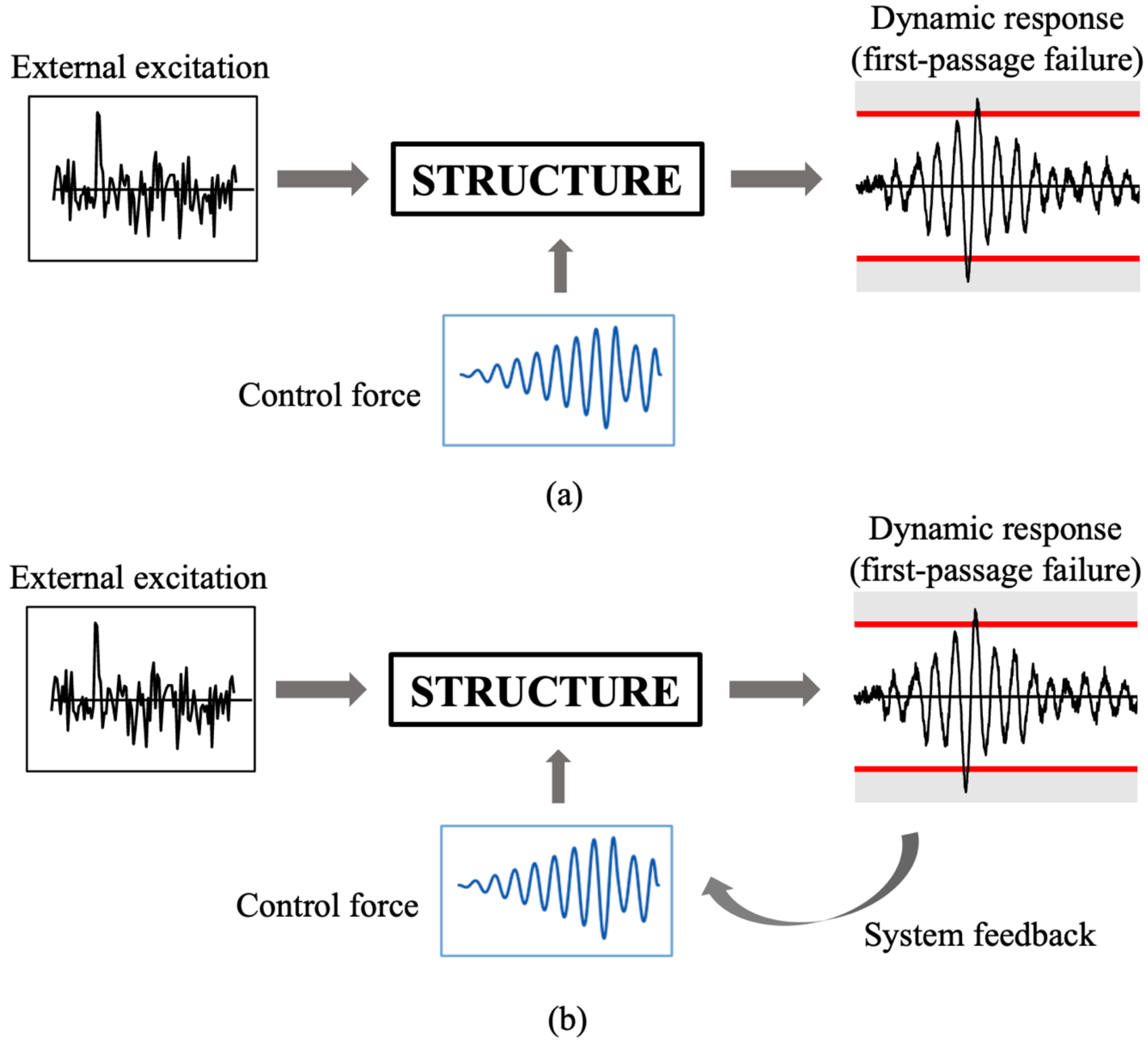

Figure 3: Structural failure in Girsanov transformation-based importance sampling, with (a) pre-determined open-loop controls.

According to Girsanov theorem [Girsanov 1960], the introduction of control forces transforms the probability measure $\mathbb{P}$ associated with the original SDE to a new measure $\mathbb{Q}$, such that $\widetilde{\boldsymbol{B}}(t)$ is a Brownian motion process with respect to $\mathbb{Q}$ and has the properties $E_{\mathbb{Q}}\left[\Delta\widetilde{\boldsymbol{B}}(t)\right] = 0$, and $E_{\mathbb{Q}}\left[\Delta\tilde{B}_i(t_1)\Delta\tilde{B}_j(t_2)\right] = C_{ij}\Delta t\delta(t_1 - t_2)$ for $t_1, t_2 \geq 0$ and $i, j = 1,2, \dots, q$. The probability space associated with the modified SDE is thus given by $(\Omega, \mathcal{F}, \mathbb{Q})$. The relationship between the two measures $\mathbb{P}$ and $\mathbb{Q}$ is defined by a scalar process known as the Radon-Nikodym derivative, and is given by

$$\frac{\mathrm{d}\mathbb{P}}{d\mathbb{Q}}(t) = \frac{R(t)}{R_0} = \exp\left(-\int_0^t \boldsymbol{u}\left[\boldsymbol{\theta}, \widetilde{\boldsymbol{X}}(\tau), \tau\right]^{\mathrm{T}} \mathbf{C}^{-1}\mathrm{d}\widetilde{\boldsymbol{B}}(\tau) - \frac{1}{2}\int_0^t \boldsymbol{u}\left[\boldsymbol{\theta}, \widetilde{\boldsymbol{X}}(\tau), \tau\right]^{\mathrm{T}} \mathbf{C}^{-1}\boldsymbol{u}\left[\boldsymbol{\theta}, \widetilde{\boldsymbol{X}}(\tau), \tau\right]\mathrm{d}\tau\right)$$

The Radon-Nikodym derivative satisfies the conditions $\mathbb{Q}\left[\frac{R(t)}{R_0} \geq 0\right] = 1$ and $E_{\mathbb{Q}}\left[\frac{R(t)}{R_0}\right] = \int \frac{\mathrm{d}\mathbb{P}}{\mathrm{d}\mathbb{Q}}(t)\mathrm{d}\mathbb{Q} = 1$, and can be obtained as a solution of the scalar SDE [Grigoriu 2002, Oksendal 2003]

$$\mathrm{d}R(t) = -R(t)\boldsymbol{u}\left[\boldsymbol{\theta}, \widetilde{\boldsymbol{X}}(t), t\right]^{\mathrm{T}} \mathbf{C}^{-1}\mathrm{d}\widetilde{\boldsymbol{B}}(t); R(0) = R_0 \tag{63}$$

Using the above transformation, the failure probability is re-expressed as

$$P_F(\boldsymbol{\theta}) = \int_{F(\boldsymbol{\theta})} \mathrm{d}\mathbb{P} = \int_{F(\boldsymbol{\theta})} \left(\frac{\mathrm{d}\mathbb{P}}{d\mathbb{Q}}(T)\right) \mathrm{d}\mathbb{Q} = E_{\mathbb{Q}}\left[\frac{R(T)}{R_0} \mathrm{I}\left\{h^* - \max_{0<t\leq T} h\left[\boldsymbol{\theta}, \widetilde{\boldsymbol{X}}(t)\right] \leq 0\right\}\right] \tag{64}$$

An estimator for evaluating the above expectation is thus obtained as

$$\hat{P}_F(\boldsymbol{\theta}) = \frac{1}{N}\sum_{k=1}^{N} \frac{R^k(T)}{R_0^k} \mathrm{I}\left\{h^* - \max_{0<t\leq T} h\left[\boldsymbol{\theta}, \widetilde{\boldsymbol{X}}^k(t)\right] \leq 0\right\} \tag{65}$$

where $\widetilde{\boldsymbol{X}}^k(t)$ and $R^k(t)$ are random draws from Equations (61) and (63), respectively. Here, the correction term $\frac{R(T)}{R_0}$ in the failure probability estimator accounts for the distortion of the original dynamical system caused by the addition of controls.

The failure probability estimator in Equation (65) is unbiased, i.e., $E_{\mathbb{Q}}\left[\hat{P}_F(\boldsymbol{\theta})\right] = P_F(\boldsymbol{\theta})$, and has a sampling variance that depends not only on *N*, but also on the control $\boldsymbol{u}\left[\boldsymbol{\theta}, \widetilde{\boldsymbol{X}}(t), t\right]$ which is yet to be selected. Hence, a prudent choice of $\boldsymbol{u}\left[\boldsymbol{\theta}, \widetilde{\boldsymbol{X}}(t), t\right]$ provides the scope to minimize $\mathrm{Var}\left(\hat{P}_F\right)$ while retaining *N* at a manageable level. It can be shown that an ideal control force, which renders $\mathrm{Var}\left(\hat{P}_F\right) = 0$, exists and is given by [Milstein 1995]

$$\boldsymbol{u}^*\left[\boldsymbol{\theta}, \widetilde{\boldsymbol{X}}(t), t\right] = \frac{1}{\zeta_F\left[\boldsymbol{\theta}, \widetilde{\boldsymbol{X}}(t), t\right]} \mathbf{C}\boldsymbol{\sigma}\left[\boldsymbol{\theta}, \widetilde{\boldsymbol{X}}(t), t\right]^{\mathrm{T}} \left(\frac{\partial \zeta_F\left[\boldsymbol{\theta}, \widetilde{\boldsymbol{X}}(t), t\right]}{\partial \widetilde{\boldsymbol{X}}(t)}\right), \tag{66}$$

where

$$\zeta_F\left[\boldsymbol{\theta}, \widetilde{\boldsymbol{X}}(t), t\right] = \mathbb{P}\left[\left\{h^* - \max_{t<r\leq T} h[\boldsymbol{\theta}, \boldsymbol{X}(r)] \leq 0\right\} \middle| \boldsymbol{X}(t) = \tilde{X}(t)\right]. \tag{67}$$

It is evident that to construct $\boldsymbol{u}^*\left[\boldsymbol{\theta}, \widetilde{\boldsymbol{X}}(t), t\right]$, one needs to determine the function $\zeta_F[\boldsymbol{\theta}, \boldsymbol{X}_s, s]$ for all values of the arguments $(s, \boldsymbol{X}_s)$ such that $s \in [0, T]$ and $h[\boldsymbol{\theta}, \boldsymbol{X}_s] < h^*$. It may be noted that $\zeta_F[\boldsymbol{\theta}, \boldsymbol{X}_0, 0] = P_F$ is the very quantity being sought to be determined. This indicates that, in practical applications of Girsanov transformation, the use of the ideal control function is infeasible, as it requires prior knowledge of the quantity to be estimated. Hence, one needs to devise sub-optimal Girsanov controls that may not achieve the ideal situation but can still help reduce the sampling variance appreciably.

#### 4.2.2.1 Open-loop Girsanov controls for first-passage failure probability

Open-loop Girsanov controls for first-passage failure probability are a deterministic control function applied to the structure. These controls are independent of the state-space trajectory of the dynamical system, and are strictly functions of time $t$, i.e., $\boldsymbol{u}\left[\boldsymbol{\theta}, \widetilde{\boldsymbol{X}}(t), t\right] = \boldsymbol{u}(\boldsymbol{\theta}, t)$. An open-loop control $\boldsymbol{u}^j(\boldsymbol{\theta}, t)$ that promotes the occurrence of an instantaneous failure event $F_j$ can be found through a procedure analogous to the FORM, by minimizing a distance function [Tanaka 1997] given by

$$\beta(\boldsymbol{\theta}, t_j) = \sqrt{\int_0^{t_j} \left(\boldsymbol{u}(\boldsymbol{\theta}, t)\right)^{\mathrm{T}} \boldsymbol{u}(\boldsymbol{\theta}, t) dt} \tag{68}$$

subject to the constraint $h^* - h\left[\boldsymbol{\theta}, \boldsymbol{Z}(t_j)\right] = 0$, where $\boldsymbol{Z}(t)$ is the solution of the hypothetical deterministic dynamical system

$$\dot{\boldsymbol{Z}}(t) = \boldsymbol{A}[\boldsymbol{\theta}, \boldsymbol{Z}(t), t] + \boldsymbol{\sigma}[\boldsymbol{\theta}, \boldsymbol{Z}(t), t]\boldsymbol{u}(\boldsymbol{\theta}, t);\ \boldsymbol{Z}(0) = \boldsymbol{X}_0;\ t \geq 0 \tag{69}$$

The Girsanov control $\boldsymbol{u}(\boldsymbol{\theta}, t)$ for the first-passage failure probability is selected from among the controls $\boldsymbol{u}^j(\boldsymbol{\theta}, t); j = 1, \ldots, n_T$ according to weights $w_j(\boldsymbol{\theta}); j = 1, \ldots, n_T$ [Macke and Bucher 2003], i.e., $\boldsymbol{u}(\boldsymbol{\theta}, t) = \boldsymbol{u}^j(\boldsymbol{\theta}, t)$ with probability $w_j(\boldsymbol{\theta})$, where

$$w_j(\boldsymbol{\theta}) = \Phi\left(-\beta(\boldsymbol{\theta}, t_j)\right)\left[\sum_{r=1}^{n_T} \Phi\left(-\beta(\boldsymbol{\theta}, t_r)\right)\right]^{-1}. \tag{70}$$

The numerical computation of the open-loop control forces is performed in a time-discretized setting by solving a high-dimensional optimization problem, with the control force values at the discrete time instants as optimization variables. An alternative, and often more efficient, means to compute the open-loop control is to convert the constrained optimization problem in Equations (68) and (69) into a two-point boundary value problem in time, as suggested by Kanjilal and Manohar [2017].

#### 4.2.2.2 Closed-loop Girsanov controls for first-passage failure probability

The methods discussed in the previous section pertain to state-independent Girsanov controls, which, by their very nature, are pre-computed and deterministic. On the other hand, the optimal zero-sampling-variance control is state-dependent (or closed-loop) and hence stochastic in nature. Thus, it appears that open-loop controls, because of their deterministic nature, have limited potential to approach ideal controls, no matter how carefully they are designed. To this end, alternative strategies for deriving closed-loop controls for estimating the first-passage failure probability have been proposed.

Early studies considered finding a closed-loop control through an approximate solution of the Kolmogorov backward equation [Schoenmakers et al., 2002], wherein the control function is given by

$$\boldsymbol{u}[\boldsymbol{\theta}, \widetilde{\boldsymbol{X}}(t), t] = \frac{1}{v[\boldsymbol{\theta}, \widetilde{\boldsymbol{X}}(t), t]} \left(\boldsymbol{\sigma}[\boldsymbol{\theta}, \widetilde{\boldsymbol{X}}(t), t]\right)^{\mathrm{T}} \frac{\partial v[\boldsymbol{\theta}, \widetilde{\boldsymbol{X}}(t), t]}{\partial \widetilde{\boldsymbol{X}}(t)}, \tag{71}$$

where $v$ is the solution of

$$\frac{\partial v[\boldsymbol{\theta}, \boldsymbol{y}, t]}{\partial t} + (\boldsymbol{A}[\boldsymbol{\theta}, \boldsymbol{y}, t])^{\mathrm{T}} \frac{\partial v[\boldsymbol{\theta}, \boldsymbol{y}, t]}{\partial \boldsymbol{y}} + \left(\boldsymbol{\sigma}[\boldsymbol{\theta}, \widetilde{\boldsymbol{X}}(t), t]\right)^{\mathrm{T}} \mathbf{H}(v) \boldsymbol{\sigma}[\boldsymbol{\theta}, \widetilde{\boldsymbol{X}}(t), t] = 0 \tag{72}$$

with $[\mathbf{H}(v)]_{jk} = \frac{\partial^2 v[\boldsymbol{\theta}, \boldsymbol{y}, t]}{\partial y_j y_k}$ and boundary conditions $v[\boldsymbol{\theta}, \boldsymbol{y} \in D_s, T] = 0$; $D_s = \{\boldsymbol{y}: h[\boldsymbol{\theta}, \boldsymbol{y}] < h^*\}$, and $v[\boldsymbol{\theta}, \boldsymbol{y} \in D_b, t] = 1; 0 < t \leq T$; $D_b = \{\boldsymbol{y}: h[\boldsymbol{\theta}, \boldsymbol{y}] = h^*\}$. In general, Equation (72) must be solved numerically, e.g., using the finite element method, which poses computational challenges for structures with many mechanical DOFs.

The works of Olsen and Naess [2006, 2007] proposed a two-step procedure for determining the control function. The procedure is based on obtaining a numerical approximation of the failure probability function $\zeta_F[\boldsymbol{\theta}, \boldsymbol{X}_s, s]$ by employing open-loop controls. In the first stage, an approximate $\zeta_F[\boldsymbol{\theta}, \boldsymbol{X}_s, s]$ is determined through simulations performed on a grid of initial values of $(\boldsymbol{X}_s, s)$ using open-loop controls. In the second stage, during numerical integration of the SDE, the value of the control function at each time point is taken from the multi-dimensional data calculated in the previous iteration. The above method has been applied to single-DOF oscillators driven by stationary Gaussian excitation. The special case of single-DOF elasto-plastic oscillators has been considered in the works of Au [2008, 2009a, 2009b], which propose heuristic control laws that increase the response amplitude [Au 2008] or the energy of the dynamical system [Au 2009a, 2009b] to match that under critical excitation.

The studies by Kanjilal and Manohar [2018, 2019] consider the equivalent series system in Equation (57) and propose a closed-loop control $\boldsymbol{u}[\boldsymbol{\theta}, \widetilde{\boldsymbol{X}}(t), t]$ for estimating the first-passage probability that is defined in terms of the optimal Girsanov control for the probabilities of the instantaneous failure events. If one considers the failure event $F_j = \{h^* - h[\boldsymbol{\theta}, \boldsymbol{X}(t_j)] \leq 0\}$ for $t_j \in [0, T]$, the optimal control for the instantaneous failure probability $\mathbb{P}[F_j(\boldsymbol{\theta})]$ becomes

$$\boldsymbol{u}_j^*[\boldsymbol{\theta}, \widetilde{\boldsymbol{X}}(t), t] = \frac{1}{\zeta_{F_j}[\boldsymbol{\theta}, \widetilde{\boldsymbol{X}}(t), t]} \mathbf{C}\boldsymbol{\sigma}[\boldsymbol{\theta}, \widetilde{\boldsymbol{X}}(t), t]^{\mathrm{T}} \left(\frac{\partial \zeta_{F_j}[\boldsymbol{\theta}, \widetilde{\boldsymbol{X}}(t), t]}{\partial \widetilde{\boldsymbol{X}}(t)}\right), \tag{73}$$

where $\zeta_{F_j}[\boldsymbol{\theta}, \widetilde{\boldsymbol{X}}(t), t] = \mathbb{P}[\{h^* - h[\boldsymbol{\theta}, \boldsymbol{X}(t_j)] \leq 0\} | \boldsymbol{X}(t) = \widetilde{\boldsymbol{X}}(t)]$ for $t \in [0, t_j)$, and $\boldsymbol{u}_j^*[\boldsymbol{\theta}, \widetilde{\boldsymbol{X}}(t), t] = 0$ for $t \geq t_j$. The Girsanov control $\boldsymbol{u}[\boldsymbol{\theta}, \widetilde{\boldsymbol{X}}(t), t]$ is selected from among the controls $\boldsymbol{u}_j^*[\boldsymbol{\theta}, \widetilde{\boldsymbol{X}}(t), t]; j = 1, \ldots, n_T$ according to weights $w_j(\boldsymbol{\theta}) = \mathbb{P}[F_j(\boldsymbol{\theta})] \left(\sum_{r=1}^{n_T} \mathbb{P}[F_r(\boldsymbol{\theta})]\right)^{-1}; j = 1, \ldots, n_T$. When the input excitation is a Gaussian process and the structural response is a linear function of the input, the function $\zeta_{F_j}[\boldsymbol{\theta}, \widetilde{\boldsymbol{X}}(t), t]$ and the instantaneous failure probabilities $\mathbb{P}[F_j(\boldsymbol{\theta})]; j = 1, \ldots, n_T$ can be derived analytically in closed form. For nonlinear structural response, approximation to the optimal control $\boldsymbol{u}_j^*[\boldsymbol{\theta}, \widetilde{\boldsymbol{X}}(t), t]$ has been suggested based on a local linearization method [Kanjilal and Manohar 2018].

### 4.2.3 Extension to structures with random parameters

Here, we consider the general case in which randomness in structural system parameters and external excitation are simultaneously present. In this situation, one can evaluate the first-passage probability by integrating the failure probability function over the outcome space of the random structural parameters:

$$P_F = \int_{\boldsymbol{\theta} \in \mathbb{R}^{n_s}} P_F(\boldsymbol{\theta}) p_{\boldsymbol{\Theta}}(\boldsymbol{\theta}) d\boldsymbol{\theta}. \tag{74}$$

In principle, one can estimate the integral as a sample average of the estimates of the probabilities $P_F(\boldsymbol{\theta}^i); i = 1, \ldots, N_s$, where $\boldsymbol{\theta}^i; i = 1, \ldots, N_s$ are independent samples from the PDF $p_{\boldsymbol{\Theta}}(\boldsymbol{\theta})$. When the function $P_F(\boldsymbol{\theta})$ exhibits large variability over the support of $p_{\boldsymbol{\Theta}}(\boldsymbol{\theta})$, many Monte Carlo samples are necessary to achieve acceptable accuracy. To address this problem, importance sampling has been applied to sample from the regions that contribute most to the integrand in Equation (74).

Application of importance sampling leads to the modified reliability integral

$$P_F = \int_{\boldsymbol{\theta} \in \mathbb{R}^{n_s}} P_F(\boldsymbol{\theta}) W(\boldsymbol{\theta}) q_{\boldsymbol{\Theta}}(\boldsymbol{\theta}) d\boldsymbol{\theta}, \tag{75}$$

where $q_{\boldsymbol{\Theta}}(\boldsymbol{\theta})$ is the ISPDF of the random structural parameters and $W(\boldsymbol{\theta}) = \frac{p_{\boldsymbol{\Theta}}(\boldsymbol{\theta})}{q_{\boldsymbol{\Theta}}(\boldsymbol{\theta})}$ is the IS weight. The estimator of the first-passage failure probability based on Equation (75) is given by

$$\hat{P}_F = \frac{1}{N_s} \sum_{i=1}^{N_s} \hat{P}_F(\boldsymbol{\theta}^i) W(\boldsymbol{\theta}^i). \tag{76}$$

In Equation (76), $\boldsymbol{\theta}^i; i = 1, \ldots, N_s$ are independent samples from the ISPDF $q_{\boldsymbol{\Theta}}(\boldsymbol{\theta})$ and $\hat{P}_F(\boldsymbol{\theta}^i)$ is the estimate of the first-passage probability for the specific realization $\boldsymbol{\Theta} = \boldsymbol{\theta}^i$ of the random structural parameters. $\hat{P}_F(\boldsymbol{\theta}^i)$ can be evaluated using any of the procedures described in Sections 4.2.1 and 4.2.2. A natural choice is an ISPDF $q_{\boldsymbol{\Theta}}(\boldsymbol{\theta})$ centred at a representative point $\boldsymbol{\theta}^*$ that maximizes the integrand $P_F(\boldsymbol{\theta}) p_{\boldsymbol{\Theta}}(\boldsymbol{\theta})$ [Jensen and Valdebenito 2007]. To tackle failure domains where $P_F(\boldsymbol{\theta}) p_{\boldsymbol{\Theta}}(\boldsymbol{\theta})$ has multiple modes; an ISPDF, which is a mixture of component PDFs centred at the local maximizers of each mode, has been suggested [Au et al., 1999]. The application of surrogate models to approximate the integrand and construct an effective ISPDF has been considered by Valdebenito et al., [2014].

The studies by Kanjilal et al., [2021, 2022] propose an adaptive multilevel sampling method to construct an ISPDF of the random structural parameters that closely resembles the optimal ISPDF

$$q_{\boldsymbol{\Theta}}^{opt}(\boldsymbol{\theta}) = \frac{1}{P_F} P_F(\boldsymbol{\theta}) p_{\boldsymbol{\Theta}}(\boldsymbol{\theta}). \tag{77}$$

The multilevel approach relies on approximating a sequence of intermediate target PDFs $q_{\boldsymbol{\Theta}}^l(\boldsymbol{\theta}); l = 1, \ldots, L$, that gradually shift and shape towards the optimal ISPDF $q_{\boldsymbol{\Theta}}^{opt}(\boldsymbol{\theta})$. The density sequence is constructed by tempering the failure probability function $P_F(\boldsymbol{\theta})$:

$$q_{\boldsymbol{\Theta}}^l(\boldsymbol{\theta}) = \frac{1}{c_l} P_F(\boldsymbol{\theta})^{\gamma_l} p_{\boldsymbol{\Theta}}(\boldsymbol{\theta}), \tag{78}$$

where $\gamma_l; l = 1, \dots, L$ are tempering parameters satisfying $0 = \gamma_0 < \gamma_1 < \cdots < \gamma_L = 1$. The approximation is sought in terms of a parametric family of PDFs $q_{\boldsymbol{\Theta}}(\boldsymbol{\theta}; \boldsymbol{v}_l); l, \dots, L$, where the parameter vectors $\boldsymbol{v}_l; l = 1, \dots, L$ are determined by sequentially minimizing the Kullback-Leibler divergence between the intermediate target PDFs in Equation (78) and the parametric family $q_{\boldsymbol{\Theta}}(\boldsymbol{\theta}; \boldsymbol{v})$. Common choices for the parametric family are the Gaussian mixture distribution and the von-Mises Fisher Nakagami distribution. The latter is comparatively robust in terms of scalability to problems with a large number of random structural parameters.

At an intermediate level of the multilevel approach, the parameter vector $\boldsymbol{v}_l$ is determined by solving the CE optimization

$$\boldsymbol{v}_l = \underset{\boldsymbol{v}}{\operatorname{argmax}}\, E_{p_{\boldsymbol{\Theta}}}\left[P_F(\boldsymbol{\theta})^{\gamma_l} \ln\left(q_{\boldsymbol{\Theta}}(\boldsymbol{\theta}; \boldsymbol{v})\right)\right], \tag{79}$$

which minimizes the Kullback-Leibler divergence $D_{KL}\left(q_{\boldsymbol{\Theta}}^l(\boldsymbol{\theta}), q_{\boldsymbol{\Theta}}(\boldsymbol{\theta}; \boldsymbol{v})\right)$. The expectation in Equation (79) is estimated by importance sampling from $q_{\boldsymbol{\Theta}}(\boldsymbol{\theta}; \hat{\boldsymbol{v}}_{l-1})$, $\hat{\boldsymbol{v}}_{l-1}$ being the estimate of $\boldsymbol{v}_{l-1}$ in the previous step. Accordingly, the sample counterpart of the CE optimization problem is given by

$$\hat{\boldsymbol{v}}_l = \underset{\boldsymbol{v}}{\operatorname{argmax}} \frac{1}{N_{CE}} \sum_{k=1}^{N_{CE}} W_l(\boldsymbol{\theta}^{l-1,k}; \hat{\boldsymbol{v}}_{l-1}, \gamma_l) \ln\left(q_{\boldsymbol{\Theta}}(\boldsymbol{\theta}^{l-1,k}; \boldsymbol{v})\right), \tag{80}$$

where $\boldsymbol{\theta}^{l-1,k}; k = 1, \dots, N_{CE}$ are independent samples drawn from $q_{\boldsymbol{\Theta}}(\boldsymbol{\theta}; \hat{\boldsymbol{v}}_{l-1})$ and $W_l(\boldsymbol{\theta}^{l-1,k}; \hat{\boldsymbol{v}}_{l-1}, \gamma_l) = \hat{P}_F(\boldsymbol{\theta}^{l-1,k})^{\gamma_l} \frac{p_{\boldsymbol{\Theta}}(\boldsymbol{\theta}^{l-1,k})}{q_{\boldsymbol{\Theta}}(\boldsymbol{\theta}^{l-1,k}; \hat{\boldsymbol{v}}_{l-1})}$. The tempering parameter $\gamma_l$ in each level is selected such that the coefficient of variation of the weights $W_l(\boldsymbol{\theta}^{l-1,k}; \hat{\boldsymbol{v}}_{l-1}, \gamma_l); k = 1, \dots, N_{CE}$ is less than a prescribed threshold. This ensures that the sampling density $q_{\boldsymbol{\Theta}}(\boldsymbol{\theta}; \hat{\boldsymbol{v}}_{l-1})$ does not differ significantly from the target PDF $q_{\boldsymbol{\Theta}}^l(\boldsymbol{\theta})$, so that $\boldsymbol{v}_l$ can be estimated accurately with a small number of samples. In a recent study by Kanjilal et al., [2026], an approach based on non-parametric regression has been proposed to reduce the variability of the Monte Carlo estimates $\hat{P}_F(\boldsymbol{\theta}^{l-1,k}); k = 1, \dots, N_{CE}$, to ensure a smooth convergence to the optimal ISPDF $q_{\boldsymbol{\Theta}}^{opt}(\boldsymbol{\theta})$.

### 4.3 Discussion on IS methods

**Discussion on IS methods for time-invariant reliability:** FORM-based importance sampling relies on locating the ISPDF directly around the LSF design point. These methods generally boast very low computational complexity when the LSF meets the necessary regularity condition. In contrast, adaptive importance sampling methods rely on iterative pre-sampling and parameter-update phases, which inherently entail a much higher computational cost. Nevertheless, if the reliability problem features a strongly nonlinear failure boundary or complex orientation of non-dominant design points, FORM-based methods can become infeasible or yield significant estimation errors. In such scenarios, the flexibility of adaptive Monte Carlo methods becomes necessary despite their higher computational costs. Adaptive IS methods seek to iteratively approximate the theoretically ideal, yet practically unattainable, ISPDF to achieve variance reduction in structural reliability analysis. Within this domain, kernel density approximation (KDA), the principle of maximum entropy (PME), the cross-entropy (CE) method, and directional importance sampling (DIS) offer distinct mathematical strategies for targeting the failure domain.

KDA is a non-parametric approach that constructs a sub-optimal ISPDF by weighting a sum of Gaussian kernels centred directly on samples from the failure region. The parameters are mathematically optimized to match the empirical cumulative distribution function (CDF) of the collected samples. Alternatively, KDA can employ wavelet density estimation using orthogonal bases. However, KDA is severely limited by the curse of dimensionality. Because its functional form can grow indefinitely, the number of training samples required to prevent large estimation errors becomes prohibitively large as the number of random variables increases. Like KDA, PME fundamentally relies on generating an initial set of samples directly from the failure domain. However, rather than using local kernel smoothing, PME builds the ISPDF by maximizing Shannon entropy, ensuring the most objective probability distribution possible. It reconstructs the target density dimension by dimension using orthogonal Legendre polynomials to rigorously satisfy joint moment constraints. Because it uses global polynomial expansions, PME avoids the specific bandwidth-scaling limitations of KDA.

The CE method diverges from KDA and PME by operating strictly within a standard Gaussian space and utilizing a pre-defined parameterized family of ISPDFs. Random variables are mapped to this space using an iso-probabilistic transformation. It optimizes the ISPDF by minimizing the Kullback-Leibler divergence (a measure of mathematical distance) between the parameterized family and the ideal ISPDF. The CE method overcomes the major inefficiency of KDA and PME: the need to initially sample extremely rare failure events. It bypasses this by systematically introducing a gradual, smooth approximation transition toward the failure domain, allowing the algorithm to start by sampling the higher-probability regions in the safe domain and step-by-step refine the parameters of the ISPDF toward the ultimate target failure event. Furthermore, CE mitigates the curse of dimensionality by adopting parametric families, such as the von Mises-Fisher distribution, where the unknown parameters scale linearly with the dimension rather than exponentially.

DIS offers a fundamentally different framework by transforming the Gaussian space into polar coordinates, separating random vectors into a radius and a direction. Because the squared radius follows a chi-squared distribution, DIS can analytically evaluate the radial component in a closed form. This mathematically eliminates radial sampling variance and reduces the actual simulation space from $n$ dimensions to a $n-1$ dimensional unit hypersphere. This dimension reduction gives DIS a massive advantage over KDA's prohibitive sample growth and PME's computationally intensive statistical moment constraints. Finally, DIS functions exceptionally well as a framework to enhance other methods. For example, CE can be deployed within DIS to optimize only the direction vector, and DIS can pair with auxiliary limit-state functions to entirely bypass the initialization hurdles that plague standalone KDA and PME.

**Discussion on IS methods for time-variant reliability:** To evaluate time-variant reliability without suffering from weight degeneracy and the curse of dimensionality, structural dynamic systems can be addressed through three primary IS methods: equivalent time-invariant reliability formulations, Girsanov transformations, and extensions for random structural parameters. Each method offers distinct advantages depending on the system's linearity, dimensionality, and excitation type.

The first approach discretizes the random excitation process into an equivalent set of random variables, transforming the time-variant problem into an equivalent time-invariant series system composed of instantaneous failure events. For linear systems with Gaussian excitations, this method is highly advantageous because the solution for the design point is available in closed form, enabling highly efficient sampling via weighted sums of truncated Gaussian probability density functions. For nonlinear dynamical systems, traditional gradient-based optimization to find these design points is computationally demanding. However, this framework overcomes this limitation by employing the critical excitation approach. By determining the minimum energy input required to drive the structural response to a threshold state, the critical excitation method can elegantly locate design points for the ISPDF without resorting to costly gradient-based algorithms.

The second primary approach utilizes the Girsanov transformation for Markovian systems, introducing artificial control forces to drive the response trajectory toward failure while maintaining an unbiased probability estimator and reducing sampling variance. The relative advantage of open-loop controls, which are deterministic and strictly a function of time, lies in their computational tractability. They can be efficiently computed by converting constrained optimization problems into two-point boundary value problems, and their deterministic nature makes them highly practical for implementation in laboratory testing procedures. Their primary drawback, however, is that their state-independent nature limits their ability to achieve the ideal zero-sampling variance.

Conversely, closed-loop controls are state-dependent and stochastic, affording them the crucial advantage of approaching the optimal zero-sampling-variance limit. While computing exact closed-loop controls via the Kolmogorov backward equation poses severe computational challenges for multi-degree-of-freedom structures, researchers have developed highly advantageous approximations. For instance, a two-step procedure can efficiently approximate control functions utilizing grid-based data gathered from open-loop simulations. Furthermore, optimal closed-loop controls can be derived analytically in closed form for instantaneous failure events in linear Gaussian systems, and local linearization techniques extend this advantage to nonlinear systems.

Finally, when structures feature simultaneous randomness in system parameters and external excitations, evaluating reliability typically requires an excessive number of Monte Carlo samples. Tailored IS methods overcome this by centering the ISPDF at representative points that maximize the integrand, or by using mixture PDFs for multimodal domains. A highly advantageous development in this area is adaptive multilevel sampling, which uses tempering parameters to smoothly guide intermediate target densities toward the optimal ISPDF. By leveraging parametric families like the von-Mises Fisher Nakagami distribution, this method guarantees robustness and scalability for problems with many random structural parameters.

## 5.0 Particle and trajectory splitting methods

The basic idea behind splitting methods is to propagate sample realizations towards the failure region ($F$) by introducing a sequence of easier-to-reach "intermediate failure regions" $F_0 \supset F_1 ... \supset F_{l_{\mathrm{fail}}} = F$. In the context of time-invariant reliability, sample realizations are vectors in multidimensional space (visualized as "particles", see Figure 4), and therefore these strategies are termed "particle splitting methods". Similarly, for time-variant reliability estimation, sample realizations are random processes (visualized as "trajectories", see Figure 5), and in this context, these strategies are called "trajectory splitting methods". As the simulation proceeds, samples that reach these intermediate failure regions are replicated or split and those that do not are discarded. The key observation here is that the conditional probability of samples reaching these intermediate regions is much higher than the probability of the samples reaching the true failure region. This makes estimation of the conditional probabilities much more computationally feasible. The key mathematical result that allows us to estimate the failure probability can be stated as follows.

$$P_F = \mathbb{P}\left[F_1\right]\mathbb{P}\left[F_2 \middle| F_1\right]...\mathbb{P}\left[F_{l_{\mathrm{fail}}} \middle| F_{l_{\mathrm{fail}}-1}\right] \tag{81}$$

Since the conditional probabilities $\mathbb{P}\left[F_l \middle| F_{l-1}\right]$ are relatively large (of the order of $10^{-1}$), their estimation requires substantially fewer samples compared to the true failure probability (which is typically of the order of $10^{-4}$ or smaller). The origin of the idea of splitting methods can be traced back to the work of Kahn and Harris [1951] pertaining to nuclear physics, where the authors were interested in the small probability that a neutron is transmitted through a shield or a barrier without being absorbed or reflected. Subsequently, the mathematics of splitting methods has been explored and developed quite extensively in the computational statistics literature. We refer to the paper by Glasserman et al., [1999] for a discussion on earlier works, the textbook by Kroese et al., [2011] and the review paper by Cerou et al., [2019] for more recent developments in this area.

The idea of trajectory splitting methods first appeared in the structural reliability literature in the form of the "double and clump" method proposed by Pradlwarter et al., [1994]. A particle splitting method was first employed in the context of structural reliability as the "subset simulation" method, which was proposed by Au and Beck [2001a] (see the book by Au and Wang [2014] for a detailed treatment of the algorithm.). Kanjilal and Manohar [2015] have investigated the application of several particle splitting methods developed in the computational statistics literature for estimation of the reliability integral. The authors compared statistical properties such as bias, sampling variance, and approximate sampling distributions of the subset simulation estimator vis-à-vis other particle splitting methods.

The present review focuses on the development of particle and trajectory splitting methods in the structural reliability literature. Accordingly, this section comprises two categories of methods, viz., particle splitting methods for time-invariant reliability and trajectory splitting methods for time-variant reliability estimation.

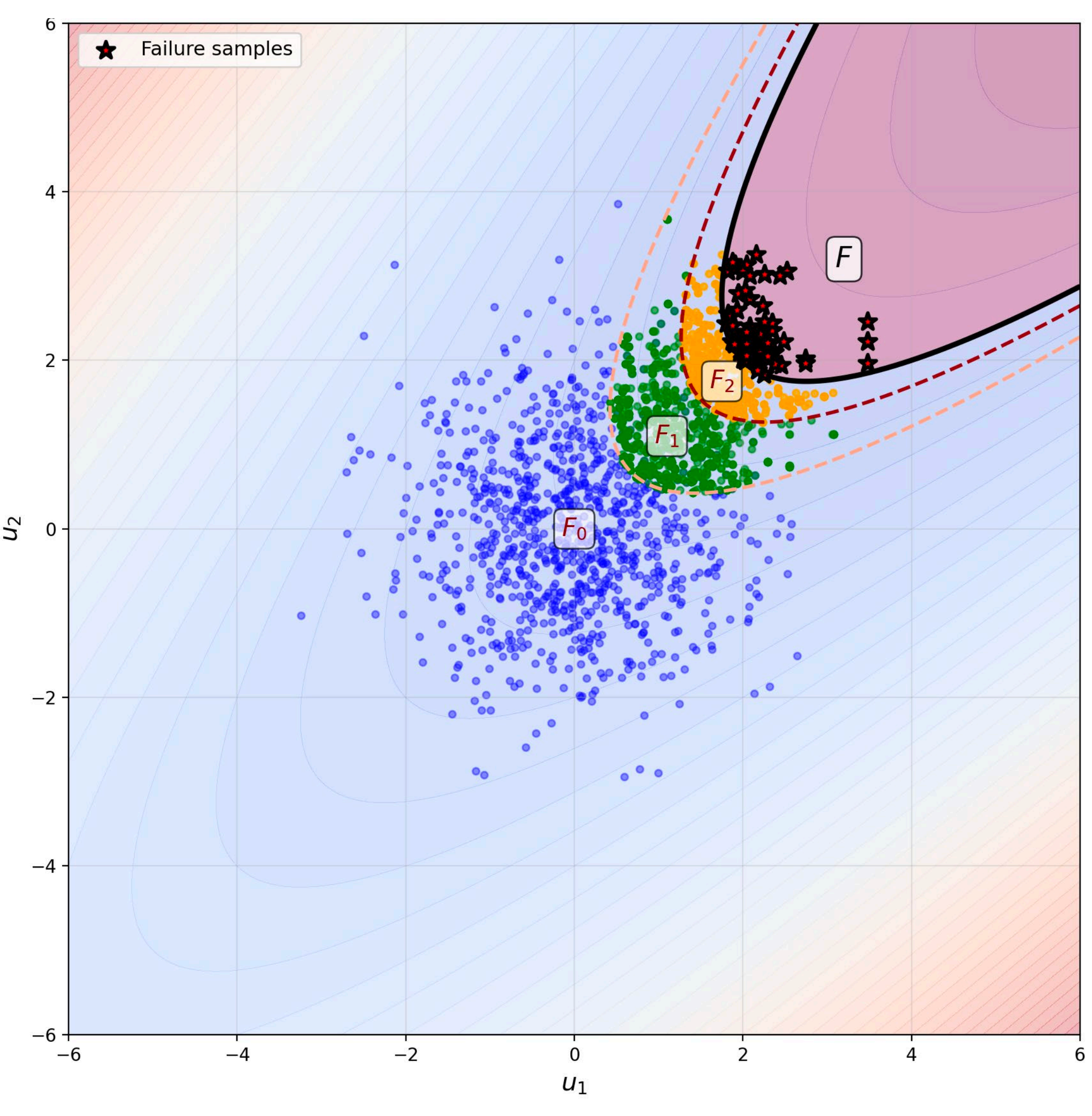

Figure 4: Visualizing particle splitting methods

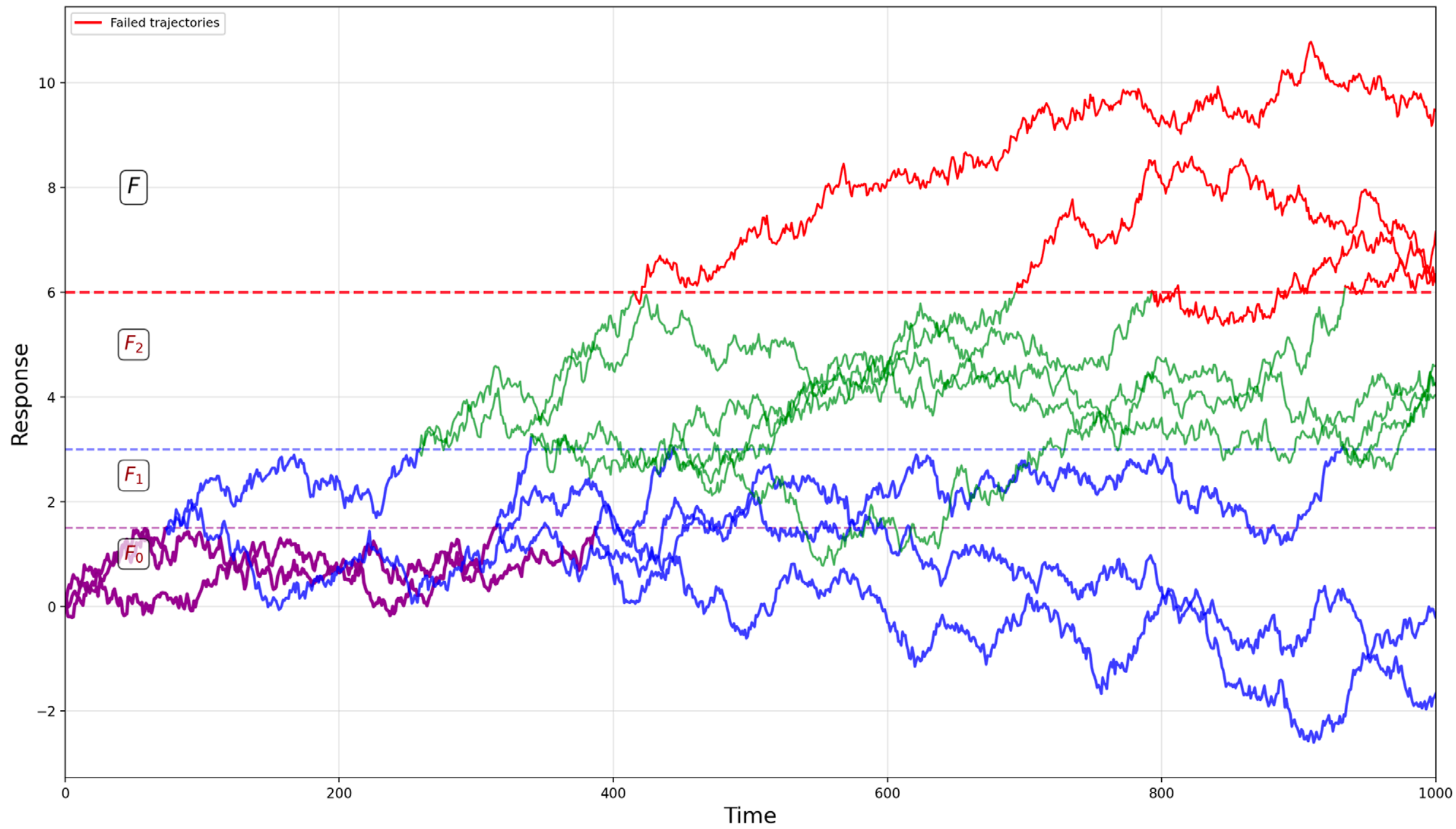

Figure 5: Visualizing trajectory splitting methods

### 5.1 Particle splitting methods for time-invariant reliability estimation

In particle splitting methods, intermediate failure regions are defined as $F_l = \left\{ \boldsymbol{u} \in \mathbb{R}^{n_d} : G(\boldsymbol{u}) \leq c_l \right\}; l = 1, 2, ..., l_{\text{fail}}$ such that $c_1 > c_2 > ... > c_{l_{\text{fail}}} = 0$ and $\mathbb{R}^{n_d} = F_0 \supset F_1 ... \supset F_{l_{\text{fail}}} = F$, where the actual failure event is defined as $F = \left\{ \boldsymbol{u} \in \mathbb{R}^{n_d} : G(\boldsymbol{u}) \leq 0 \right\}$. Following this, the failure probability $P_F$ is obtained as $P_F = \mathbb{P}[F_1]\mathbb{P}[F_2 | F_1]...\mathbb{P}\left[F_{l_{\text{fail}}} \middle| F_{l_{\text{fail}}-1}\right]$, and we re-emphasize that $P_F$ is now expressed as a product of larger conditional probabilities which are easier to estimate with a substantially smaller number of samples. Research in this area has been driven by the following considerations.

#### 5.1.1 Choice of the intermediate failure regions

The mechanism of choosing the intermediate failure regions is central to any particle splitting algorithm. These regions may be chosen a priori by the analyst either (a) through a simplified analysis of the problem, as in Kanjilal and Manohar [2015], or (b) via a preliminary small-scale simulation as in the study by Sharma and Manohar [2023]. However, in the reliability literature, it is more common to determine these regions adaptively when implementing the subset simulation method and its variants [Au and Beck 2001a]. This is done by specifying the value of the conditional probabilities $\mathbb{P}\left[F_l \middle| F_{l-1}\right] = p_0$ for each $l \geq 1$ , and the intermediate threshold value $c_l$ is chosen adaptively to ensure that $\mathbb{P}\left[F_l \middle| F_{l-1}\right] = p_0$. This is done by generating $N_S$ samples $\boldsymbol{u}^i, i = 1, 2, ..., N_S$ from $p_U\left(\boldsymbol{u} \middle| F_{l-1}\right) = I\left[\boldsymbol{u} \in F_{l-1}\right] \frac{p_U\left(\boldsymbol{u}\right)}{\mathbb{P}\left(F_{l-1}\right)}$ for each $l \geq 1$. The subsequent intermediate threshold value $c_l$ is chosen by arranging the values $G\left(\boldsymbol{u}^i\right),\ i = 1, 2, ..., N_S$ in ascending order and selecting $c_l = G\left(\bar{\boldsymbol{u}}\right)$, where $G\left(\bar{\boldsymbol{u}}\right)$ is the $N_S p_0{}^{\text{th}}$ value in the ascending sequence. This procedure is repeated until $c_l \leq 0$ , and we set $l_{\text{fail}} = l$. The failure probability estimator is then computed as $\hat{P}_F = p_0^{l_{\text{fail}}-1} \frac{N_F}{N_S}$, where $N_F$ is the number of failed samples at the end of the simulation. In this setting, the value of $p_0$ is an important algorithmic parameter. For a given number of samples per level, if $p_0$ is too small, then the intermediate failure thresholds would exhibit large sampling fluctuations, leading to high estimator variance. On the other hand, if $p_0$ is too large, then the number of intermediate failure thresholds required to reach the failure region would be too high, which would be computationally demanding. The study by Zuev et al., [2012] showed that choosing $p_0 \approx 0.2$ minimizes the coefficient of variation of the subset simulation estimator, and noted that values ranging from 0.1 to 0.3 lead to similar performance. The choice of $p_0 = 0.1$ is commonly used in practice.

### 5.1.2 Estimation of the conditional probabilities:

Estimation of $\mathbb{P}\left[F_l \middle| F_{l-1}\right]$ requires generating samples from the multidimensional conditional Gaussian pdf $p_U\left(\boldsymbol{u} \middle| F_{l-1}\right) = I\left[\boldsymbol{u} \in F_{l-1}\right] \frac{p_U\left(\boldsymbol{u}\right)}{\mathbb{P}\left(F_{l-1}\right)}$. In engineering applications, the indicator function $I\left[\boldsymbol{u} \in F_{l-1}\right]$ would typically need to be computed based on an implicitly defined performance function. Furthermore, depending on the details of the particle splitting algorithm, the normalizing constant $\mathbb{P}(F_{l-1})$ may not be available to the analyst. Given this setting, Markov chain Monte Carlo (MCMC) sampling algorithms appear to be the most suitable way forward. Au and Beck [2001a] showed that the acceptance rate for the vanilla Metropolis sampler tends to zero as the number of dimensions increases. The authors therefore proposed the modified Metropolis algorithm to generate conditional samples in high dimensions to resolve this issue. The scaling of the proposal pdf is central to the success of any MCMC sampler and therefore of a particle splitting method. This governs the correlation between samples of a chain which, in turn, contributes to the sampling variance of the failure probability estimate. In this light, Zuev et al., [2012] suggested a heuristic strategy where variance of the proposal pdf in the modified Metropolis algorithm is chosen such that the acceptance rate is between 30% and 50%. A recent study by Sepúlveda et al., [2025] suggested that the acceptance rate should be 35-45%. Au and Zhou [2025] have shown through a mathematical analysis of a simplified reliability problem that for the choice of $p_0 = 0.1$, which is commonly adopted in subset simulation, the optimal acceptance rate asymptotically tends to 44.8%. A more recent study by Au [2026] derives an alternative measure for correlation between samples, termed the "failure mixing rate". Development of advanced MCMC samplers focusing on their ability to explore the state space, on reducing correlation between successive samples, and on improving the acceptance rate has remained a central theme of research in this area. Section 5.1.6 provides a review of MCMC methods that have been applied in the context of particle splitting methods. Alternative strategies of generating such samples have also been explored in the literature.

Song et al., [2009] proposed using an ISPDF to generate the intermediate failure samples instead of MCMC samplers. This is done for each intermediate failure region by choosing, among the samples generated, the sample with the highest pdf value. The ISPDF is then centred at this point and used to generate the subsequent intermediate failure samples. The authors also propose two methods for sensitivity of the failure probability with respect to the parameters of the distribution of the basic random variables. The study by Wang et al., [2025] combined the ISPDF-based technique to generate intermediate failure samples with the probability reanalysis method proposed by Nikolaidis et al., [2009]. Helal and Elvira [2026] have proposed an adaptive importance sampling-based approach where, similar to subset simulation, intermediate samples are guided towards the failure region based on a mixture of importance sampling densities instead of MCMC sampling.

In contrast to importance sampling, Jia et al., [2017] have proposed an alternative method of generating independent samples from the intermediate failure region $F_l$ by employing rejection sampling using a proposal density obtained via adaptive kernel sampling density. In a subsequent study, Jia et al., [2021] developed a method called the density extrapolation approach, which uses the kernel density estimation method within the framework of subset simulation by rewriting the failure probability as $\hat{P}_F = \tilde{P}\left(F_m \middle| F_n\right)\prod_{k=1}^{n}\mathbb{P}\left(F_k \middle| F_{k-1}\right)$ where $\tilde{P}\left(F_m \middle| F_n\right)$ is the tail approximation of the conditional probability, which is estimated using kernel density estimation.

**5.1.3 Reliability estimation for multiple performance functions**

It may be of interest in structural engineering applications to compute failure probabilities for multiple performance functions associated with different limit states. In such cases, it would be computationally beneficial to estimate these probabilities with a single run of a particle splitting method. The study by Hsu and Ching [2010] proposed the parallel subset simulation where the authors propose an algorithm to simultaneously compute failure probability corresponding to multiple performance functions $G_i(\boldsymbol{X}), i = 1,2,...,N_{SYS}$. The main idea is to define a principal variable $\tilde{H}$ as $\tilde{H}\left(\boldsymbol{U}\right) = \frac{1}{N_{SYS}}\sum_{i=1}^{N_{SYS}} G_i\left(\boldsymbol{U}\right)$ and then execute subset simulation using $\tilde{H}\left(\boldsymbol{U}\right)$ as the performance function. The failure probability for each performance function is computed based on the following expression.

$$\begin{aligned}
P_F^i &= \mathbb{P}\left(G_i\left(\boldsymbol{U}\right) < 0 \middle| \tilde{H}\left(\boldsymbol{U}\right) > c_1\right)\mathbb{P}\left(\tilde{H}\left(\boldsymbol{U}\right) > c_1\right) + \\
&\sum_{k=1}^{l_{\text{fail}}-1}\mathbb{P}\left(G_i\left(\boldsymbol{U}\right) < 0 \middle| c_k \ge \tilde{H}\left(\boldsymbol{U}\right) \ge c_{k+1}\right)\mathbb{P}\left(c_k \ge \tilde{H}\left(\boldsymbol{U}\right) \ge c_{k+1}\right) \\
&+\mathbb{P}\left(G_i\left(\boldsymbol{U}\right) < 0 \middle| \tilde{H}\left(\boldsymbol{U}\right) > c_{l_{\text{fail}}}\right)\mathbb{P}\left(\tilde{H}\left(\boldsymbol{U}\right) > c_{l_{\text{fail}}}\right),\ i = 1,2,...,N_{SYS}
\end{aligned} \tag{82}$$

A subsequent study by Du et al., [2019] proposed a parallel subset simulation framework for time-variant reliability estimation. Studies by Li et al., [2015] and Bansal and Cheung [2017] also contend with the problem of estimating failure probability with multiple performance functions by modifying the original subset simulation algorithm. In these studies, a unified intermediate failure region is used by taking the union of the individual intermediate failure region and generating samples that simultaneously approach the failure region corresponding to each performance function. This allows the computation of failure probability corresponding to each performance function with a single subset simulation run. Xia and Liao [2025] have recently combined multiple importance sampling with the aforementioned method proposed by Li et al., [2015] by rewriting the failure probability as the weighted sum $P_F = \sum_{i=1}^{M-1} \int \alpha_i(\boldsymbol{u}) p_U(\boldsymbol{u}) I(\boldsymbol{u} \in F)\, d\boldsymbol{u}$ where the weight functions $\alpha_i(\boldsymbol{u})$ are chosen such that $\sum_{i=1}^{M-1} \alpha_i(\boldsymbol{u}) = 1$ when $p_U(\boldsymbol{u}) I(\boldsymbol{u} \in F) \neq 0$. With a suitable choice of the weight functions, the authors derive the following estimator for failure probability $\hat{P}_F = \sum_{i=0}^{M} \frac{1}{N} \sum_{k=1}^{N} \frac{I(\boldsymbol{u}_{i,k} \in F)}{\sum_{i=0}^{M-1} \hat{p}_{i_1}^{-1} I(\boldsymbol{u}_{i,k} \in F_{i_1})}$. More recently, Naderi and Jia [2026] have proposed a subset simulation framework handling multi-fidelity models. The authors treat the fidelity of the model as a random variable, solve the extended reliability problem using subset simulation, and subsequently use Bayes' theorem to derive the failure probability estimate corresponding to the high-fidelity model.

### 5.1.4 Reliability estimation for performance functions with geometric complexities

Several studies by Breitung [2019, 2021] have pointed out a number of examples where vanilla subset simulation method underestimates the probability of failure due to the underlying the MCMC sampler being unable to draw representative samples from the conditional pdf $p_U\left(\boldsymbol{u} \middle| F_{l-1}\right) = I\left[\boldsymbol{u} \in F_{l-1}\right] \frac{p_U(\boldsymbol{u})}{\mathbb{P}\left(F_{l-1}\right)}$. In particular, the author showed that in certain cases with multiple or disconnected regions of failure, possibly with sharp changes or discontinuities in the performance function, the samples generated during a particle splitting algorithm may converge to a less important failure region. This difficulty arises because as the Markov chain samples propagate to subsequent intermediate failure regions, the geometry of these regions may change such that the samples fail to detect and sufficiently populate the failure regions of interest. These studies prompted several developments attempting to address this challenge in the subset simulation framework, leading to two broad approaches: (a) improving the exploration capacity of the underlying MCMC sampler being used so that the important failure region is effectively identified and sampled from, or (b) modifying the particle splitting algorithm so that the exploratory ability of the resulting method is improved. Strategy (a) has been explored by Sharma and Manohar [2023] and Sharma et al., [2026], where they adopt a modified replica exchange MCMC sampler within a particle splitting framework (see Section 5.1.6). The following paragraphs focus on methods that adopt strategy (b), which has received more attention in the recent literature.

Abdollahi et al., [2020] have proposed a modification to the subset simulation algorithm, called the subset control variate. The idea behind the subset control variate is to consider an intermediate failure region, say $\overline{F}$ , and rewrite the failure probability as

$$\begin{aligned} P_F &= \int_F p_U(\boldsymbol{u})\,d\boldsymbol{u} = \alpha\int_{\overline{F}} p_U(\boldsymbol{u})\,d\boldsymbol{u} + \int_F p_U(\boldsymbol{u})\,d\boldsymbol{u} - \alpha\int_{\overline{F}} p_U(\boldsymbol{u})\,d\boldsymbol{u} \\ &= \alpha P_{\overline{F}} + \int_{\mathbb{R}^N}\left[I\left(G(\boldsymbol{u})\le 0\right) - \alpha I\left(G(\boldsymbol{u})\le c\right)\right]p_U(\boldsymbol{u})\,d\boldsymbol{u}, \text{ where } c>0 \end{aligned} \tag{83}$$

The authors rewrite this equation as

$$P_F = \alpha P_{\overline{F}} + \int_{\mathbb{R}^{n_d}} P_{\overline{F}}\left[I\left(G(\boldsymbol{u})\le 0\right) - \alpha I\left(G(\boldsymbol{u})\le c\right)\right]h^*(\boldsymbol{u})\,d\boldsymbol{u} \tag{84}$$

Here, $h^*(u)$ denotes the instrumental pdf that generates samples that lie in the intermediate failure region $\overline{F}$. The coefficient $\alpha$ is determined by setting the second term of Equation (84) to zero. This leads to the following expression for $\alpha$.

$$\alpha = \frac{\int_{\mathbb{R}^N} I\left(G(u)\le 0\right)h^*(u)\,du}{\int_{\mathbb{R}^N} I\left(G(u)\le c\right)h^*(u)\,\mathrm{du}} \tag{85}$$

The coefficient $\alpha$ is computed using an MCMC sampling technique. The resulting failure probability estimator is $P_F = \alpha P_{\overline{F}}$. The authors implement the above procedure in an iterative fashion that leads to the failure probability estimator $P_F = \prod_{i=1}^{n-1}\alpha_i P_{\overline{F}_i}$ where $\overline{F}_i$ are the intermediate failure regions. The control variate-based simulation method was also explored in a previous study by Rashki [2018]. Additionally, based on the control variate technique, Rashki [2021] has developed the sequential space conversion (SESC) method for reliability estimation. Subsequently, the SESC method has been integrated with active learning based on kriging surrogate modeling, leading to the AK-SESC method [Ameryan et al., 2022], and has also more recently been combined with the multi-armed bandit algorithm for time-variant reliability estimation by Aliahmad et al., [2026]. The study by Abdollahi et al., [2021] proposed four alternative procedures to select seeds for initiation of the MCMC samplers in intermediate failure regions in the context of subset simulation. These procedures are rooted in the procedures for sample selection in genetic algorithms. The procedures assign a weight to each intermediate sample $\boldsymbol{U}$ by computing the joint pdf values in the standard normal space $p_U(\boldsymbol{u})$, say $W = \{W_1, W_2, ..., W_{N_S}\}$. The four proposed procedures for seed selection are as follows:

a) Proportional probability: Here, the seeds lying in the intermediate failure region are selected with probability proportional to their corresponding weights.
b) Roulette wheel I and II: In this procedure, two sets of weights are defined for each sample: (i) $W_I^k = \frac{W_k}{\sum_{j=1}^{N_S} W_j}$ for Roulette wheel I and (ii) $W_{II}^k = \frac{W_k}{\max(W)}$ for Roulette wheel II. The corresponding cumulative probability masses $q_I^i = \sum_{k=1}^{i} W_I^k$ and $q_{II}^i = \sum_{k=1}^{i} W_{II}^k$ are defined for each sample. A sample of $r \sim$ U(0,1) is generated. If $r < q_I^1$, $\boldsymbol{U}^i$ is chosen as the seed. Otherwise, $\boldsymbol{U}^i$ is chosen as the first seed where $q_I^{i-1} < r < q_I^i$. This is repeated until $N_S$ seeds are chosen.
c) Tournament: Here, a pre-chosen number of seeds are selected uniformly at random, and the one with the highest weight value is saved. This procedure is repeated until $N_S$ seeds are obtained.

These seeds are used to initiate the MCMC sampler in order to generate the intermediate failure samples.

Kinnear and DiazDelaO [2025] proposed the niching subset simulation, which draws from the area of evolutionary multimodal optimization to address the multimodality that arises due to geometric complexities such as multiple regions of failure or rapid changes in performance functions. The authors employ a niching strategy, which has been explored in the context of evolutionary optimization, and adaptively partition the space of random variables using support vector machines to deal with these geometric challenges. The authors also propose a novel niching strategy called the hill valley graph that performs well in high dimensions. Kinnear and DiazDelaO [2026] have also recently proposed an importance sampling approach based on the niching strategy for such performance functions.

Li et al., [2025] proposed the relaxed subset simulation where a subset simulation algorithm is implemented on a modified indicator function for each intermediate failure region defined as $I_{\widetilde{F}_i}(\boldsymbol{u}) = \begin{cases} 1 & \text{if } \boldsymbol{u} \in \widetilde{F}_i \\ 0 & \text{if } \boldsymbol{u} \notin \widetilde{F}_i \end{cases}$, where $\widetilde{F}_i = \{\boldsymbol{u} \in \mathbb{R}^{n_d}: G(\boldsymbol{u}) \leq b_i \text{ or } p_U(u) \leq \phi_i\}$. This hybrid indicator function is designed to promote the exploration of the input space so that the algorithm is able to detect relevant failure regions even when the underlying performance function has geometric complications.

### 5.1.5 Additional methods and perspectives on particle splitting methods

A different perspective on reliability estimation involves a Bayesian interpretation of $\hat{P}_F$ by modeling it as a stochastic variable and deducing the distribution of $\hat{P}_F$ as opposed to a single point estimate [Zuev et al., 2012]. In this study, each conditional probability is assigned a uniform prior and is updated using Bayes' theorem based on the conditional failure samples generated by the MCMC sampler. The study by Bect et al., [2017] proposed the Bayesian subset simulation, where subset simulation is used in conjunction with a stepwise uncertainty reduction algorithm based on Gaussian processes. Lei et al., [2025] combined subset simulation with the variational Bayesian Monte Carlo framework, where the intermediate conditional density is obtained by maximizing the evidence lower bound.

The study by Sharma and Manohar [2023] used the bootstrap algorithm to derive the approximate sampling distribution and sampling variance of the failure probability estimator noting that approximations and asymptotic assumptions involved in analytically deriving sampling variance of such estimates may not be satisfied in practice. The authors use a circular block bootstrap strategy to take into account the sample correlations that inevitably arise in MCMC sampling. In a study by Liao et al., [2024a], the authors empirically compared lognormal and beta distributions as candidates for the sampling distribution of $\hat{P}_F$ by considering the Kolmogorov-Smirnov (KS) statistic against the empirical distribution of $\hat{P}_F$ obtained from independent runs of subset simulation. For the numerical examples considered in the study, the lognormal distribution produced lower values of the KS statistic and is seen to approximate the empirical distribution more closely. Recently, Miao and Low [2025] derived a sampling variance estimator for subset simulation by taking into account the correlation between Markov chains induced by their dependence on correlated samples from preceding levels.

Sen and Bhattacharya [2015] have examined variance reduction techniques from a Pareto optimality perspective. The authors note that reliability estimation involves two competing objectives: (a) reducing the error of the failure probability estimate, and (b) reducing the computational effort required to obtain an estimate. Since improvement in one of the objectives hinders the other, the authors formulate the problem as a multi-objective stochastic optimization problem as follows.

$$\min_{\boldsymbol{x} \in D} F(\boldsymbol{x}) = \left[\mathbb{E}\left(F_1(\boldsymbol{x})\right) \quad \mathbb{E}\left(F_2(\boldsymbol{x})\right)\right]^{\mathrm{T}} \tag{86}$$

Here, $\boldsymbol{x}$ denotes the vector of algorithmic parameters, $D$ is the feasible parameter space, and the first and second components denote the expected error and the expected computational effort involved in each variance reduction scheme. The authors define the first objective as $\frac{\mathbb{E}[(\hat{P}_F - P_F)^2]}{P_F^2}$, and the second objective as the average number of performance function calls. In the numerical implementation, the expectations are computed based on 5 independent runs of each algorithm. The multi-objective stochastic optimization is solved using NSGA-II, an evolutionary multi-objective optimization algorithm. Based on the results obtained from the optimization problem, for each problem, the authors identify the algorithmic parameter combinations associated with Pareto optimal performance and conclude that in design point-based IS, the design point is not necessarily the optimal point to centre the ISPDF, and that the proposal pdf in subset simulation affects the quality of failure probability estimates.

Ullmann and Papaioannou [2015] proposed a generalization of subset simulation and derived a multilevel estimator for failure probability broadly based on the idea of multilevel Monte Carlo. The study by Giovanis and Shields [2022] uses subset simulation along with information-theoretic multi-model selection when the probability models for the basic random variables are uncertain. Hristov and DiazDelaO [2023] propose a version of subset simulation where the outputs of the performance function evaluation are considered as random variables, enabling the use of subset simulation in the presence of computational uncertainty.

The study by Cheng et al., [2022] proposed a generalized subset simulation algorithm. An auxiliary probability density function is defined, $p_\sigma(u)$ which is the standard normal pdf with a magnified standard deviation $\sigma$. The authors show that the failure probability can be rewritten as $P_F = P_1 \prod_{i=1}^{k} S_i$ where $P_1 = \int_{\mathbb{R}^N} I\left(G(u) \le b_1\right) p_{\sigma_1}(u) du$ and $S_i = \mathrm{E}_{h_i}\left(w_i(u)\right)$ where $\sigma_1$ and $b_1$ are to be chosen by the analyst, $h_i(u) = \frac{I\left(G(u) \le c_i\right) p_{\sigma_i}(u)}{P_i}$, and $w_i(u) = \frac{I\left(G(u) \le c_{i+1}\right)}{I\left(G(u) \le c_i\right)} \frac{p_{\sigma_{i+1}}(u)}{p_{\sigma_i}(u)}$ which is estimated at each conditional level by executing an MCMC algorithm with target pdf $h_i(u) = \frac{I\left(G(u) \le b_i\right) p_{\sigma_i}(u)}{P_i}$. The authors show that if the magnification factors in this algorithm are equal to unity, then this algorithm reduces to the subset simulation algorithm.

Chan et al., [2022] have considered the problem of reliability estimation when the performance function is discontinuous and/or the basic random variables are discrete in nature. The authors have proposed an adaptive effort-based modification to the subset simulation algorithm and a novel independent Metropolis-Hastings sampler for drawing samples from the intermediate failure regions for such problems. Chan et al., [2024] have subsequently combined this method with the annealed particle integration method. The study by Vořechovský [2022] also considers the problem of discrete input variables or discontinuous performance functions, and the author proposes an adaptive sequential sampling strategy along with a measure of sensitivity of failure probability with respect to the input random variables.

Breitung [2024] investigated the relationship between the FORM/SORM-based approaches, several variance reduction techniques, and Gaussian process-based surrogate modeling techniques under the asymptotic condition that the Hasofer-Lind reliability index $\beta_{HL} \to \infty$. The methods considered include subset simulation method [Au and Beck 2001a], cross-entropy-based importance sampling [Geyer et al., 2019], density extrapolation method [Jia et al., 2021], line sampling [Koutsourelakis et al., 2004], and importance sampling techniques based on Gaussian mixture [Kurtz and Song, 2013], kernel density estimation [Au and Beck, 1999], and surrogate modelling with Gaussian processes [Dubourg et al., 2013]. The author has shown that under the asymptotic condition $\beta_{HL} \to \infty$, the samples generated during the course of these methods approach the neighbourhood of the design points. Therefore, the author emphasizes that under this asymptotic condition, the neighbourhoods of the design points are the only important regions of failure. Rashki et al., [2025] also recently studied the relationship between SORM and subset simulation-based approaches. The authors propose a reliability estimation framework based on extreme value theory which is used to extrapolate the results obtained from subset simulation.

### 5.1.6 A review of MCMC techniques used in particle splitting methods

In this section, we present a brief summary of the MCMC samplers that have been implemented in the context of particle splitting methods. For the purposes of the following discussion, the possibly unnormalized target density is assumed to be a $n_d$-dimensional pdf $p(\boldsymbol{x})$. For each MCMC sampler, we outline the transition from the current state $\boldsymbol{x}_i$ to $\boldsymbol{x}_{i+1}$.

i. Modified Metropolis algorithm [Au and Beck 2001a]: Given the current state $\boldsymbol{x}_i = \left[x_i(1) \quad x_i(2) \quad \ldots \quad x_i(n_d)\right]$, a proposed sample for each component is separately generated using a symmetric proposal pdf $x_i'(j) \sim q_j\left(x \middle| x_i(j)\right)$, and for each component, the acceptance probability is $\alpha_j\left(x_i(j), x_i'(j)\right) = \min\left(1, \frac{p_j\left(x_i'(j)\right)}{p_j\left(x_i(j)\right)}\right)$. The resulting sample is accepted if it lies in the intermediate failure region.

ii. Modified Metropolis-Hastings with delayed rejection [Zuev and Katafygiotis 2011]: Given the current state $\boldsymbol{x}_i = \left[x_i(1) \quad x_i(2) \quad \ldots \quad x_i(n_d)\right]$, a proposed sample for each component is separately generated $\xi_i(j) \sim q_j\left(\xi \middle| x_i(j)\right)$, and the acceptance probability is $\alpha_j = \min\left(1, \frac{p_j\left(\xi_i(j)\right) q_j\left(x_i(j) \middle| \xi_i(j)\right)}{p_j\left(x_i(j)\right) q_j\left(\xi_i(j) \middle| x_i(j)\right)}\right)$. If the proposed sample lies in the failure region, accept the sample. Otherwise, generate a new proposed state $\bar{\xi}_i(j) \sim \bar{q}_j\left(\xi \middle| x_i(j), \xi_i(j)\right)$. The acceptance probability is $\bar{\alpha}_j = \min\left(1, \frac{p_j\left(\bar{\xi}_i(j)\right) q_j\left(\xi_i(j) \middle| \bar{\xi}_i(j)\right) \bar{q}_j\left(x_i(j) \middle| \xi_i(j), \bar{\xi}_i(j)\right) \alpha_j\left(\bar{\xi}_i(j), \xi_i(j)\right)}{p_j\left(x_i(j)\right) q_j\left(\xi_i(j) \middle| x_i(j)\right) \bar{q}_j\left(\bar{\xi}_i(j) \middle| \xi_i(j), x_i(j)\right) \alpha_j\left(x_i(j), \xi_i(j)\right)}\right)$.

iii. Regenerative adaptive MCMC [Miao and Ghosn 2011]: Given the current state $\boldsymbol{x}_i = \left[ x_i(1) \quad x_i(2) \quad \dots \quad x_i(n_d) \right]$, a proposed sample is generated $y_i(j) \sim q_j\left(y \middle| x_i(j)\right)$. If the modified Metropolis criterion is satisfied, a regeneration step is executed, the detailed steps for which have been described in the study by Miao and Ghosn [2011]. If the sample fails the modified Metropolis criterion, the delayed rejection algorithm outlined earlier is performed.

iv. Modified Metropolis-Hastings algorithm with reduced chain coefficient [Santoso et al., 2011]: Given the current state $\boldsymbol{x}_i$ of the chain, a proposed sample is generated according to the proposal pdf $\boldsymbol{x}' \sim q\left(\boldsymbol{x} \middle| \boldsymbol{x}_i\right)$. The acceptance probability is computed as $\alpha = \min\left(1, \dfrac{p(\boldsymbol{x}')q\left(\boldsymbol{x} \middle| \boldsymbol{x}'\right)}{p(\boldsymbol{x})q\left(\boldsymbol{x}' \middle| \boldsymbol{x}\right)}\right)$. If the sample is rejected, generate a different proposed sample $\boldsymbol{x}' \sim q\left(\boldsymbol{x} \middle| \boldsymbol{x}_i\right)$ and repeat until acceptance.

v. Conditional sampling in $\boldsymbol{U}$-space [Papaioannou et al., 2015]: Here, we assume the target density to be $p_U\left(\boldsymbol{u} \middle| F\right)$. Given the current state $\boldsymbol{x}_i = \left[ x_i(1) \quad x_i(2) \quad \dots \quad x_i(n_d) \right]$, generate a proposed sample $x_i'(j) \sim N\left(\rho_j x_i(j), \sqrt{1-\rho_j^2}\right)$ for each component. If the resulting sample lies in the failure region, the sample is accepted. By considering a simplified reliability problem, Au and Zhou [2025] show mathematically that the correlation parameter should adapt with the conditional levels to keep the acceptance rate satisfactory. Papaioannou et al., [2015] have also proposed an adaptive conditional sampling algorithm that adjusts the correlation parameter $\rho_j$, where a separate parameter $p_a$ determines the number of times the correlation parameter is adjusted. A value of $p_a = 0.1$ is suggested by Sepúlveda et al., [2025] based on several case studies considered in their work.

vi. Parallel tempering [Xiao et al., 2019]: Given the current state, $\boldsymbol{x}_i$ define a set of tempered pdfs $p_j(\boldsymbol{x}) \propto \left[p(\boldsymbol{x})\right]^{1/T_j}, j = 1, 2, \dots, M_T$. Parallelly run $M_T$ MCMC samplers. At each time step, choose two chains corresponding to $p_k(\boldsymbol{x})$ and $p_l(\boldsymbol{x})$ with states $\boldsymbol{x}_k$ and $\boldsymbol{x}_l$, respectively. Exchange the states between the chains with probability $\alpha^{PT}(\boldsymbol{x}_k, \boldsymbol{x}_l) = \min\left(1, \frac{p_k(\boldsymbol{x}_l) p_l(\boldsymbol{x}_k)}{p_l(\boldsymbol{x}_k) p_k(\boldsymbol{x}_l)}\right)$.

vii. Affine invariant ensemble MCMC [Shields et al., 2021]: Here, $n_C$ chains with current states $\boldsymbol{x}^i, i = 1, 2, ..., n_C$ are initiated. A complementary ensemble is defined as $\boldsymbol{x}^{\sim i} = \{\boldsymbol{x}^i, i \neq j\}$. To propagate state $\boldsymbol{x}^i$, select $\boldsymbol{x}*$, a randomly chosen state from $\boldsymbol{x}^{\sim i}$. A proposed sample is generated as $\boldsymbol{y} = \boldsymbol{x}* + z(\boldsymbol{x}^i - \boldsymbol{x}*)$ where $z$ is distributed as $p(z) \propto \frac{1}{\sqrt{z}}, z \in \left[\frac{1}{a}, a\right]$, $a = 2$. The acceptance probability is $\bar{\alpha}(\boldsymbol{x}^i, \boldsymbol{y}) = \min\left(1, z^{n_d - 1} \frac{p(\boldsymbol{y})}{p(\boldsymbol{x}^i)}\right)$.

viii. Hamiltonian Monte Carlo [Wang et al., 2019, Chen et al., 2022, Thaler et al., 2024]: The current state $\boldsymbol{x}_i$ of the chain is interpreted as the position vector $\boldsymbol{q}$ in a Hamiltonian system. The momentum $\boldsymbol{p}$ is distributed according to the multivariate Gaussian pdf. The Hamiltonian is defined by $H(\boldsymbol{q}, \boldsymbol{p}) = U(\boldsymbol{q}) + K(\boldsymbol{p})$, and the Hamiltonian dynamics are given by $d\boldsymbol{q}/dt = \partial H / \partial \boldsymbol{p}$, and $d\boldsymbol{p}/dt = -\partial H / \partial \boldsymbol{q}$, where $U(\boldsymbol{q}) = -\ln(p(\boldsymbol{q}))$ and $K(\boldsymbol{p}) = \frac{1}{2}\boldsymbol{p}\mathbf{M}^{-1}\boldsymbol{p}$. The mass matrix $\mathbf{M}$ is generally assumed to be a multiple of the identity matrix. In the Riemannian manifold Hamiltonian Monte Carlo [Chen et al., 2022], the mass matrix is adapted using the Hessian of the target pdf. The proposal state $(\boldsymbol{q}*, \boldsymbol{p}*)$ is obtained by integrating the Hamiltonian equations using the leapfrog integrator, and the acceptance probability is given as $\bar{\alpha} = \min\left[1, \exp\left(H(\boldsymbol{q}, \boldsymbol{p}), H(\boldsymbol{q}*, \boldsymbol{p}*)\right)\right]$. In the study by Thaler et al., [2024], the authors combine Hamiltonian Monte Carlo with Hamiltonian neural networks, which reduces the computational effort required in integrating the Hamiltonian equations.

ix. Replica exchange [Sharma and Manohar 2023]: In addition to the target pdf $p_U(\boldsymbol{x}|F_l)$, an additional conditional standard normal pdf called the explorer pdf, $p_U(\boldsymbol{y}|C_l)$ is defined. The idea behind replica exchange is to initiate two chains: (a) one drawing samples from the target pdf, called the sampler chain, and the other (b) drawing samples from the explorer pdf, called the explorer chain. The states of these two chains (say $\boldsymbol{x}_i$ and $\boldsymbol{y}_i$) are occasionally exchanged by computing the ratio $\alpha^{RE} = \min\left[1, \frac{p_U(\boldsymbol{y}_i|F_l)p_U(\boldsymbol{x}_i|C_l)}{p_U(\boldsymbol{x}_i|F_l)p_U(\boldsymbol{y}_i|C_l)}\right]$. This strategy improves the exploratory ability of the MCMC sampler and is able to correctly sample from the conditional pdf even when the performance function exhibits difficulties outlined in Section 5.1.4.

x. Elliptical slice sampling [Liao et al., 2024b]: Here, given a current state $\boldsymbol{x}_i$, $v \sim N(\boldsymbol{0}, \mathbf{I})$ is generated and $R_m = \sqrt{x_{i,m}^2 + v_m^2}$ is computed. The angle $\boldsymbol{\beta}^{(0)}$ is computed so that $\boldsymbol{u} = \boldsymbol{R}\cos\boldsymbol{\beta}^{(0)}$ where $\boldsymbol{R} = [R_1 \quad R_2 \quad \cdots \quad R_{n_d}]$ and $\boldsymbol{\beta}^{(0)} = \left[\beta_1^{(0)} \quad \beta_2^{(0)} \quad \ldots \quad \beta_{n_d}^{(0)}\right]$, and $\beta_m^{(0)} \in [0, \pi]$. With $\beta_m^t \sim U\left(\beta_m^{(min)}, \beta_m^{(max)}\right)$ (initially, $\beta_m^{(min)} = 0$ and $\beta_m^{(max)} = \pi$) for $m = 1, 2, \ldots, n_d$, candidate sample $\boldsymbol{\xi} = \boldsymbol{R}\cos\boldsymbol{\beta}$ is generated. If $\boldsymbol{\xi}$ lies in the intermediate failure region, the next sample is set as $\boldsymbol{x}_i = \boldsymbol{\xi}$. Otherwise, the maximum and minimum angles are modified as $\beta_m^{(min)} = \beta_m^t$ if $\beta_m^t < \beta_m^{(0)}$, and $\beta_m^{(min)} = \beta_m^t$ if $\beta_m^t > \beta_m^{(0)}$.

xi. Intrepid sampler [Sharma et al., 2026]: Chakroborty and Shields [2026] recently proposed the Intrepid MCMC sampler where the Markov transition is governed by a mixture of two kernels: (a) a local kernel which focuses on drawing samples from the mode that the chain currently resides in, and (b) a global kernel which aims to explore the parameter space in search of previously undetected modes. The local kernel is taken to be a component-wise Metropolis sampler in the Cartesian components [Au and Beck 2001a], and the global kernel is taken to be a full-dimensional Metropolis sampler in the hyperspherical components. Sharma et al., [2026] used a cyclic component-wise version of the Intrepid sampler in conjunction with subset simulation and showed that the sampler improves exploration in the standard normal space when faced with complexities such as multiple failure surfaces or sharply varying and potentially discontinuous performance functions.

### 5.1.7 Discussion on MCMC samplers in particle splitting methods

Developments in MCMC techniques for particle splitting methods have primarily aimed to improve the acceptance rate, enhance the exploratory capability, reduce inter-chain and intra-chain correlation, or achieve a combination of these objectives. Initial works by Zuev and Katafygiotis [2011], Miao and Ghosn [2011], and Santoso et al., [2011] proposed several variations of the modified Metropolis algorithm to improve acceptance rates and consequently reduce correlation among samples in Markov chains. The elliptical slice sampling [Liao et al., 2024b] similarly seeks to reduce the intra-chain and inter-chain correlation during sampling. The conditional $\boldsymbol{U}$-sampling proposed by Papaioannou et al., [2015] is a variant of the preconditioned Crank-Nicolson algorithm [Cotter et al., 2013] applied to subset simulation. Here, a correlation is imposed between the current and the next state of the chain and a notable feature of this method is that samples are never repeated.

More recent efforts have focused on improving the exploratory ability of Markov chains for performance functions whose failure domains exhibit challenging geometries. The parallel tempering sampler [Xiao et al., 2019] focuses on drawing samples from disconnected failure regions, whereas the affine invariant sampler [Shields et al., 2021] is designed to improve sampling when the conditional distribution is strongly anisotropic or degenerate. Hamiltonian Monte Carlo and its variants [Wang et al., 2019, Chen et al., 2022, Thaler et al., 2024] exploit gradient-informed, physics-based proposal dynamics leading to better exploration of the standard normal space.

While these samplers improve exploration in the standard normal space in various settings, they may still perform inadequately when performance functions possess sharp changes or discontinuities, especially near one or more important regions. In such cases, the samplers could be misled by the local geometry to sample from a less important failure region. In this context, the replica exchange sampler proposed by Sharma and Manohar [2023] was explicitly designed to address discontinuities and rapid changes in performance functions and avoid being misled by the local geometry of the limit surface. Although the replica exchange sampler was able to tackle these challenges in relatively lower-dimensional problems, a more recent work by Sharma et al., [2026] developed the cyclic component-wise version of the Intrepid sampler that addresses these difficulties in higher dimensional problems.

## 5.2 Trajectory splitting methods for time-variant reliability estimation

For time-variant reliability estimation, it is possible, in principle, to discretize the random excitation into a large-dimensional vector of random variables, which leads to an equivalent time-invariant reliability problem. Subset simulation, or indeed any particle splitting method, can subsequently be used to estimate the failure probability (see, for example, studies by Au and Beck [2001a, 2003a], Schueller and Pradlwarter [2007], and Du et al., [2019]). However, in employing this strategy, the underlying MCMC sampler is required to traverse and adequately explore a large-dimensional parameter space. The input space may contain geometric complexities such as multimodal conditional pdfs or sharp variations in performance functions (these challenges and the literature pertaining to them are described in Section 5.1.4). In the presence of such difficulties in high dimensions, samplers may fail to detect and sample from the most important failure regions. This motivates trajectory splitting methods, which do not require reformulating the problem as a high-dimensional reliability integral. In these methods, response trajectories that are deemed unlikely to reach the failure domain are terminated, whereas those that are more likely to do so are multiplied or split. This leads to a collection of trajectories being nudged towards the failure domain. Computational effort is conserved by terminating trajectories that are unlikely to enter the failure region.

### 5.2.1 Trajectory splitting methods for time-variant reliability estimation of structures subjected to random excitations

In the context of structural engineering, Pradlwarter et al., [1994] proposed an algorithm called "double and clump" which begins by initiating $N_S$ response trajectories $\boldsymbol{X}^k(t), k=1,2,...,N_S$, each associated with a weight $w_k(t)$ such that $w_k(t)\geq 0$ for all $k$ and $\sum_{k=1}^{N} w_k(t)=1$. Each trajectory is assigned an importance factor denoted as $c_k(t)$ and defined as

$$c_k(t)=\left(E_{k,P}(t)+E_{k,K}(t)\right)P_{k,E}(t)\left[w_k(t)\right]^{\beta} \tag{87}$$

Here, $E_{k,P}(t)$ denotes the potential energy and $E_{k,K}(t)$ denotes the kinetic energy of the structural system. $P_{k,E}(t)$ represents the power of the input excitation and $\beta$ is an algorithmic parameter that takes a value between 0.1-0.5. At time step *t*, a specified fraction of the most important trajectories (determined based on the value of $c_k(t)$) are "doubled". This involves generating an identical copy of the selected response trajectory and assigning half of the initial weight $w_k(t)$ to each trajectory. Simultaneously, a certain fraction of the least important trajectories are "clumped" where two trajectories are merged into a single trajectory with the weight equal to the sum of the individual weights, i.e., $\hat{P}_F = \sum_{k=1}^{N_S} I\left[h^* - h\left[\boldsymbol{X}_k(t)\right] \le 0\right] w_k(t)$.
The failure probability is obtained by summing up the weights of the trajectories that entered the failure region. In a subsequent study, Pradlwarter and Schueller [1999] proposed an alternative distance-controlled procedure to identify important and unimportant response trajectories.

In studies by Pradlwarter and Schueller [1997, 1999] and Melnik-Melnikov and Dekhtyaruk [2000], the "Russian Roulette and splitting" technique is used to estimate failure probabilities. In the study by Pradlwarter and Schueller [1997, 1999], at time instant $t$, the following functions are defined.

$$J_n(t) = \begin{cases} 1 & \text{with probability } P_n(t) \\ 0 & \text{with probability } 1 - P_n(t) \end{cases} \tag{88}$$

$$\overline{w}_n(t) = \frac{w_n(t)}{P_n(t)}, \text{ where } P_n(t) \propto p_n(t) = 1 - \left(\frac{c_{\max}(t) - c_n(t)}{c_{\max}(t)}\right)^2 \tag{89}$$

The failure probability is accordingly written as follows.

$$\hat{P}_F^{RR} = \sum_{n=1}^{N} J_n(t) I\left[h^* - h\left[\boldsymbol{X}_n(t)\right] \le 0\right] \overline{w}_n(t) \tag{90}$$

In the study by Melnik-Melnikov and Dekhtyaruk [2000], a number of intermediate failure domains between 0 and $h^*$ are defined a priori. $N_S$ response trajectories $\boldsymbol{X}_k(t), k = 1, 2, ..., N_S$ are generated. Each time a response trajectory up-crosses the intermediate failure domain, the trajectory is split into $n_C$ copies and each trajectory is assigned a weight $\overline{w} = \frac{w}{n_C}$, where $w$ is the initial weight of the trajectory. This procedure is called "splitting". Similarly, each time a response trajectory down-crosses an intermediate failure domain, the trajectory is terminated with probability $1 - \frac{1}{n_C}$, and the trajectory is continued with an updated weight of $\overline{w} = w n_C$ with probability $\frac{1}{n_C}$. This procedure is called "Russian Roulette". The failure probability estimate is given by $\hat{P}_F = w_{N_l} \sum_{k=1}^{N_S} I_k$, where $w_{N_l}$ is the weight of the samples that reach the failure domain.

Ching et al., [2005a] have proposed a version of the subset simulation method called subset simulation with splitting. Here, intermediate failure samples are generated by splitting the response trajectories that have reached an intermediate threshold value into multiple offspring instead of using MCMC sampling as in [Au and Beck 2001a]. Consider a response trajectory $x(t)$ that reaches an intermediate threshold value $\tilde{h}^*$ at time instant $t_k$, and denote the past portion of the response trajectory as $x(t) = x^-(t)$ for $t < t_k$. Denote the corresponding excitation trajectory as $u^-(t)$ for $t < t_k$ and $u^+(t)$ for $t > t_k$. $u^+(t)$ is generated according to $p\left(u^+ \middle| u^-\right) = \frac{p\left(u^+, u^-\right)}{p\left(u^-\right)}$. In a subsequent study, Ching et al., [2005b] proposed a hybrid subset simulation method that combines the original subset simulation and the splitting method. The study by Wang et al., [2014] proposed a version of subset simulation with splitting where the input process over a given time duration of interest, which is high dimensional, is divided into a sequence of shorter duration random processes, which are low dimensional.

### 5.2.2 Splitting methods for time-variant reliability estimation of randomly parametered structures subjected to random excitations

Here, we assume that the system parameters $\boldsymbol{\Theta}$ are modeled as a set of random variables with joint density $p_{\boldsymbol{\Theta}}(\boldsymbol{\theta})$ which leads to the following governing SDE.

$$d\boldsymbol{X}(t) = \boldsymbol{A}\left[\boldsymbol{\Theta}, \boldsymbol{X}(t), t\right]dt + \sigma\left[\boldsymbol{\Theta}, \boldsymbol{X}(t), t\right]d\boldsymbol{B}(t); \boldsymbol{X}(0) = \boldsymbol{X}_0 \tag{91}$$

The failure probability and its direct Monte Carlo estimator are, respectively,

$$\begin{aligned} P_F &= 1 - \mathbb{P}\left(h\left[\boldsymbol{\Theta}, \boldsymbol{X}(t)\right] < h^* \forall t \in [0,T]\right) = E_{\mathbb{P}}\left(I\left[h^* - \max_{t\in[0,T]} h\left[\boldsymbol{\Theta}, \boldsymbol{X}(t)\right]\right] \le 0\right) \\ \hat{P}_F^{DMC} &= \frac{1}{N_S}\sum_{i=1}^{N_S} I\left[h^* - \max_{t\in[0,T]} h\left[\boldsymbol{\Theta}^i, \boldsymbol{X}^i(t)\right] \le 0\right] \end{aligned} \tag{92}$$

Subset simulation and its variants [Au and Beck 2001a, Ching et al., 2005b, Katafygiotis and Cheung 2007, Li et al., 2019] can indeed be extended to the computation of time-variant reliability of dynamical systems with uncertain system parameters (see the benchmark study by Schueller and Pradlwarter [2007]). This can be achieved by augmenting the random system parameters with the random vectors associated with the random excitations and employing subset simulation on the extended random vector. The realizations $\boldsymbol{\Theta}^i$ and $\boldsymbol{X}^i(t)$ are drawn from the joint pdf $p_{\Theta}(\boldsymbol{\theta})$ and the governing equation. In the study by Sundar and Manohar [2014a], the authors used Girsanov transformation in conjunction with subset simulation in order to estimate the unconditional time-variant reliability of a nonlinear dynamical system under random excitations. Here, instead of computing the Girsanov control for each sample realization of the system parameter, a representative control is established for each intermediate failure region. This method is especially efficient for linear systems for which the state-independent Girsanov control is analytically available, leading to a substantial decrease in the computational cost involved in estimating failure probability.

### 5.3 Discussion on particle and trajectory splitting methods

Particle splitting methods have emerged as an important class of algorithms for structural reliability estimation since their introduction to the literature by Au and Beck [2001a] in the form of subset simulation. The method was demonstrated to handle high-dimensional, nonlinear, and implicitly defined performance functions. These advantages motivated extensive research on the algorithm, leading to substantial developments, improvements, and extensions of the original framework over the past 25 years.

A major focus in the development of particle splitting methods has been the improvement of the MCMC samplers employed in the algorithm. Advanced samplers have been proposed to improve sampling variability, proposal dynamics, exploration ability, and/or sampling from complicated target pdfs arising from highly nonlinear performance functions. However, computational issues pertaining to sampling in high dimensions persist, particularly when the failure domain has complex geometry. The problem of estimating multiple failure probabilities in a single particle splitting run for the same structure has also been tackled, although this area has received comparatively limited attention and there remains scope for developing more robust and versatile methodologies. Similarly, particle splitting methods incorporating multi-fidelity models have only recently gained attention in the literature [Naderi and Jia 2026]. This remains a relatively unexplored area where methodological developments could be valuable, especially for computationally expensive structural models.

For time-variant reliability with random excitations and deterministic system parameters, a natural approach is to discretize the excitation into a large-dimensional random vector and employ the particle splitting framework on this random vector. This approach, however, has a clear limitation: the dimension of the random vector can become very large, leading to ineffective MCMC sampling. Trajectory splitting methods address this issue by operating on the response trajectories rather than the random excitations. These methods involve solving the governing equation of motion to generate response trajectories which are either discontinued or multiplied based on their proximity to failure. Within the context of splitting methods, time-variant reliability estimation in which both excitation and system parameters are random has received comparatively limited attention. Once again, similar to the case of deterministic system parameters, a straightforward approach is to discretize the excitations, augment this high-dimensional random vector with the system parameters, and apply the particle splitting framework. This runs into the same issue of ineffective high-dimensional MCMC sampling. Trajectory splitting methods developed for this class of problems include the hybrid subset simulation [Ching et al., 2005b] and Girsanov transformation combined with subset simulation [Sundar and Manohar, 2014a].

An outstanding issue common to all three classes of problems viz., time-invariant reliability, time-variant reliability with deterministic system parameters, and time-variant reliability with stochastic system parameters, is that of the geometry of the limit state and the failure domain. Since these computations are generally carried out in large dimensions and with implicitly defined performance functions, the analyst has no prior information regarding the topology of the failure domain. There may exist multiple failure domains, and the performance function may possess discontinuities or rapid variations over the parameter space. Consequently, these algorithms may miss important failure regions and produce incorrect failure probability estimates. This remains a major epistemic barrier for accurate structural reliability estimation. Although Breitung [2019, 2021] presented these challenges in the context of subset simulation, it must be noted that such challenges are ultimately rooted in the difficulty of exploring large-dimensional spaces to identify potentially multiple regions of failure. Therefore, such complicating features would pose similar exploratory challenges for FORM/SORM-based methods, surrogate modelling-based methods, and variance reduction methods more generally.

In FORM/SORM-based methods, this difficulty appears in the form of locating multiple global and/or local minima in a large-dimensional space. Similarly, in surrogate modelling methods, the challenge is in identifying all the regions in the parameter space to refine the performance function surrogate models. It is worth noting that this limitation of subset simulation had been pointed out in the original subset simulation work by Au and Beck [2001a]. Given these difficulties, Breitung [2021, 2024] advocates for FORM/SORM or design point-based methods based on four observations: (a) the optimization literature is vast with an extensive array of possible methods to be used, even when complicated geometries are encountered, (b) in the asymptotic limit $\beta_{HL} \to \infty$, Hohenbichler's lemma [Breitung 2024] shows that probability content is concentrated near design points, and (c) the SORM estimate is asymptotically exact in the limit $\beta_{HL} \to \infty$, and (d) design point-based methods provide useful information for sensitivity analysis and failure-mode interpretation. While the asymptotic and sensitivity results are valuable, the practical significance of these observations depends on how well the asymptotic results hold for potentially highly nonlinear performance functions when the asymptotic limit is not reached. Indeed, $\beta_{HL}{\sim}O(1)$ in engineering contexts, and the applicability of the asymptotic results in this regime warrants further mathematical investigation.

Recent literature has attempted to address this difficulty either by improving the MCMC sampler [Sharma and Manohar 2023, Sharma et al., 2026], by modifying the particle splitting algorithm through scaling of the input space (for example, [Rashki 2021]), and by smoothening the indicator function [Li et al., 2025]. Similar geometric issues can be expected to arise in the context of time-variant reliability of deterministic and stochastic dynamical systems undergoing bifurcation. The challenge becomes more severe when the bifurcation parameter is uncertain, leading to qualitatively different dynamical behaviour depending on the parameter value. Although these difficulties have been identified in particle splitting methods in the recent literature, such potential challenges have remained unexplored in settings involving trajectory splitting methods. Overall, splitting methods have largely addressed the issue of estimating small probabilities efficiently and the challenges posed by high dimensions in parameter and/or excitation space when the failure regions and the limit state functions have reasonably smooth geometry. However, the challenge of discovering new, possibly important, failure regions in large dimensions remains an outstanding challenge. This issue is likely to be central to future developments, and it continues to be an active area of research.

## 6.0 Other sampling variance reduction methods for time-invariant reliability estimation

This section discusses variance reduction techniques that do not fall into the categories enumerated in the previous sections.

### 6.1 Stratified sampling:

Stratified sampling involves partitioning the space spanned by the random variables into $N_{ST}$ mutually exclusive and exhaustive sets $D_i, i = 1,2,...,N_{ST}$ i.e., $D_i \bigcap D_j = \phi$ for all $i \neq j$ and $\bigcup_{i=1}^{N_{ST}} D_i = \mathbb{R}^{n_d}$. The failure probability estimator and its sampling variance are given by the following equations [Song and Kawai 2023a].

$$\hat{P}_F^{SS} = \sum_{m=1}^{N_{ST}} \lambda_m \hat{P}_m \tag{93}$$

$$\text{Var}\left(\hat{P}_F^{SS}\right) = \sum_{m=1}^{N_{ST}} \frac{\left(\lambda_m \hat{\sigma}_m\right)^2}{N_m} \tag{94}$$

Here, $\hat{P}_m = \frac{1}{N_m} \sum_{k=1}^{N_m} I\left(G\left(\boldsymbol{X}_{m,k}\right) \leq 0\right)$, $\hat{\sigma}_m^2 = \frac{1}{N_m} \sum_{k=1}^{N_m} I\left(G\left(\boldsymbol{X}_{m,k}\right) \leq 0\right) - \hat{P}_m^2$, $n = \sum_{m=1}^{N_{ST}} N_m$, where $\boldsymbol{X}_{m,k}, k = 1,2,...,N_m$ are random vectors drawn according to the pdf $\frac{p_X\left(\boldsymbol{x}\right) I\left(\boldsymbol{x} \in D_m\right)}{\lambda_m}$ and $\lambda_m = \text{E}_p\left[I\left(\boldsymbol{X} \in D_m\right)\right]$. Reliability analyses that employ stratified sampling can be found in the studies by Feng et al., [2010], Zuniga et al., [2011], Shields et al., [2015], Arunachalam and Spence [2023], Song and Kawai [2023a], and Song and Kawai [2023b]. Notably, the spherical subset simulation, which was proposed by Katafygiotis and Cheung [2007] is also founded on the idea of stratifying the failure region into non-overlapping hyperspherical annuli and sequentially sampling from these regions using specially designed MCMC samplers. Wang and Song [2018] proposed a hyperspherical extrapolation method based on decomposing the failure region into hyperspherical caps and by exploiting the geometric nature of high-dimensional spaces pointed out by Katafygiotis and Zuev [2008] and Katafygiotis and Cheung [2007].

### 6.2 Line sampling:

In the line sampling technique [Koutsourelakis 2004, Koutsourelakis et al., 2004], the performance function is expressed as

$$G(\boldsymbol{U}) = G_0(\boldsymbol{U}_{-1}) - U_1;\ \boldsymbol{U}_{-1} = \begin{bmatrix} U_2 & U_3 & \dots & U_p \end{bmatrix} \tag{95}$$

Here, $G_0$ is a function such that the equation $F = \left\{ \boldsymbol{u} \in \mathbb{R}^{n_d} : \boldsymbol{u}_1 \in G_0(\boldsymbol{u}_{-1}) \right\}$ is satisfied. The failure probability is written as

$$\begin{aligned} P_F &= \int_{\mathbb{R}^p} I\left[G(\boldsymbol{u}) \le 0\right] \prod_{i=1}^{p} p_{U_i}(u_i)\, du \\ &= \int_{\mathbb{R}^{p-1}} \left\{ \int_{\mathbb{R}} I\left[G(\boldsymbol{u}) \le 0\right] p_{U_1}(u_1) \right\} \prod_{i=1}^{p} p_{U_i}(u_i)\, d\boldsymbol{u}_{-1} \\ &= \int_{\mathbb{R}^{p-1}} \hat{I}_{\left[G(\boldsymbol{u}) \le 0\right]} \prod_{i=2}^{p} p_{U_i}(u_i)\, d\boldsymbol{u}_{-1} \\ &= \mathrm{E}_{\boldsymbol{U}_{-1}}\left(\hat{I}_{\left[G(\boldsymbol{U}) \le 0\right]}\right) \end{aligned} \tag{96}$$

The failure probability estimate is then given by

$$\hat{P}_F = \frac{1}{N} \sum_{i=1}^{N} \hat{I}^{i}_{\left[G(\boldsymbol{U}_{-1}) \le 0\right]} \tag{97}$$

An advanced line sampling algorithm has been proposed by de Angelis et al., [2015], where the authors propose an adaptive strategy to choose the important direction. A different adaptive strategy for the selection of the important direction was proposed by Shayanfar et al., [2017]. The study by Papaioannou and Straub [2021] has developed combination line sampling, which is a generalization of the advanced line sampling proposed by de Angelis et al., [2015]. Valdebenito et al., [2021b] have developed a multi-domain line sampling for series system reliability estimation of problems involving Gaussian random variables and linear performance functions. Wei et al., [2023b] use stochastic collocation in conjunction with line sampling in order to estimate the conditional failure probability function. Yuan et al., [2024] propose a line sampling scheme for time-variant reliability estimation where the problem is expressed as an equivalent time-invariant reliability problem associated with a series system, after which the proposed adaptive combined line sampling scheme is used for failure probability estimation. Zhang [2026] proposed a line sampling algorithm with improved accuracy in the context of highly nonlinear performance functions. Here, the line sampling method is used in conjunction with a cross-entropy-based importance sampling scheme using Gaussian mixture and von Mises-Fisher-Nakagami mixture distributions. Kanjilal et al., [2026] have also integrated the cross-entropy-based importance sampling with multi-domain line sampling in order to estimate the first-passage probability of a randomly parametered linear dynamical system subjected to Gaussian excitations. Line sampling has also been extensively used to compute sensitivity of the failure probability with respect to the input random variables (see, for example, studies by Song et al., [2007], Lu et al., [2008], Valdebenito et al., [2018], Zhang et al., [2020b], Valdebenito et al., [2021a, 2021b], Valdebenito et al., [2024]).

**7.0 Experimental testing for reliability**

It is standard practice to test critical components in aerospace and automotive applications for vibratory loads wherein the loads are specified in terms of a power spectral density function (MIL-STD-810G, ISO16750-3). Even though the test load is specified as a random process, the acceptance criteria here do not involve limits on probability of failure. For instance, if the target failure probabilities are in the range of $10^{-5}$ to $10^{-4}$, an experimental demonstration of this fact would require approximately $10/P_f$ number of tests, which becomes practically infeasible to conduct. A possible way out here would be to incorporate ideas from sampling variance reduction schemes (developed in the context of computational reliability modelling) into experimental protocols (see the studies by Sundar and Manohar [2014b], Nayek and Manohar [2015], Sundar et al., [2015], and Sonal et al., [2018]). At the outset, it needs to be noted that the class of structures that require qualification via testing are invariably complex in nature with components such as moving parts, free play, inherently nonlinear elements, active control units, and fluid-structure interactions. This means that any strategy in this context to intelligently test for reliability must not depend on the formulation of any updated finite element models for the structure. The study by Sundar and Manohar [2014b] explored the application of Girsanov control-based ideas to time-variant reliability analysis of linear structures being tested on an earthquake shake table. It was noted in this study that the state-independent suboptimal control and Radon-Nikodym derivative as proposed by Macke and Bucher [2003] can be derived based on a set of impulse response functions of the system. The key idea here was that these functions can be measured experimentally without recourse to any mathematical modelling of the test structure. The authors considered testing of a five-storey bending-torsion coupled building frame on a multi-axes earthquake simulator with the load modelled as a nonstationary Gaussian random process. The study demonstrated that probabilities of failure as low as $10^{-5}$ could be acceptably estimated with about 1750 samples. Nayek and Manohar [2015] extended this study to building frames with active base isolation systems (viz., MR damper-based isolator). The damper here behaved nonlinearly, and the study clarified that the active control strategy implemented to mitigate the response and the Girsanov control-based strategy that aimed to promote failure do not work at cross purposes. The study by Sundar et al., [2015] extended the study to consider time-variant system reliability and demonstrated the performance of the method with respect to the testing of a four-wheeled automotive system tested on a four-post rig for road-induced vibrations. The study by Sonal et al., [2018] tackled testing for time-variant component/system reliability by embedding particle splitting-based methods into the test protocol.

## 8.0 Machine learning tools in structural reliability

The mathematical methods which underlie the subjects of machine learning (ML) and structural reliability modelling share substantial common ground. Thus, the tools of probability, statistics, matrix algebra, and optimization form the backbone of both these subjects. Monte Carlo simulation-based methods (augmented by strategies for sampling variance control) combined with FE modelling of structural behaviour offer a powerful framework for modelling structural reliability. The developments in this framework have been outlined in the preceding sections. Given these developments, it is useful to ask what the ML-based tools could offer to advance the subject further.

To answer this question, it may be noted that there exist two primary categories of difficulties associated with simulation-based methods. Firstly, the effort needed to tackle the problem increases with the size of the computational models, the number of basic random variables, the possibility of nonlinear structural behaviour, the need to model dynamic behaviour of the structure, uncertainties present both in applied dynamic loads and system parameters, and rarity of the failure event. These difficulties can be viewed as soft barriers which can be tackled by using greater computational effort and numeric strategies. The second source of difficulties pertains to the lack of prior knowledge about convexity/nonconvexity of the failure surface, disconnected or multiply connected failure regions, multiple regions of comparable importance and their locations in the standard normal space, and highly nonlinear limit surfaces. This epistemic barrier prevents satisfactory exploration of the failure surface, thereby potentially leaving parts of the failure surface undetected, leading to unintended bias in the estimates of the failure probability. Methods such as those based on importance sampling and subset simulations succeed well in locally exploiting information that is gathered as simulation proceeds but struggle in globally exploring the input random variable space and in discovering multiple/well-separated important regions contributing significantly to the failure probability.

In problems of time-variant reliability, typically involving random dynamic excitations and system parameter uncertainties, one could envisage similar difficulties. One class of challenges, again, pertains to computational demands on evaluation of the performance functions, and the other involves a more subtle set of challenges. The latter class of challenges includes the following: (a) analytical solutions based on the assumption of a Poisson model for level crossing, an exponential model for first passage time, and a Gumbel model for extremes offer tractable means to develop reliability models at least for a class of problems. However, these results are only asymptotically valid, and it is hard to judge how close a given problem is to the asymptotic limits. (b) When system parameters are random, their support may contain critical values linked to system bifurcations. Thus, tracking bundles of trajectories leading to failure may be difficult if the system is prone to such bifurcations. (c) The time-variant reliability problems can be posed as equivalent time-invariant reliability problems by introducing extreme value theory. The theory of extremes is largely limited to univariate distributions originating from independent and identically distributed (IID) sequences, while one often needs to deal with nonstationary responses. The max operation implied here introduces non-smooth failure surfaces.

Thus, the question of what ML-based methods can offer in this milieu provides a useful lens to examine how these methods have been integrated with the task of reliability modelling. Accordingly, this section is devoted to reviewing literature related to the application of ML methods in reliability estimation. We refer readers to the works of Murphy [2012, 2022, 2023], Bishop and Bishop [2024], Strang [2019], Brunton and Kutz [2022], Alpaydin [2020], Prince [2023], Goodfellow et al., [2016], and Sutton and Barto [2018] for a comprehensive account of the ML methodologies. The paper by Karniadakis et al., [2021] provides extensive background to physics-informed machine learning. We do not repeat the details of these tools; instead, we focus on how these tools have been applied in the field of structural reliability modelling.

At the outset we note that these applications can be categorized into five broad areas: (a) development of surrogate models to alleviate computational burden in repeated calculations of the performance function, (b) in devising strategies for reducing sampling variance in simulation-based estimation of probability of failure, (c) unsupervised partition-based methods, (d) simulation of loads/material properties based on limited data using deep generative modelling tools, and (e) reliability model updating of instrumented structures.

### 8.1 Surrogate models

#### 8.1.1 Time-invariant reliability models

Evaluation of probability of failure using sampling methods involves repeated evaluation of the performance function $G(\boldsymbol{u})$. In realistic examples, this calculation invariably involves expensive calls to long-running black-box computer codes (termed as oracles) such as those residing in professional finite element models. The idea of surrogate modelling is to replace these codes by approximate models $\hat{G}(\boldsymbol{u})$, termed as meta-models (models for models), and use them in repeated calculations while estimating the reliability. At the outset, one could note the following points:

1. The determination of $\hat{G}(\boldsymbol{u})$ needs systematic strategies for sampling from $p_U(\boldsymbol{u})$. The question here is whether one aims to obtain a global fit to $G(\boldsymbol{u})$ (problem of regression) or aims to detect if, for a given $\tilde{\boldsymbol{u}}$, whether $G(\boldsymbol{u}) \leq 0$ or not (problem of binary classification). What makes the problem specific to reliability modelling is that, even when the problem of regression is considered, the resulting model needs to perform well near the limit surface $G(\boldsymbol{u})$=0, as the model for $G(\boldsymbol{u})$ enters reliability estimation only via the indicator function $\mathrm{I}\left[G(\boldsymbol{u}) \leq 0\right]$.
2. From the perspective of ML methods, the two problems mentioned above (viz., regression and classification) belong to the class of supervised learning problems. This problem can be tackled using established methods such as Gaussian process regression (kriging), support vector machines, or artificial neural networks (ANNs). These models have the mathematical guarantee that they are potentially universal function approximators and therefore have the capability to model $G(\boldsymbol{u})$ with a desired accuracy even when it has one or more of the geometric complexities mentioned above. One could thus expect that these surrogate models will retain all these difficult-to-model complexities and thus would become easy to compute and yet powerful proxies for $G(\boldsymbol{u})$. The establishment of these proxies facilitates implementation of direct Monte Carlo simulations (typically with no need for sampling variance reduction) to estimate probability of failure.

3. A point worth noting is that the guarantees provided by the universal approximation theorem, while tackling the regression problem, do not automatically extend to the problem of binary classification implicit in the estimation of $P_F$. One of the basic difficulties here is that the event $\{G(\boldsymbol{U}) \leq 0\}$ is typically going to be rare, and this creates imbalance in the classes, which poses notable challenges.
4. For the surrogate modelling approach to be effective, it is evident that the computational effort needed to establish the surrogate model needs to be substantially less than what is needed for estimating $P_F$ based on repeated evaluation of $G(\boldsymbol{u})$. The computational effort here is linked to producing a labelled dataset to train the ML model and the optimization effort involved in determining the learning parameters. It is evident that the cost of producing sample realizations of inputs is trivial (as it involves obtaining random draws from a standard Gaussian pdf) compared to the cost of obtaining the label (which involves running the long-running computational code).
5. The problem of obtaining a suitable labelled dataset remains as the main challenge. For the models to be effective, the labelled dataset needs to contain points in the failure region. Any attempt to obtain such points would encounter the hard epistemic barrier of lack of prior knowledge about the nature of the limit surface.
6. A model for the error of representation $|G(\boldsymbol{u}) - \hat{G}(\boldsymbol{u})|$ needs to be provided.
7. When a new data point, $\boldsymbol{u}^*$ is queried, suitable guidelines need to be available on deciding if the surrogate prediction $\hat{G}(\boldsymbol{u}^*)$ is acceptable or one needs to go back to the true model (oracle) for finding $G(\boldsymbol{u}^*)$. This entails further questions on strategies for updating the surrogate model $\hat{G}(\boldsymbol{u})$.

One of the early conceptual demonstrations of ANN-based surrogates for reliability estimation can be found in the work of Papadrakakis et al., [1996]. The authors use fully connected feed-forward ANNs to approximate critical load factors of elasto-plastic frames under static loads. The surrogate models thus created are combined with importance sampling-based Monte Carlo

Simulations to estimate the probability of failure. The examples considered were modest (up to 3 random variables, ANNs with about ~50 learning parameters, and a labelled dataset of size ~20), and the focus of the study was on demonstrating how ANNs can enter the reliability computation steps. The early works of Hurtado and Alvarez, [2000, 2001], Hurtado [2001], and the monograph by Hurtado [2004] are perhaps the earliest efforts to view the problem of reliability estimation through the lens of the theory of statistical learning. The monograph considers tools such as support vector machines and feed-forward neural networks to study limit state functions either as a problem in regression analysis (obtain a global surrogate model for the performance function) or as a problem in binary classification (separate the safe and unsafe regions). The work also discusses problems of model reduction using tools such as principal component analysis for reducing the dimension of the input and (or) output spaces. These studies laid the conceptual foundations for later studies involving adaptive sampling and active learning. A study aimed at benchmarking alternative surrogate modelling frameworks (viz., polynomials, moving least squares, radial basis functions, and ANN) is reported by Bucher and Most [2008]. The size of the chosen examples, however, is modest (up to 6 random variables and FE models with dofs up to 200).

The idea of approximating the function $G(\boldsymbol{u})$ as a Gaussian random field evolving in the space spanned by $\boldsymbol{u}$ underlies kriging/Gaussian process regression. Kaymaz [2005], Kaymaz and McMahon [2005], and Panda and Manohar [2008] investigated the application of kriging models as surrogates for the performance function in structural reliability studies. The idea here is to represent $G(\boldsymbol{u})$ as

$$G(\boldsymbol{u}) = \boldsymbol{\beta}^{\mathrm{T}} \boldsymbol{F}(\boldsymbol{u}) + Z(\boldsymbol{u}) \tag{98}$$

where the $\boldsymbol{\beta}$ is a $p \times 1$ vector of unknown parameters, $\boldsymbol{F}(\boldsymbol{u})$ is a $p \times 1$ vector of known functions (to be chosen by the analyst), and $Z(\boldsymbol{u})$ is a zero-mean, stationary Gaussian random field evolving in the $n$-dimensional space spanned by the vector $\boldsymbol{u}$. Furthermore, it is assumed that the form of the covariance function $\mathrm{E}\left[Z(u_r), Z(u_s)\right] = \sigma_Z^2 R\left[|u_r - u_s|, \boldsymbol{\theta}\right]$ is known, with the scalar $\sigma_Z^2$ and the $q \times 1$ vector $\boldsymbol{\theta}$ being unknowns. Thus, the unknowns to be determined are given by $\boldsymbol{\beta}$, $\sigma_Z^2$, and $\boldsymbol{\theta}$. The method employs the dataset $\left[\boldsymbol{u}^j, G(\boldsymbol{u}^j); j = 1, 2, \cdots, l\right]$ generated by careful sampling to obtain the maximum likelihood estimates $\hat{\boldsymbol{\beta}}$, $\hat{\sigma}_Z^2$, and $\hat{\boldsymbol{\theta}}$. The metamodel for the performance function at an unseen data point $\boldsymbol{u}^*$ is obtained as $\hat{G}(\boldsymbol{u}^*) = \mathrm{E}\left[G(\boldsymbol{u}^*) \middle| G(\boldsymbol{u}^1, \boldsymbol{u}^2, \cdots, \boldsymbol{u}^l)\right]$. The study by Panda and Manohar [2008] employed space-filling Latin-hypercube sampling to choose the sampling points $\boldsymbol{u}^1, \boldsymbol{u}^2, \cdots, \boldsymbol{u}^l$. In the estimation of reliability, one replaces the true model $G(\boldsymbol{u})$ by the surrogate $\hat{G}(\boldsymbol{u})$ and obtains the estimate for the probability of failure as

$$\hat{P}_F = \frac{1}{N}\sum_{i=1}^{N}\hat{G}(\boldsymbol{u}^{*j}) \text{ with } \boldsymbol{u}^{*j} \sim p_U(\boldsymbol{u}) \tag{99}$$

As has been already noted, when a new data point, $\boldsymbol{u}^*$ is queried, one has the option of using the surrogate $\hat{G}(\boldsymbol{u}^*)$ or the original $G(\boldsymbol{u}^*)$. Clearly the cost of determining $\hat{G}(\boldsymbol{u}^*)$ is far less than the cost of calling the original code to determine $G(\boldsymbol{u}^*)$. However, $\hat{G}(\boldsymbol{u}^*)$ is essentially an approximation and if one can assess if the approximation is acceptable, one can confidently proceed to use $\hat{G}(\boldsymbol{u}^*)$ instead of the costly task of finding $G(\boldsymbol{u}^*)$. This concern opens the doors for the application of active learning strategies in reliability estimation.

Active learning strategies in machine learning have been developed to deal with situations in which labelling is expensive, time-consuming, and requires expert intervention. A preliminary surrogate model for the function under study is first obtained. The main idea here is to adaptively select further data points that require expert labelling based on the notion of a learning function. This function signals whether a new data point requires expert labelling or not. If it indeed does, the label data thus obtained is used to update the initial surrogate model. The learning process is stopped once the learning function indicates that the updated surrogate model is sufficiently accurate, implying that no additional calls to the original high-fidelity function are required.

The work of Hurtado [2007] contains some of the features of active learning schemes. This study begins by choosing an importance sampling pdf (e.g., a Gaussian pdf centred at the FORM-based design point). An SVM-based classifier is constructed based on an initial labelled dataset (either drawn from the ISpdf or by other engineering insights). Samples are then drawn from the ISpdf, and the original code for evaluating $G(\boldsymbol{u})$ is called only if the sampled point lies within the SVM margin, else the current SVM classifier is used to label the data point.

Whenever a call is made to the oracle, the SVM is updated. Thus, the SVM serves here as a filter to decide if a call to the oracle is needed or not. In this scheme, the SVM does not contribute to any refinement in the choice of the ISpdf, and the burden of choosing ISpdf still rests with the analyst. The study by Bichon et al., [2008] develops surrogates for the performance function by using Gaussian Process regression and uses the measure of uncertainty provided by the model to decide when to trust the label provided by the surrogate and when to call the oracle code. Whenever a call to the oracle is made, the data thus generated is used to update the surrogate model, thereby facilitating informed decisions on newer samples. In this model, one approximates the true performance function, given a limited dataset, to be a random function given by $G(\boldsymbol{u}) \sim \mathrm{N}\left[\mu_G(\boldsymbol{u}), \sigma_G(\boldsymbol{u})\right]$ with $\mu_G(\boldsymbol{u})$ serving as the surrogate. The study introduces the notion of rgw expected feasibility function

$$\mathrm{EF}\left[\hat{G}(\boldsymbol{u})\right] = \int_{-\infty}^{\infty} \left[\varepsilon - |\overline{z} - g|\right] p_G(g) dg \quad \text{with}\, p_G(g) = \text{with}\ \mathrm{N}\left[\mu_G(\boldsymbol{u}), \sigma_G(\boldsymbol{u})\right] \tag{100}$$

where $G(\boldsymbol{u}) - \overline{z} = 0$ is the limit surface (note the departure from standard notations) and $\varepsilon$ is a parameter that helps to define a band around the limit surface. A new sampling point which maximizes $\mathrm{EF}\left[\hat{G}(\boldsymbol{u})\right]$ is chosen as the candidate point for calling the oracle and facilitates the further updating process.

The problem of reliability modelling of structures prone to bifurcations has been considered by Basudhar et al., [2008]. In this class of problems, the failure surface cannot be easily defined a priori by smooth performance functions, as the response tends to be discontinuous and failure regions tend to be disconnected. The authors employ clustering algorithms to first identify the geometry of failure regions and then use SVM classifiers with data-based kernels for reliability assessment. This model is further used within the reliability-based design optimization tasks. The labelled data points needed for the initial clustering step is obtained by employing space-filling Latin hypercube sampling (LHS). Illustrative examples include studies on the snap-through behaviour of a randomly parametered arch. An SVM-based binary classifier for reliability estimation has been studied by Hurtado and Alvarez [2010]. To choose the labelled dataset, the authors generate sample points using Sobol sequence-based quasi-random numbers and further refine the points by minimizing the function $G^2(\boldsymbol{u})$. The idea is that this function has a local minimum all along the limit surface, and the optimization ensures that the training points are chosen to lie in the neighbourhood of the limit surface. The optimization task is accomplished using particle swarm optimization and the dataset thus generated is used to train an SVM classifier. This model is subsequently employed in a direct Monte Carlo simulation loop to estimate the failure probability.

The paper by Echard et al., [2011] introduces a new method termed as AK-MCS (Active learning reliability method combining kriging and Monte Carlo Simulations). The surrogate model here is based on Gaussian Process regression. The study introduces a new learning function called the U-function, which aims to measure the probability that the kriging model fails to correctly classify a new point. The performance function here is modelled as $G(\boldsymbol{u}) \sim \mathrm{N}\left[\mu_G(\boldsymbol{u}), \sigma_G^2(u)\right]$ and the kriging surrogate is taken to be $\mu_G(\boldsymbol{u})$. At a given (new) point $\boldsymbol{u}$, the probability of misclassification can be shown to be given by

$$\mathrm{P}\left[\mu_G(u) G(u) < 0\right] = \Phi\left[-\frac{|\mu_G(u)|}{\sigma_G(u)}\right] \tag{101}$$

The ratio $\frac{|\mu_G(u)|}{\sigma_G(u)}$ is called the U-function. The approach begins by simulating a fixed amount of Monte Carlo population from $p_U(\boldsymbol{u})$. A randomly chosen subset of this population is used to train the classifier. The resulting surrogate is used to estimate the probability of failure based on the samples from the fixed population. Next, for each member of the population, the learning function is evaluated, and the sample point which has the highest probability of misclassification is identified. The label for this point is obtained using the oracle, and this data point is added to the existing set of labelled dataset and the GP classifier is retrained. The process is repeated till the minimum value of the U-function for the population exceeds a prescribed threshold. The ability of the method to adequately capture geometric details of the failure surface depends upon the choice of fixed Monte Carlo population and its intersection with important regions contributing to the failure probability. An extension of this approach to problems of system reliability modelling can be found in the work of Fauriat and Gayton [2014].

Richard et al., [2012] build on the earlier work of Bucher and Bourgund [1990] and introduce SVM-based regression models in the reliability calculations instead of the polynomial-based models. The study relies on Hasofer-Lind reliability index-based framework and proposes an adaptive experimental design scheme. The method begins by adopting a star sampling scheme in the standard normal space and building an SVM regressor for $G(\boldsymbol{u})$. A FORM calculation is performed on this model, leading to the identification of the design point. In the following step, the centre of sampling is translated towards the design point and, more importantly, the sampling axes are rotated to coincide with the gradient vector (computed analytically from the surrogate) at the design point. A new SVM regressor is now developed, leading to further improvements in the location of the design point, with the iterations stopping when the design point converges. The gradient computations needed for implementing FORM calculations here are based on the surrogate and not on calls to the oracle. The method thus can be viewed as an enhanced tool for FORM calculations when the performance function is defined via long-running codes.

For problems involving nonlinear incremental structural analysis (as in pushover analysis, for example), the study by Jiang et al., [2015] proposes an SVM-based approach for constructing the failure surface. Here, the authors begin by obtaining a set of input variables by adopting a uniform sampling strategy to ensure space-filling coverage. For each realization of the input vector, the incremental nonlinear analysis is performed with a view to identify $\lambda_L$, the limiting load factor at which the failure occurs. Based on this, the study picks a sample pair on safe and failure load closest to $\lambda_L$. A collection of such pairs for all the input data points forms the labelled dataset to train an SVM classifier in an augmented space of basic variables and the load parameter. The method thus ensures that all samples are deliberately generated close to the true failure surface, thereby enhancing the quality of the SVM classifier. Furthermore, the SVM model implicitly defines the relationship between the input vector and $\lambda_L$ , which forms the basis for further reliability calculations (with no further calls to the oracle). Clearly, the method provides the description of the failure surface in a global sense, while such a detailed model may not be needed if interest is limited to only reliability computations (in which case, details of the function deep into the failure region is of little value). Jiang et al., [2017] propose a heuristic approach in which the standard normal space is partitioned into a set of sectors and an SVM-based classifier is constructed within each sector. The piecewise representation of the failure surfaces is combined in a subsequent simulation step to estimate $P_F$. The method involves iterative steps to ensure correct classification of points within each sector. The study by Guimaraes et al., [2018] employs classical polynomial-based surrogates and contains several steps that are adopted in ML implementation (e.g., adaptations and cross-validation steps). A deterministic local sensitivity analysis guides an iterative strategy for model reduction and location of regions where surrogate models are developed. Once an acceptable surrogate is obtained, a FORM analysis is performed, and a further enrichment step involving sampling around the design point and calls to the oracle is implemented before the final model is accepted. The work relies strongly on intuitive local sensitivity analyses with no direct reference to sensitivities with respect to probability of failure.

Improvements to the AK-MCS approach developed earlier by Echard et al., [2011] are suggested in the work of Lelievre et al., [2018]. The suggested improvements are in three aspects: (a) adoption of a sequential Monte Carlo simulation strategy (instead of a single large population) that alleviates demands on computer memory and enables sequential processing in small batches, (b) simultaneous enrichment using multiple samples (as opposed to a single point enrichment), and (c) introduction of a stopping criterion based on the probability

$$\mathrm{P}\left[\text{all } n \text{ points are correctly classified}\right] = \prod_{i=1}^{n} \Phi\left[-\frac{\left|\mu_G\left(u^j\right)\right|}{\sigma_G\left(u^j\right)}\right] \tag{102}$$

and not just on the values of the U-function. The study also discusses issues related to parallelization of the computations.

Xu et al., [2020] combine adaptive kriging regressors with sampling strategies that resemble those in subset simulation method. They partition the domain of integration into a set of annular hyper-shells ($D_i, i = 1, 2, \cdots, m$) of increasing inner radii where the radii are determined implicitly to satisfy a prescribed conditional probability level. Samples from the pdf $p_U\left(\boldsymbol{u} \middle| \boldsymbol{u} \in \bigcup_{r=1}^{i} D_r\right)$ are drawn by using MCMC samplers for $i = 1, 2, \cdots, m$. It needs to be noted that this sampling is based purely on geometric considerations, and no calls are made to the oracle in this step. This pool of samples is held fixed in further computations. A kriging surrogate is made by calling the oracle using samples chosen uniformly randomly from the population of samples already generated. An active learning step using U-function is next implemented to find the sample point with the lowest U value and the corresponding label using the oracle. The kriging model is updated with this additional data point. The process is repeated till stable estimate of the failure probability is obtained. It may be noted that this scheme does not aim to reduce sampling variance, notwithstanding the fact that ideas from subset simulations are borrowed. The MCMC step that the authors use could as well be replaced by a more direct simulation strategy. An improvement to the active learning scheme used in AK-MCS is discussed in the work of El Haj and Soubra [2021]. As in the AK-MCS method, a fixed Monte Carlo population is used and a global kriging model for the performance function is initially made. The study treats $\mathrm{P}\left[\hat{G}(\boldsymbol{u}) \leq 0\right]$ as a random variable with

$$
\begin{aligned}
&\mathrm{E}\left[P_f\right]=\frac{1}{N}\sum_{i=1}^{N} e_i \text{ with } e_i=\mathrm{E}\left\{\mathrm{P}\left[\hat{G}(u_i)\leq 0\right]\right\}=\Phi\left[-\frac{\mu_i}{\sigma_i}\right] \\
&\mathrm{Var}\left[P_f\right]=\frac{1}{N^2}\left[\sum_{i=1}^{N} e_i(1-e_i)+2\sum_{i<j}^{N}\left(e_{ij}-e_ie_i\right)\right] \text{ with } e_{ij}=\mathrm{E}\left\{\mathrm{P}\left[\hat{G}(u_i)\leq 0\cap\hat{G}(u_j)\leq 0\right]\right\}
\end{aligned}
\tag{103}
$$

The contribution to $\mathrm{Var}\left[P_f\right]$ by the $i^{\text{th}}$ points is shown to be given by

$$
c_i=e_i\left(1-e_i\right)+\sum_{j\neq i=1}^{N}\left(e_{ij}-e_ie_i\right) \tag{104}
$$

The study proposes $c_i$ be used as a pointwise learning function such that the calls to the oracle are made only for those points for which $c_i$'s are high. By doing so, the expected reduction in the uncertainty of the failure probability is maximized.

The idea of applying Bayesian support vector regression (SVR) in the context of reliability modelling has been explored by Cheng and Lu [2021], and Wang et al., [2021]. These studies do not employ MCMC samplers: instead, they assume a Gaussian model for the predictive posterior (despite the likelihood being non-Gaussian), a Gaussian model for the prior, and combine this assumption with MAP-based approximation for system parameters to arrive at measures of uncertainty in the model predictions. The likelihood function in the $\varepsilon$-insensitive region is assumed to be flat. The study by Cheng and Lu [2021] rely largely on the active learning scheme as adopted in the AK-MCS method, while the study by Wang et al. [2021], proposes a multi-factor learning function to implement the active learning step, which represents refinements to a similar function used in AK-MCS. The factors considered here include: proximity to the failure function, uncertainty linked to the prediction, the value of $p_U(\boldsymbol{u})$ at the sampling point, and a measure of diversity of the samples.

The study by Luo et al., [2022] builds on an idea proposed in the earlier study by Naess and Gaidai [2009]. Here, a scaling parameter $\lambda$ such that $0<\lambda<1$, which multiplies the term representing the structural capacity, is introduced so that the probability of failure can be written as $P_f(\lambda)=\mathrm{P}\left[\lambda R(\boldsymbol{U})-S(\boldsymbol{U})\leq 0\right]$. Clearly, the smaller the value of $\lambda$ is, the larger will be $P_f(\lambda)$. The idea is to arrive at a set of labelled data $\left[\lambda_i,\hat{P}_F(\lambda_i)\right], i=1,2,\cdots,n_\lambda$ using direct Monte Carlo simulations, using which an ANN is trained. The ANN output for $\lambda=1$ is taken as the estimate for the probability of failure. The surrogate model here is not for the performance function but for $P_f(\lambda)$. The authors propose that the training of the ANN be based on only those data points for which the coefficient of variation of the estimate of $P_f(\lambda)$ lies in the range $(0,0.5)$ so that the confidence in the ANN model is enhanced. The success of this method crucially hinges on the admissibility of the extrapolation of the ANN model for the case of $\lambda=1$. Also, the method assumes that the performance function can be written explicitly in the form $G(\boldsymbol{U})=R(\boldsymbol{U})-S(\boldsymbol{U})$, which may not always be possible (e.g., consider performance functions related to failure due to buckling or occurrence of resonance). The work of Guo et al., [2025] develop this idea of introducing a scaling parameter further by introducing active kriging-based surrogates and SVM-based tools for extrapolation of results for $\lambda=1$.

Yin and Du [2022] embed ideas from sufficient dimension reduction and introduce the framework of generalized sliced inverse regression (GSIR) [Lee et al., 2013] into the framework of reliability modelling. The study involves an initial application of FORM to locate the design point and to define a shifted Gaussian ISPDF. A set of $N_s$ random draws, termed IS population, is drawn from this ISPDF. An LHS strategy is used to create a cloud of initial data points around the origin. A set of labelled data points $\left[\boldsymbol{u}^i,G(\boldsymbol{u}^i)\right], i=1,2,\cdots,n_d$ is thus obtained. The GSIR is applied to this dataset, which leads to the model reduction $\boldsymbol{U}\in\mathbf{R}^n\rightarrow f_{GSIR}(\boldsymbol{U})\in\mathbf{R}^d$ such that $G(\boldsymbol{U})\approx\hat{G}\left[f_{GSIR}(\boldsymbol{U})\right]$ where the surrogate $\hat{G}\left[f_{GSIR}(\boldsymbol{U})\right]$ is taken to be a Gaussian random process. At the implementation step, the study assumes $d=1$. An active learning step based on *U*-function is next implemented to choose the most uncertain point among the ISPDF samples generated above. A call to the oracle is made at this point, and the GSIR and GP steps are repeated. The estimate for the probability of failure is obtained as

$$\hat{P}_F = \frac{1}{N}\sum_{i=1}^{N}\mathbf{I}\left[\hat{G}_{GSIR}\left(f\left(\hat{U}^i\right)\right)\leq 0\right]\frac{\phi_U\left(\hat{U}^i\right)}{\phi_U^{ISPDF}\left(\hat{U}^i\right)} \tag{105}$$

The method does not reduce the original $\boldsymbol{U}$-space: instead, it obtains a reduced functional representation for the failure surface. The method works if the problem admits a low-dimensional sufficient representation, typically associated with limit surfaces possessing only one design point.

The study by Liu et al., [2022] introduce refinements to the AK-MCS by restricting the repeated efforts involved in evaluating $U$-function for the members of the candidate sampling pool (CSP). As in the AK-MCS, the study begins by creating a CSP, choosing a few points from this CSP and arriving at an initial kriging model for $G(\boldsymbol{U})$. Based on this, the members of CSP are grouped into two categories: one for which $U(u)\geq u^*$ (representing points which are confidently classified) and the second for which $U(u)<u^*$ (representing points which are deemed poorly classified), where $u^*$ is a user-defined threshold. The first set of points are now removed from the CSP. Among the surviving points in the reduced pool, the most uncertain point is now added to the training set, and an updated kriging model is obtained. The process is repeated till $\min U(u)\geq 2.$ In a further refinement, the method recalculates the $U$-function at all the points in the original CSP and checks if any of the points which were removed from the CSP are no longer confidently classified. To correct for any such missteps, the method increments $u^*$ and the whole process is repeated. Lieu et al., [2022] propose an ANN-based surrogate model which is adaptively refined. The method begins with the generation of a Monte Carlo population and creating an ANN-based surrogate for the performance function based on a labelled dataset from a select sample from the population. The surrogate is next evaluated at all the points in the MCS population and those points which lie within a specified band around the failure surface are chosen for oracle calls. This results in additional labelled data points with which the surrogate is retrained. As the iterations conclude, the less important data points are removed, and the final surrogate is made by using only those points close to the limit surface. The study does not accord any greater importance to points close to the origin where probability masses are significant; instead, it arrives at a global fit to the performance function near the limit surface. Since the ANN based surrogate typically provides no measures of uncertainty (unlike kriging, for instance), the criterion used for adapting the surrogate remains deterministic.

The idea of applying Bayesian support vector regression (SVR) in the context of reliability modelling has been explored by Wang et al., [2021]. The study does not employ MCMC samplers: instead, it assumes a Gaussian model for the predictive posterior (despite the likelihood being non-Gaussian), a Gaussian model for the prior, and combines this assumption with MAP based approximation for system parameters to arrive at measures of uncertainty in the model predictions. The likelihood function in the $\varepsilon$-insensitive region is assumed to be flat. The study proposes a multi-factor learning function to implement the active learning step, which represents refinements to a similar function used in AK-MCS. The factors considered here include: proximity to the failure function, uncertainty linked to the prediction, the value of $p_U(\boldsymbol{u})$ at the sampling point, and a measure of diversity of the samples. The scope of this core idea is further expanded in Wang et al., [2023, 2026] by introducing variance reduction strategies (subset simulation and importance sampling), dimension reduction, along with methodological refinements at the active learning and convergence control steps.

Wang et al., [2022] consider the adaptive kriging framework and introduce a new learning function that considers four factors, namely, (a) proximity to the limit surface, (b) variability in the kriging model, (c) value of $p_U(\boldsymbol{u})$, and (d) distance to the existing set of points. Furthermore, the study embeds a bootstrap-based sampling step to formulate the stopping criterion, which respects the convergence of estimates of $P_F$ across kriging model adaptation, surrogate model misclassification, and the sample size of the Monte Carlo pool of samples.

Zhang et al., [2023] employ a Bayesian ANN formulation to develop probabilistic ANN surrogates for the performance function. The study begins by obtaining a labelled dataset (300-500) by calls to the oracle. A MAP estimate for the ANN parameters is obtained based on which a Gaussian approximation to the posterior pdf of the parameters is obtained by centring the distribution around the MAP estimate. An ensemble of ANNs is subsequently generated based on which a probabilistic model for $P_F$. The assumption of a Gaussian posterior here sidesteps the need for expensive MCMC samplers but nevertheless makes the formulation ad hoc. The study by Dang et al., [2025] begins with an initial GPR surrogate by using a small number of samples drawn uniformly within a *d*-dimensional ball in standard normal space. A Gaussian Process Regression model is made for the performance function based on this initial sample. This leads to the estimate of the mean $P_f$. To evaluate a measure of dispersion, instead of evaluating the variance of $P_f$, the authors propose to use an upper bound on the mean absolute deviation of $P_f$. This measure is shown to be computable using a *d*-dimensional quadrature (instead of 2*d* dimensional integral if variance was used as a measure of dispersion). The authors proceed to formulate an active learning strategy and stopping criteria based on this measure of dispersion.

**8.1.2 Time-variant reliability models**

Problems of randomly parametered systems under deterministic excitations have been considered by Hu and Du [2015]. These authors develop kriging models $\hat{G}(\boldsymbol{u},t) \sim \mathrm{N}\left[\mu_G(\boldsymbol{u},t), \sigma(\boldsymbol{u},t)\right]$ and consider the problem of determining the probability

$$\mathrm{P}_F = 1 - \mathrm{P}\left[\hat{G}(\boldsymbol{u},t) \le g^* \, \forall t \in (0,T)\right] = 1 - \mathrm{P}\left[\max_{t \in (0,T)} \hat{G}(\boldsymbol{u},t) \le g^*\right] \tag{106}$$

This is accomplished by using direct Monte Carlo simulations on the surrogate combined with active learning guided calls to the oracle. The study proposes a modified learning function given by

$$\mathrm{EIF}(\boldsymbol{u},t) = \mathrm{E}\left[\max\left(\left\{G(\boldsymbol{u},t) - y^*\right\}, 0\right)\right] \tag{107}$$

where $y^*$ is the currently available extreme response. A simple way to decide upon the sample for which an oracle call is to be made among a pool of samples is to pick the sample which maximizes $\mathrm{EIF}(\boldsymbol{u},t)$. Hu and Mahadevan [2016] consider a more general problem in which, in addition to allowing for randomness in system parameters, the external excitations could also be modelled as a vector of random processes. In this case the random processes are discretized using Karhunen-Loeve expansion and thus represented as a set of random variables. Kriging model is now developed as a function of time as well as the composite vector of random variables involving system parameters and discretization of excitations. The active learning step here is based on the use of a modified *U*-function.

Garg et al., [2022] build on the work of Lu et al., [2021] and develop DeepONet (Deep Operator Network) based surrogate models for randomly excited nonlinear dynamical systems and employ them for estimating time-variant reliability based on Monte Carlo simulations on the surrogate. The surrogate is a map $G:\boldsymbol{f}(\bullet)\rightarrow\boldsymbol{x}(\bullet)$ where $\boldsymbol{f}(t)$ is the input and $\boldsymbol{x}(t)$ the output functions $\left[x(t)\right]$ and consists of a branch neural network (NN) (encoding the input forcing function) and a trunk NN (encoding time) leading to the model of the form $x(t)=\sum_{k=1}^{p}\underbrace{b_k\left[f(t)\right]}_{\text{Branch}}\underbrace{t_k(t)}_{\text{Trunk}}$. These two components are akin to data-driven modal decomposition, where $t_k(t)$ are the temporal basis functions and $b_k\left[f(t)\right]$ are input-dependent modal weights. The training of the NNs here is based labelled dataset obtained by solving the governing equations for different realizations of the input process with $p$ being the dimension of the latent space, a model hyperparameter.

Navaneeth and Chakraborty [2022] build on NN-based Koopman learning [Yeung et al., 2019, Lusch et al., 2018] to develop surrogate models for randomly parametered dynamical systems under deterministic excitations. The focus of Koopman operator here is to learn from data a higher-dimensional linear representation of a given nonlinear dynamical system of the form $\boldsymbol{x}_{k+1}=\boldsymbol{f}\left[\boldsymbol{x}_k,\boldsymbol{u}_k,\boldsymbol{\theta}\right];\boldsymbol{x}_{k=0}=\boldsymbol{x}_0$. The NN architecture here consists of an encoder, with an input layer consisting of $\boldsymbol{x}_k$ and $\boldsymbol{\theta}$ and an output layer $\boldsymbol{z}_k=N_1\left[\boldsymbol{x}_k,\boldsymbol{\theta},\boldsymbol{\alpha}\right]$ and an evolution step $\boldsymbol{z}_{k+1}=\mathbf{K}\boldsymbol{z}_k$. The decoder accepts $\boldsymbol{z}_{k+1}$ as input and delivers $\hat{\boldsymbol{x}}_{k+1}=N_2\left[\boldsymbol{z}_{k+1},\boldsymbol{\beta}\right]$ as the output. The learning parameters here are $\boldsymbol{\alpha},\boldsymbol{\beta}$, and $\mathbf{K}$. The training is performed on a labelled dataset of pairs of system parameters $\boldsymbol{\theta}$ and states $\boldsymbol{x}_k$ (computed from the oracle). A straightforward Monte Carlo simulation of the surrogate thus developed leads to the estimates of failure probability. It must be noted that this architecture assumes that the excitation remains fixed and therefore, the uncertainty here is restricted to only the system parameters. A challenge here is to choose the size of the latent vector, which represents the size of the enhanced linear operator, as this choice essentially implies an ad hoc closure on the model size. Tripura et al., [2024] consider randomly excited nonlinear dynamical systems and develop surrogates based on the idea of wavelet neural operators developed earlier by the same authors [Tripura and Chakraborty 2023]. The input forcing vector $\boldsymbol{f}(t)$ is lifted to a larger vector $\boldsymbol{v}_0(t)$ through a fully connected FFNN. This is followed by a sequence involving wavelet transform, learned linear filtering in wavelet domain, inverse transform, followed by a nonlinear activation function, resulting in the output at the final layer. $\boldsymbol{v}_L(t)$. This serves as the input to a multi-variate regression step leading to the time history of the response vector. This model takes into account system nonlinearity and nonstationarity in the input as well as the output and allows for non-Gaussian models for the input random process.

Das and Tesfamariam [2024] and Guo et al., [2024] consider problems of time-variant reliability analysis of dynamical systems in which all uncertainties (including system parameters and excitations) can be represented by a finite-dimensional vector of random variables. The authors employ a PDEM-based formulation to derive the governing equation for the transport of the pdf and use an ANN-based collocation method to approximate the solution. The study does not explicitly enforce the condition on non-negativity of the pdf and the requirement on the area under the curve to be unity. The study by Gong and Cheung [2025] is relevant to surrogate modelling of systems governed by stochastic differential equations, although the study itself does not address questions on rare event simulations.

## 8.2 ML tool embedding in sampling variance reduction strategies

### 8.2.1 Time-invariant reliability models

The role of tools such as particle splitting and importance sampling pdf-based methods in controlling sampling variance has already been discussed. Incorporating ML-based methods into these frameworks could either aim to reduce the number of calls to the oracle or aid in the formulation of the ISPDF strategy.

Li and Xiu [2010] and Li et al., [2011] consider approaches in which surrogates are used as global regressors for the performance function. Based on rigorous theoretical arguments, the first study highlights that an accurate global regressor need not lead to accurately classifying points into failure and safe regions. At the stage of estimating $P_F$ the study proposes the use of a surrogate only for sufficiently safe regions and makes calls to the oracles when the point is close to the failure region. In the subsequent study, the authors introduce ideas from importance sampling into the framework. The ISPDF here is determined by minimizing the Kullback-Leibler divergence between a parametrized family of pdfs (Gaussian in this case) and the ideal ISPDF. While implementing the optimization in this step, the study uses the surrogates when the points are in sufficiently safe regions and calls are made to the oracle only when points lie close to the limit surface.

The idea of introducing SVM-based classifiers at every intermediate failure level within the subset simulation framework was developed by Bourinet et al., [2011] and the method was styled as $^{2}$SMART (Subset simulation by Support-vector Margin Algorithm for Reliability esTimation). The study retains all the essential elements of subset simulations as developed by Au and Beck [2001a] but introduces a step involving training a SVM-based binary classifier (with Gaussian RBF kernels) at every intermediate failure level using labelled data already available. This classifier is then used to label the new data points, with the oracle being called only when the data lies within a prescribed neighbourhood of the classifier boundary. The method becomes advantageous when individual calls to the oracle is computationally demanding. The study by Huang et al., [2016] adopt adaptive kriging modelling incorporating U-function based active learning to obtain a regression model for the performance function. The probability of failure is estimated by adopting a subset simulation step on the surrogate model developed.

Dubourg et al., [2013] considered kriging-based regressors for the performance function. By noting that $\hat{G}(\boldsymbol{u}) \sim \mathrm{N}\left[\mu_G(\boldsymbol{u}), \sigma_G(\boldsymbol{u})\right]$, the authors introduce a probabilistic classification function given by $\tilde{p}(\boldsymbol{u}) = \mathrm{P}\left[\hat{G}(\boldsymbol{u}) \le 0 \middle| \boldsymbol{U} = \boldsymbol{u}\right]$, leading to the notion of an augmented probability of failure given by

$$P_F^{\mathrm{aug}} = \int_{-\infty}^{\infty} \tilde{p}(\boldsymbol{u}) p_U(\boldsymbol{u}) d\boldsymbol{u} = \mathrm{E}_{p_U(\boldsymbol{u})}\left[\tilde{p}(\boldsymbol{U})\right] \tag{108}$$

Clearly, $\tilde{p}(\boldsymbol{u})$ is not a probability density function. However, the function, termed the quasi-optimal instrumental pdf, given by

$$\hat{h}(\boldsymbol{u}) = \frac{\tilde{p}(\boldsymbol{u}) p_U(\boldsymbol{u})}{\int_{-\infty}^{\infty} \tilde{p}(\boldsymbol{u}) p_U(\boldsymbol{u}) d\boldsymbol{u}} = \frac{1}{P_F^{aug}} \tilde{p}(\boldsymbol{u}) p_U(\boldsymbol{u}) \tag{109}$$

is a valid pdf. The probability of failure is now evaluated using

$$
\begin{aligned}
P_F &= \int_{-\infty}^{\infty} \mathrm{I}\left[G(\boldsymbol{u}) \leq 0\right] p_U(\boldsymbol{u}) d\boldsymbol{u} = \int_{-\infty}^{\infty} \frac{\mathrm{I}\left[G(\boldsymbol{u}) \leq 0\right]}{\hat{p}(\boldsymbol{u})} \hat{p}(\boldsymbol{u}) p_U(\boldsymbol{u}) d\boldsymbol{u} \\
&= \int_{-\infty}^{\infty} \frac{\mathrm{I}\left[G(\boldsymbol{u}) \leq 0\right]}{\hat{p}(\boldsymbol{u})} P_F^{aug} \hat{h}(\boldsymbol{u}) d\boldsymbol{u} = P_F^{aug} \mathrm{E}_{\hat{h}(\boldsymbol{u})}\left[\frac{\mathrm{I}\left[G(\boldsymbol{U}) \leq 0\right]}{\hat{p}(\boldsymbol{U})}\right] = P_F^{aug} \alpha_{cor}
\end{aligned}
\tag{110}
$$

While $P_F^{\text{aug}}$ can be estimated by sampling from $p_U(\boldsymbol{u})$, the estimation of $P_F$ involves sampling from $\hat{h}(\boldsymbol{u})$ for this purpose, the authors use MCMC samplers. The study demonstrates that the estimator thus obtained is unbiased.

One of the earliest strategies for developing importance sampling pdf-based methods is to employ a Gaussian pdf shifted to the design point calculated using FORM [Melchers 1990]. The study by Echard et al., [2013] builds on this idea and introduces kriging models with active learning capabilities to enhance the approach. The study begins by the application of FORM to approximately locate the design point. This involves several calls to the oracle, and the available data is next used to build a kriging-based classifier. New samples are now drawn from a Gaussian pdf centred at the design point. These newly drawn samples are now labelled using the kriging model with a U-function based active learning step, which identifies the sample with maximum classification uncertainty for which a call to the oracle is made. The kriging model is retrained using the new data point labelled using the oracle. At every iteration step, the provisional $P_f$ is estimated using importance weights. The iteration stops when the kriging model classifies all the sample points with sufficient confidence.

Cadini et al., [2014] combine ideas from the meta-model-based ISPDF framework (meta IS) of Dubourg et al, [2013] and the adaptive kriging importance sampling (AK-IS) method of Echard et al., [2013]. The new method they develop avoids the use of ideas from FORM, uses meta IS method to construct an ISPDF located in the failure region carry forward this information into the AK-IS to obtain an estimate of failure probability. Tong et al., [2015] combine subset simulations, kriging, active learning, and importance sampling to develop a reliability estimation scheme. The importance sampling here is done at every subset level, thereby replacing MCMC samplers. The active learning scheme at every subset level uses a modified U-function defined for every subset level $b_k$ given by

$$U\left(\boldsymbol{u},b_k\right)=\frac{\left|\mu_G\left(\boldsymbol{u}\right)-b_k\right|}{\sigma_G\left(\boldsymbol{u}\right)} \tag{111}$$

enabling a population-level stopping criterion that is less conservative than the pointwise criterion used in AK-IS.

The study by Chojaczyk et al., [2015] compares the relative performance of ANN-based reliability models and models based on FORM + importance sampling, on a large-sized problem (141450 dof nonlinear FE model with 15 random variables). The study also discusses several ANN-based alternatives for reliability modelling (e.g., ANN surrogates combined with direct Monte Carlo simulations, FORM, importance sampling, and adaptive strategies).

An adaptive strategy that combines surrogate-based regression with ideas from subset simulations is developed by Bourinet [2016]. The author uses SVM regressor, which is adaptively refined by first focusing on the interior of the safe region and then progressing towards the limit surface. An initial set of labelled data of size *N* is obtained by calling the oracle. The resulting $G(\boldsymbol{u})$'s are rank-ordered, and the median point $G^m$ is identified. A modified performance function $G(\boldsymbol{u})-G^m$ is defined, and a subset simulation-based estimate of violation of this limit state is obtained. From the pool of data points produced by the subset simulation run, m points in the failure region are randomly selected and calls to the oracle are now made at these failure points. This leads to a pool of *N*+*m* data points labelled by the oracle. From this pool, the lowest *N* points are selected, and its median is used to define the next failure level. This process is repeated till the samples migrate towards the true failure surface. The main advantage of this method vis-à-vis is the reduction in calls to the oracle: in a subset simulation framework approximately *N* calls are made to oracle at every intermediate failure level, whereas here the number of calls is *m*. The study by Lv et al., [2015] introduces an entropy-based learning function to be used in conjunction with a kriging-based regressor for $G(\boldsymbol{u})$.The intuition here is that the contribution to the entropy associated with the kriging model $\hat{G}(\boldsymbol{u})\sim \mathrm{N}\left[\mu_G(\boldsymbol{u}),\sigma_G(\boldsymbol{u})\right]$ in the neighbourhood of the limit surface provides a measure of uncertainty associated with the classification task. Thus, if the quantity

$$H(\boldsymbol{u}) = \left| -\int_{G^-}^{G^+} p_{\hat{G}}(g) \log p_{\hat{G}}(g) dg \right| \tag{112}$$

with $G^{\pm} \approx \pm 2\sigma_{\hat{G}}$ is evaluated at a set of input points, its value will be high for points for which uncertainty is high. These points thus become candidate points for calling the oracle. Based on this learning function, the initial kriging model is refined, and a line sampling scheme is further applied to estimate $P_F$.

The study by Cui and Ghosn [2019] integrates five distinct and complementary tools, viz., subset simulations, kriging, active learning, K-means clustering, and FORM to address several challenges inherent in problems of system reliability. The tools are orchestrated to tackle multiple issues such as automatic identification of multiple dominant failure modes, low probabilities of failure, performance functions defined implicitly via nonlinear FEM codes, and relative importance of different variables. Roy et al., [2019] note the limitations of conventional hyperparameter tuning strategies while developing SVM regressor based surrogates for reliability modelling and propose alternatives to address related issues.

Xiang et al., [2020] use a reinforcement learning framework to choose sampling points to build the surrogate model. The study primarily focuses on 2-dimensional problems and divides the input space into a uniformly spaced grid over $[-4,4]^2$ which defines the states in reinforcement learning. The reward function is designed such that points close to the limit surface receive higher rewards. The sampling points thus generated are used to build an ANN-based surrogate which in turn is used to estimate the probability of failure. The study is limited by consideration of only two-dimensional problems with the anticipation that a dimension reduction strategy would enable the treatment of higher dimensional problems. The study by Zhan et al., [2022] builds on the earlier work of El Haj and Soubra [2021]. Here, an initial kriging model is first made and an ISPDF based on this model is constructed by choosing points lying in the failure region of the surrogate and points close to the failure surface. A further set of samples from this ISPDF is obtained and grouped into *k* clusters. From each of these clusters, one point is actively chosen based on its contribution to reducing the uncertainty in failure probability and calls to the oracle are made at these chosen points. The steps are repeated so that the ISPDF is developed in an adaptive manner. This strategy permits parallelization of computations related to calls to the oracle.

Ma et al., [2022] consider multiple performance functions $G_i(\boldsymbol{U}), i=1,2,\cdots,m$ associated with a common oracle and consider the problem of concurrent evaluation of associated probabilities of failure $P_{EFi}, i=1,2,\cdots,m$. Given the common oracle and the common random input random variables, $G_i(\boldsymbol{U}), i=1,2,\cdots,m$ would clearly be dependent. A single call to the oracle would result in evaluation of all $G_i(\boldsymbol{U}), i=1,2,\cdots,m$ at minor incremental cost. This leads to the expectation that all $P_{EFi}, i=1,2,\cdots,m$ can be simultaneously estimated with a common sampling strategy, thereby avoiding $m$ independent estimations of $P_{EFi}, i=1,2,\cdots,m$. Accordingly, the study considers simultaneous development of $m$ kriging models $\hat{G}_i(\boldsymbol{U}), i=1,2,\cdots,m$ and a common active learning strategy to optimally call the oracle. Associated with each of the kriging models, $\hat{G}_i(\boldsymbol{U}), i=1,2,\cdots,m$ the study computes the associated *U*-functions and EFFs. These are combined via aggregation (using a min or mean-type criterion) to select the single point that is most informative across all performance functions and where a call to the oracle is made. The estimation of $P_{EFi}, i=1,2,\cdots,m$ is subsequently carried out by using either direct Monte Carlo simulations or by the use of a generalized subset simulation strategy based on the kriging models developed. The generalization here enables concurrent estimation of all $P_{EFi}, i=1,2,\cdots,m$ using common subset simulation steps. It is to be noted here that dependence among $G_i(\boldsymbol{U}), i=1,2,\cdots,m$ provides no guarantees that the important regions that contribute to individual $P_{EFi}, i=1,2,\cdots,m$ need to geometrically lie close to each other, thereby limiting the advantages of a common sampling strategy.

Jafari-Asl et al., [2025] build on the earlier study by Rashki [2018], which introduced the idea of control variates in reliability estimation. The idea here is to rewrite the probability integral as

$$P_F = \int_{-\infty}^{\infty} \mathrm{I}\left[G(\boldsymbol{u}) \le 0\right] k(\boldsymbol{u})\, d\boldsymbol{u} + \int_{-\infty}^{\infty} \mathrm{I}\left[G(\boldsymbol{u}) \le 0\right]\left[p_U(\boldsymbol{u}) - k(\boldsymbol{u})\right] d\boldsymbol{u} \tag{113}$$

where $k(\boldsymbol{u})$ is a valid pdf which could have the same features as $p_U(\boldsymbol{u})$ but with increased standard deviation (e.g., =2). The idea is to evaluate the first integral here using subset simulations and the second integral using importance sampling, leading to

$$P_F = P_F^{SS} + \int_{-\infty}^{\infty} \mathrm{I}\left[G(\boldsymbol{u}) \le 0\right] \frac{\left[p_U(\boldsymbol{u}) - k(\boldsymbol{u})\right]}{h(\boldsymbol{u})} h(\boldsymbol{u})\, d\boldsymbol{u} \tag{114}$$

The study proposes $h^*(\boldsymbol{u}) = \frac{\mathrm{I}\left[G(\boldsymbol{u}) \le 0\right] k(\boldsymbol{u})}{P_F^{SS}}$ to be the ISPDF leading to

$$P_F = P_F^{SS} \mathrm{E}_{h^*}\left[\frac{p_U(\boldsymbol{u})}{k(\boldsymbol{u})}\right] \tag{115}$$

The present study proposes that the estimation of $P_F^{SS}$ and hence the model for $h^*(\boldsymbol{u})$ be developed based on a surrogate model for $G(\boldsymbol{u})$. The study by Wang et al., [2026] integrates several mature concepts, viz., Bayesian support vector regression, active learning, subset simulation, kernel density importance sampling, and FORM-based dimension reduction to enhance the computational efficiency of surrogate-based reliability estimation.

### 8.2.2 Time-variant reliability models

The study by Sharma and Manohar [2026] reformulates the problem of determining state-dependent Girsanov controls as a problem in constrained stochastic optimal control. The objective function here minimizes the expected square magnitude of the Girsanov control which now is state dependent (unlike the earlier formulation by Macke and Bucher [2003] wherein it was state-independent). The authors further tackle this problem within the framework of reinforcement learning methods using the deep deterministic policy gradient (DDPG) approach.

## 8.3 Unsupervised partition-based methods

A heuristic method based on agglomerative clustering of a set of *N* samples drawn from is developed by Yin and Kareem [2016]. A complete agglomerative cluster would have *N* levels, with the bottommost layer having a singleton in each cluster and the topmost cluster having all the sample points. The proposed method retains the bottommost layer but not all the upper layers. The calculations begin at a user-defined coarse cluster level, and for each cluster the medoid is identified. The oracle calls are made only at these medoids, and the resulting label is assigned to all the members of the cluster. Clusters which are classified as belonging to the safe region are removed from further calculations. In the next step of calculations, only the daughter clusters resulting from the unsafe father clusters are retained. The process of calling the oracle at the medoids of these daughter clusters and further pruning of clusters is repeated, and this is carried on till the bottom-most layer is reached. The oracle is called for all the surviving clusters in the bottom most layer. Once this process is completed, the algorithm would have assigned a label to each of the sampled points, leading to the estimation of $P_F$. Importantly, not all these assignments are based on calls to the oracle but from the cluster level decisions. It may be observed that the clustering adopted here at the input level essentially replaces input as a vector of continuous random variables into an equivalent hierarchy of discrete random variables.

### 8.4 Generative AI for earthquake load modelling

In the context of aseismic structural reliability studies, the need for an ensemble of earthquake ground motions arises naturally. A challenge here is that recorded near-field earthquake ground motions are scarce for most parts of the world, especially for large-magnitude earthquakes and near-field ground motions. Also, the assumption of IID inputs is essentially impossible to satisfy in earthquake ground motion modelling. Given the complexities associated with earthquake phenomena, such as source mechanism, wave propagation through intervening inhomogeneous media, dominant nonlinear effects for large-magnitude and near-field ground motions, and time-frequency nonstationarity, a parametric model is likely to introduce idealizations [Lin and Cai 1995] which may compromise their fidelity. In this context, database-based generative ML approaches offer potential alternatives for simulating the desired ensemble.

The review paper by Marano et al., [2024] provides an overview of applications of generative adversarial networks (GAN) in the fields of earthquake engineering and seismology. Matinfar et al., [2023] employ a CNN-based generative adversarial network (GAN) to develop an image-based modelling of recorded earthquake ground motions. The authors consider a set of reference ground motions which are compatible with a specified smooth design response spectrum. The study begins by preprocessing steps which modify each record to have the same length $(n)$ and sampling interval and are further adjusted to be compatible with the specified response spectrum. A number $N$ is further introduced such that $N^2 \geq n$. Each record is assembled into a $N \times N$ square matrix by resorting to padding with zeros if needed. This matrix is now interpreted as a hypothetical image and serves as the data for training the CNN-based GAN. All the available dataset is used to train the GAN, and the optimization iterations during training is stopped when the response spectrum corresponding to the GAN-produced time histories matches the target response spectrum in an acceptable manner. The study is largely heuristic, with an ad hoc approach to represent the inputs as images, no systematic embedding of the target response spectrum into the training loss function and no adoption of the usual splitting of the input data into training, validation, and generalization sets. The study by Matsumoto et al., [2023] also use GAN to model earthquake ground motions. The study begins by collecting a set of recorded ground motions (that could include records at different stations originating from the same event), and these are pre-processed to assign uniform length and nondimensionalization by their own peak. The data is augmented by considering two horizontal components and components obtained by 45-degree rotation. This dataset clearly does not satisfy the basic requirement that the data for training the GAN form an IID sequence. The GAN architecture is taken to be made of 1D CNNs. The authors employ a vector of descriptors consisting of measures of duration, spectral centroid, zero-crossing rate, and bandwidth measure and formulate the KL divergence between pdfs of these quantities for the training data and the GAN outputs and use it for hyperparameter selection. This study also remains heuristic, with no systematic steps to deal with earthquake records as samples of a nonstationary random process. The authors advocate the strategy to generate a database of ground motions in the context of performance-based earthquake engineering, but its application for systematic time-variant reliability analysis remains unexplored. The application of GAN to produce earthquake records with specified intensity measures has been developed by Huang et al., [2024]. The authors use a vector of features consisting of peak values of acceleration, velocity, and displacement, cumulative absolute velocity, and integral square velocity for evaluation of the generated samples. Issues related to avoiding over-training by adopting measures that are functions of PGA are also discussed. The study uses fully connected feed-forward NNs for the generator and discriminator. The model does not explicitly learn the time-frequency nonstationarity features that are present in recorded ground motion. The study by Masoudifar et al., [2025] use GAN to model 3D ground motions conditioned on a vector of features which includes magnitude, epicentral distance, site condition, and faulting mechanism. The GAN uses ground motions after each component is normalized by its own PGA. In addition to the GAN, the study develops an additional CNN-based NN to arrive at the PGA to be used for scaling the output. The study uses a CNN-based NN architecture tailored for treating time series data. A GAN-based study involving an array of 3D ground motions for a specific seismic event, in which ground motions at virtual instrument locations are estimated, has been conducted by Chen et al., [2025]. The recorded data capture the dependency of ground motion on source-to-site distance, wave propagation path, and local soil variability. While training the GAN, this recorded data is augmented by synthetic data at virtual sensor locations by using a scheme that involves interpolation of response spectra at different locations. This scheme is essentially heuristic with no wave mechanics-based justifications. This also means that the basic requirement that the data needed to train a GAN need to be an IID sequence is altogether ignored here. If one focuses on generating data for ground motion at a given location, this study indeed produces an ensemble, but variability here is enforced by the modelling approach used and does not reflect actual variability that one might expect in ground motions. Insofar as using this ensemble for modelling reliability is concerned, the approach seems to be limited in scope. The study by Ren et al., [2026] consider two components of horizontal recorded ground motions from the San Francisco Bay area with $M \leq 4$ and source to site distance $\leq 100$ km for the period 1990-2022. For training the ML model, each record is represented via the amplitude of the short-time Fourier transform along with data on magnitude, focal depth, and coordinates of the source and the measurement site. These data are used to train a variational autoencoder, which results in a probabilistic model for the latent variable vector. In the simulation step, samples of the latent variables are generated and passed through the trained decoder, resulting in the desired ensemble of ground motions. The restriction $M \leq 4$ here is not a requirement from a methodological perspective. The proposed framework seems to be better suited for aseismic structural reliability studies among the other studies considered here.

In summary, it may be noted that the generative ML-based methods offer more flexible means to simulate an ensemble of ground motions than the parametric pdf-based methods. Features such as time-frequency nonstationary and multi-component nature of ground motion are likely to be captured more naturally in the ML based approach, as the pdf-based approach may need to enforce parametric models to capture these features. In both approaches, the basic difficulties related to scarcity of data and the inevitable need to use data that is not strictly an IID sequence remain. Despite the widespread success of diffusion models in generative ML, their application to earthquake load modelling appears largely unexplored. Generative ML tools indeed can be applied to other loading scenarios such as winds, road roughness, and wave loads. However, in these situations, the scarcity of data, as witnessed in problems in earthquake engineering, does not form a defining bottleneck.

**8.5 Discussion on ML-based methods**

The following observations can be made based on the discussions presented in the preceding sections.

1. The application of ML tools in reliability modelling is much less governed by the ability of these tools to learn input features but more by their ability to obtain computationally cheap surrogates to achieve repeated evaluation of the performance function. Integrating these models with active learning steps has been the main theme of research in this field.
2. Various families of surrogate models, including kriging, SVM, ANN, Bayesian ANN, Deeponet, and neural operators, have been explored. Apart from questions of accuracy of these models, the scope for developing active learning strategies is also linked intimately with the development of these models. Kriging models are inherently capable of providing measures of predictive uncertainty, which eminently enables the formulation of active learning strategies. SVM and ANNs coupled with Bayesian framework also provide similar opportunities. Studies on time-variant reliability using tools such as Deeponet and neural operators are still emerging.
3. The considerations that have governed while developing surrogate models have been the choice of sampling points, formulation of learning functions, enrichment strategies, and stopping criteria. These tools have not been effective in addressing questions on global exploration of the input space to identify important regions that contribute significantly to failure probability.
4. Development of surrogate models for dynamical systems and developing suitable active learning strategies have remained largely unexplored.
5. Many of the ML-based methods incorporate sampling variance reduction schemes (such as subset simulations and importance sampling). However, the ML tools themselves have served only limitedly as the primary drivers to discover newer efficient variance reduction schemes.
6. The application of generative models based on generative autoencoders for simulating ensembles of time histories of earthquake-induced ground motions seems promising. Integrating these strategies with time-variant reliability modelling also has remained unexplored. The application of diffusion process models has remained unexplored.

**9.0 Closure and suggested future directions**

This review has primarily focused on methods for estimation of component structural reliability and discussed four family of approaches: (a) Analytical solutions based on FORM/SORM (for time-invariant reliability analysis) and level crossing approaching (for time-variant reliability problems), (b) sampling variance reduction by importance sampling including the Girsanov transformation method, (c) sampling variance reduction by particle/trajectory splitting, and (d) tools that embed machine learning based approaches. Research in this field is driven by two basic motivations: (A) to address computational problems in dealing with large sized problems (large mechanical dofs and large number of random variables), computation of performance functions based on long running codes needed specially to deal with nonlinear and/or dynamical systems, and (B) to address epistemic difficulties in dealing with lack of knowledge about geometry of failure surface in terms of high nonconvexity, presence and location of multiple regions of comparable importance and disconnected regions. The methods that this review discusses aspire to address one or more of these concerns, with overcoming the computational bottlenecks being the main research achievement and dealing with the hard epistemic barriers remaining a yet unresolved research challenge.

Analytical solutions-based FORM/SORM and outcrossing theory are the earliest approaches that offered interpretable and conceptually rich frameworks to tackle the problem of structural reliability assessment. The notions of reliability index and design point transformed the problem of evaluation of a multi-fold integral over an irregular domain into a problem of constrained nonlinear optimization. These ideas are arguably one of the most original and influential contributions to this field of research and provided a distinct identity to the subject of structural reliability. They also provided a systematic framework to develop reliability-based design and structural optimization, design code development, and treatment of problems of system reliability. Monte Carlo simulation-based methods, in principle, offer powerful and versatile frameworks to tackle the problem of reliability estimation. However, for these methods to be workable, they need to be reinforced by incorporating carefully designed sampling variance strategies. Much of the research here has been driven by the development of these strategies. No matter whether one employs reliability index-based methods or the variance reduction enabled simulation-based methods, one still needs to perform repeated calculations of the performance functions. Each such calculation often requires calls to be made to expensive computational codes. Research in the field of surrogate modelling aims to make this aspect of computation manageable. The emergence of machine learning-based tools in recent years has spurred research activities in this field.

A broader way of viewing these methods has recently been proposed by Rashki and Faes [2023] through the lens of the no-free-lunch (NFL) theorem as applied to reliability estimation. The authors argue that no reliability estimation algorithm can be expected to perform uniformly well across all classes of problems. By viewing reliability estimation as a search for important failure regions or neighbourhoods of design points and invoking the original NFL theorem for search and optimization developed by Wolpert and Macready [1995, 1997], they emphasize that algorithmic performance is problem-dependent and that analyst bias in method selection can affect the quality of failure probability estimates. While this provides a useful conceptual guideline, the original NFL theorem of Wolpert and Macready [1995, 1997] was established for finite countable state spaces, and as discussed by Auger and Teytaud [2010], extension of NFL to continuous state spaces is nontrivial. Thus, while NFL arguments caution against universal claims of algorithmic superiority, there is scope for a mathematically rigorous reliability-specific NFL result in standard normal or general parameter spaces.

The preceding sections have aimed to critically review the advancements in the field of structural reliability modelling. We conclude the review by offering the following suggestions for future research.

1. In studies on randomly excited dynamical systems with random parameters, the problem of reliability estimation can be tackled by combining importance sampling (to deal with system parameter randomness) with Girsanov control-based strategies (to deal with randomness in excitations). A sequential double-loop approach in which the probability of failure conditioned on system parameters is first estimated which followed by a quadrature over the pdf of the system parameters, can be easily conceived. However, the determinisation of a coupled importance sampling pdf and Girsanov control could lead to a reliability estimation procedure that involves a single computational loop.
2. Deterministic nonlinear dynamical systems are known to undergo bifurcations which could potentially lead to response amplifications. The occurrence of phenomena such as jumps, entrainment, sub- and super-harmonic resonances, and transition to chaos are well studied. The influence of random perturbations, either in excitations or in system parameters, can lead to challenging questions on estimation of reliability against failures caused by possible bifurcations. Thus, designing workable importance sampling pdfs and Girsanov controls for this class of problems can offer significant challenges and lead to important advances from a methodological perspective.

3. The development of surrogate models in the context of reliability modelling is heavily focused on static systems and time-invariant reliability estimation. A promising avenue for future research would be to explore machine learning-based surrogate meta-modelling for large computational models coupled with appropriate active learning strategies for nonlinear dynamical systems. Tools based on autoencoders and physics-guided machine learning seem to offer viable means for surrogate modelling. However, the development of suitable learning functions for active learning is much less obvious.
4. Section 7.0 has summarized experimental protocols that embed importance sampling strategies for estimating reliability of complex systems. Two promising directions to extend these studies are as follows: (a) the proposed procedure can be embedded into hybrid simulation-based methods (such as real time substruction) to perform reliability qualification tests on critical subsystems which interact with relatively large sized super structures, and (b) one could consider randomness not only in excitations but also in system parameters; here importance sampling based methods can be combined with additive manufacturing techniques to produce an intelligently designed small-sized sample of structural systems which could be tested with manageable effort.
5. Much of the development in modern machine learning is driven by tools which employ deep neural networks. The main strength of these networks is that they enable automated learning of feature vectors in complex situations in which the input data is multimodal (e.g., a combination of audio, video, and text). The direct applicability of these developments seems to be of limited when it comes to problems of reliability estimation. Here, both the inputs are well structured and so are the outputs. One could still employ ANNs for tasks such as surrogate model development. However, the alternatives such as GPR and SVM models seem to be much better suited for this purpose. It is thus interesting to inquire into areas of reliability modelling which can meaningfully exploit the power of developments in the field of deep learning. Two directions seems promising: (a) development of reduced order models and surrogate models based on application of autoencoders in the context of treatment of large-sized nonlinear dynamical systems, and (b) the application of tools of generative AI (such as diffusion process models and generative adversarial networks) to simulate samples of loads such as those due to earthquake where the available data is sparse and arise from complex nonlinear processes.

6. Numerical solutions of stochastic differential equations form a central part of computing reliability for randomly excited nonlinear systems. The field of numerical analysis of stochastic differential equations is well developed, and there exist a host of alternatives to tackle the problems (see, for example, the seminal monograph by Kloeden and Platen [1992]). The area of structural reliability modelling has not fully exploited these available tools and techniques. For instance, many structural dynamical systems are known to possess numerical stiff properties, and special considerations are needed to deal with the solutions of the governing equations of such systems. The recent study by Nisha and Manohar [2022] has shown how implicit schemes for solving stiff stochastic differential equations can lead to notable computational advantages in the context of analysis of Bayesian state space models. Similar efforts, in the context of time-variant reliability modelling (e.g., development of Girsanov controls), could lead to more efficient reliability estimation schemes.

**Acknowledgements**

Dr Oindrila Kanjilal's work here has been partially funded by the program France 2030 ANR-23-IACL-0006 through the MIAI-BALTEEC project.

**References**

1. Chojaczyk AA, Teixeira AP, Neves LC, Cardoso JB, Soares CG (2015) Review and application of artificial neural networks models in reliability analysis of steel structures. Struct Saf 52:78-89.
2. Guo H, Luo C, Zhu SP, You X, Yan M, Liu X (2025) Machine learning-based enhanced Monte Carlo simulation for low failure probability structural reliability analysis. Structures 74:108530.
3. Basudhar A, Missoum S, Sanchez AH (2008) Limit state function identification using support vector machines for discontinuous responses and disjoint failure domains. Probab Eng Mech 23(1):1-11.
4. Der Kiureghian A (2022) Structural and system reliability. Cambridge University Press, Cambridge.
5. Haldar A (ed) (2006) Recent advances in reliability-based civil engineering. World Scientific, Singapore.

6. El Haj AK, Soubra AH (2021) Improved active learning probabilistic approach for the computation of failure probability. Struct Saf 88:102011.
7. Naess A, Gaidai O (2009) Estimation of extreme values from sampled time series. Struct Saf 31(4):325-334.
8. Roy A, Manna R, Chakraborty S (2019) Support vector regression based metamodeling for structural reliability analysis. Probab Eng Mech 55:78-89.
9. Sharma A, Manohar CS (2026) Determination of state-dependent Girsanov's control via solution of a stochastic optimal control problem using reinforcement learning. Struct Saf:102714.
10. Auger A, Teytaud O (2010) Continuous lunches are free plus the design of optimal optimization algorithms. Algorithmica 57(1):121-146.
11. Sen D, Bhattacharya B (2015) On the Pareto optimality of variance reduction simulation techniques in structural reliability. Struct Saf 53:57-74.
12. Sharma A, Chakroborty P, Shields MD (2026) SuSIE: Subset Simulation with Intrepid Exploration. arXiv preprint arXiv:2607.22989.
13. Chakroborty P, Shields MD (2026) Intrepid MCMC: Metropolis-Hastings with exploration. Comput Methods Appl Mech 448:118402.
14. Abdollahi A, Azhdary Moghaddam M, Hashemi Monfared SA, Rashki M, Li Y (2021) Subset simulation method including fitness-based seed selection for reliability analysis. Eng Comput 37:2689-2705.
15. Abdollahi A, Moghaddam MA, Monfared SA, Rashki M, Li Y (2020) A refined subset simulation for the reliability analysis using the subset control variate. Struct Saf 87:102002.
16. Afshari SS, Enayatollahi F, Xu X, Liang X (2022) Machine learning-based methods in structural reliability analysis: A review. Reliab Eng Syst Saf 219:108223.
17. Aliahmad M, Miri M, Rashki M (2026) Sequential space conversion method with multi-armed bandit algorithm for time-variant reliability and dynamic project management. Reliab Eng Syst Saf 265:111540.
18. Ameryan A, Ghalehnovi M, Rashki M (2022) AK-SESC: a novel reliability procedure based on the integration of active learning kriging and sequential space conversion method. Reliab Eng Syst Saf 217:108036.
19. Ang AHS, Tang WH (2007) Probability concepts in engineering: Emphasis on applications to civil & environmental engineering. John Wiley & Sons, New York.

20. Ang GL, Ang AHS, Tang WH (1992) Optimal importance-sampling density estimator. J Eng Mech 118(6):1146-1163.
21. Arunachalam S, Spence SM (2023) An efficient stratified sampling scheme for the simultaneous estimation of small failure probabilities in wind engineering applications. Struct Saf 101:102310.
22. Au SK (2006a) Critical excitation of SDOF elasto-plastic systems. J Sound Vib 296:714-733.
23. Au SK (2006b) Sub-critical excitations of SDOF elasto-plastic systems. Int J Non-Linear Mech 41:1095-1108.
24. Au SK (2008) First passage probability of elasto-plastic systems by importance sampling with adapted processes. Probab Eng Mech 23(2-3):114-124.
25. Au SK (2009a) Importance sampling for elasto-plastic systems using adapted processes with deterministic control. Int J Non-Linear Mech 44:190-199.
26. Au SK (2009b) Stochastic control approach to reliability of elasto-plastic structures. Struct Eng Mech 32(1):21-36.
27. Au SK (2026) Optimality conditions for MCMC in rare event risk analysis. Reliab Eng Syst Saf 265:111539.
28. Au SK, Beck JL (1999) A new adaptive importance sampling scheme for reliability calculations. Struct Saf 21:135-158.
29. Au SK, Beck JL (2001a) Estimation of small failure probabilities in high dimensions by subset simulation. Probab Eng Mech 16:263-277.
30. Au SK, Beck JL (2001b) First excursion probabilities for linear systems by very efficient importance sampling. Probab Eng Mech 16:193-207.
31. Au SK, Beck JL (2003a) Subset simulation and its applications to seismic risk based on dynamic analysis. J Eng Mech 129(8):901-917.
32. Au SK, Beck JL (2003b) Important sampling in high dimensions. Struct Saf 25:139-163.
33. Au SK, Papadimitriou C, Beck JL (1999) Reliability of uncertain dynamical systems with multiple design points. Struct Saf 21:113-133.
34. Au SK, Wang Y (2014) Engineering risk assessment with subset simulation. John Wiley & Sons, Singapore.
35. Au SK, Zhou X (2025) Adaptive proposal length scale in Subset Simulation. Reliab Eng Syst Saf 261:111069.

36. Echard B, Gayton N, Lemaire M (2011) AK-MCS: an active learning reliability method combining Kriging and Monte Carlo simulation. Struct Saf 33(2):145-154.
37. Echard B, Gayton N, Lemaire M, Relun N (2013) A combined importance sampling and kriging reliability method for small failure probabilities with time-demanding numerical models. Reliab Eng Syst Saf 111:232-240.
38. Bichon BJ, Eldred MS, Swiler LP, Mahadevan S, McFarland JM (2008) Efficient global reliability analysis for nonlinear implicit performance functions. AIAA J 46(10):2459-2468.
39. Lusch B, Kutz JN, Brunton SL (2018) Deep learning for universal linear embeddings of nonlinear dynamics. Nat Commun 9(1):4950.
40. Richard B, Cremona C, Adelaide L (2012) A response surface method based on support vector machines trained with an adaptive experimental design. Struct Saf 39:14-21.
41. Baber TT, Noori MN (1985) Random vibration of degrading, pinching systems. J Eng Mech 111(8):1010-1026.
42. Bansal S, Cheung SH (2017) On the evaluation of multiple failure probability curves in reliability analysis with multiple performance functions. Reliab Eng Syst Saf 167:583-594.
43. Bect J, Ling L, Vazquez E (2017) Bayesian subset simulation. SIAM/ASA J Uncertainty Quantif 5(1):762-786.
44. Bjerager P (1988) Probability integration by directional simulation. J Eng Mech 114(8):1285-1302.
45. Breitung K (1984) Asymptotic approximation for multinormal integrals. J Eng Mech 110(3):357-366.
46. Breitung K (2015) 40 years FORM: Some new aspects? Probab Eng Mech 42:71-77.
47. Breitung K (2019) The geometry of limit state function graphs and subset simulation: Counterexamples. Reliab Eng Syst Saf 182:98-106.
48. Breitung K (2021) SORM, design points, subset simulation, and Markov chain Monte Carlo. ASCE-ASME J Risk Uncertain Eng Syst Part A Civ Eng 7(4):04021052.
49. Breitung K (2024) The return of the design points. Reliab Eng Syst Saf 247:110103.
50. Bucher CG (1988) Adaptive sampling—an iterative fast Monte Carlo procedure. Struct Saf 5(2):119-126.
51. Bucher CG, Most T (2008) A comparison of approximate response functions in structural reliability analysis. Probab Eng Mech 23(2-3):154-163.

52. Dai H, Zhang H, Wang W (2012) A support vector density-based importance sampling for reliability assessment. Reliab Eng Syst Saf 106:86-93.
53. Dai H, Zhang H, Wang W (2016) A new maximum entropy-based importance sampling for reliability analysis. Struct Saf 63:71-80.
54. Dai H, Zhang H, Rasmussen KJ, Wang W (2015) Wavelet density-based adaptive importance sampling method. Struct Saf 52:161-169.
55. Dang C, Zhou T, Valdebenito MA, Faes MG (2025) Yet another Bayesian active learning reliability analysis method. Struct Saf 112:102539.
56. Bucher CG, Bourgund U (1990) A fast and efficient response surface approach for structural reliability problems. Struct Saf 7(1):57-66.
57. Luo C, Keshtegar B, Zhu SP, Taylan O, Niu XP (2022) Hybrid enhanced Monte Carlo simulation coupled with advanced machine learning approach for accurate and efficient structural reliability analysis. Comput Methods Appl Mech 388:114218.
58. Bishop CM, Bishop H (2024) Deep learning foundations and concepts. Springer Nature, Switzerland.
59. Tong C, Sun Z, Zhao Q, Wang Q, Wang S (2015) A hybrid algorithm for reliability analysis combining Kriging and subset simulation importance sampling. J Mech Sci Technol 29(8):3183-3193.
60. Wang C (2021) Structural reliability and time-dependent reliability. Springer, Cham, Switzerland.
61. Xu C, Chen W, Ma J, Shi Y, Lu S (2020) AK-MSS: An adaptation of the AK-MCS method for small failure probabilities. Struct Saf 86:101971.
62. Yin C, Kareem A (2016) Computation of failure probability via hierarchical clustering. Struct Saf 61:67-77.
63. Cai GQ, Elishakoff I (1994) Refined second-order reliability analysis. Struct Saf 14:267-276.
64. Cerou F, Guyader A, Rousset M (2019) Adaptive multilevel splitting: Historical perspective and recent results. Chaos 29(4):043108.
65. Chan J, Papaioannou I, Straub D (2022) An adaptive subset simulation algorithm for system reliability analysis with discontinuous limit states. Reliab Eng Syst Saf 225:108607.
66. Chan J, Paredes R, Papaioannou I, Duenas-Osorio L, Straub D (2024) Adaptive Monte Carlo methods for estimating rare events in power grids. ASCE-ASME J Risk Uncertain Eng Syst Part A Civ Eng 11(1):04024082.

67. Chen JB, Li J (2005) Dynamic response and reliability analysis of non-linear stochastic structures. Probab Eng Mech 20:33-44.
68. Chen JB, Lyu MZ (2022) Globally-evolving-based generalized density evolution equation for nonlinear systems involving randomness from both system parameters and excitations. Proc R Soc A 478(2264):20220356.
69. Chen W, Wang Z, Broccardo M, Song J (2022) Riemannian Manifold Hamiltonian Monte Carlo based subset simulation for reliability analysis in non-Gaussian space. Struct Saf 94:102134.
70. Cheng K, Lu Z, Xiao S, Lei J (2022) Estimation of small failure probability using generalized subset simulation. Mech Syst Signal Process 163:108114.
71. Cheng K, Papaioannou I, Lu Z, Zhang X, Wang Y (2023) Rare event estimation with sequential directional importance sampling. Struct Saf 100:102291.
72. Ching J, Au SK, Beck JL (2005a) Reliability estimation for dynamical systems subject to stochastic excitation using subset simulation with splitting. Comput Methods Appl Mech Eng 194:1557-1579.
73. Ching J, Beck JL, Au SK (2005b) Hybrid subset simulation method for reliability estimation of dynamical systems subject to stochastic excitation. Probab Eng Mech 20:199-214.
74. Chreng RH, Wen YK (1994) Reliability of uncertain non-linear trusses under random excitation- II. J Eng Mech 120(4):748-757.
75. Cotter SL, Roberts GO, Stuart AM, White D (2013) MCMC methods for functions: modifying old algorithms to make them faster. Stat Sci 28(3):424-446.
76. de Angelis M, Patelli E, Beer M (2015) Advanced line sampling for efficient robust reliability analysis. Struct Saf 52:170-182.
77. Der Kiureghian A (2000) The geometry of random vibrations and solutions by FORM and SORM. Probab Eng Mech 15:81-90.
78. Der Kiureghian A, Lin HZ, Hwang SJ (1987) Second-order reliability approximations. J Eng Mech 113(8):1208-1225.
79. Der Kiureghian A, Stefano MD (1991) Efficient algorithm for second-order reliability analysis. J Eng Mech 117(12):2904-2923.
80. Ditlevsen O, Bjerager P (1989) Plastic reliability analysis by directional simulation. J Eng Mech 115(6):1347-1362.
81. Ditlevsen O, Bjerager P, Olesen R, Hasofer AM (1988) Directional simulation in Gaussian processes. Probab Eng Mech 3(4):207-217.

82. Drenick RF (1970) Model-free design of aseismic structures. J Eng Mech 96:483-493.
83. Drenick RF (1977) The Critical Excitation of Nonlinear Systems. J Appl Mech 44(2):333-336.
84. Du W, Luo Y, Wang Y (2019) Time-variant reliability analysis using the parallel subset simulation. Reliab Eng Syst Saf 182:250-257.
85. Dubourg V, Sudret B, Deheeger F (2013) Metamodel-based importance sampling for structural reliability analysis. Probab Eng Mech 33:47-57.
86. Alpaydin E (2020) Introduction to machine learning, 4th edn. MIT Press, Cambridge, Massachusetts.
87. Yeung E, Kundu S, Hodas N (2019) Learning deep neural network representations for Koopman operators of nonlinear dynamical systems. In: 2019 American Control Conference (ACC). IEEE, pp 4832-4839.
88. Ellingwood B, Maes M, Bartlett FM, Beck AT, Caprani C, Der Kiureghian A, Duenas-Osorio L, Galvao N, Gilbert R, Li J, Matos J, Mori Y, Papaioannou I, Parades R, Straub D, Sudret B (2025) Development of methods of structural reliability. Struct Saf 113:102474.
89. Engelund S, Rackwitz R (1993) A benchmark study on importance sampling techniques in structural reliability. Struct Saf 12(4):255-276.
90. Cadini F, Santos F, Zio E (2014) An improved adaptive kriging-based importance technique for sampling multiple failure regions of low probability. Reliab Eng Syst Saf 131:109-117.
91. Cui F, Ghosn M (2019) Implementation of machine learning techniques into the subset simulation method. Struct Saf 79:12-25.
92. Feng Z, Zhenzhou L, Lijie C, Shufang S (2010) Reliability sensitivity algorithm based on stratified importance sampling method for multiple failure modes systems. Chin J Aeronaut 23(6):660-669.
93. Ferrante FJ, Arwade SR, Graham-Brady LL (2005) A translation model for non-stationary, non-Gaussian random processes. Probab Eng Mech 20(3):215-228.
94. Firouzi A, Yang W, Li CQ (2018) Efficient solution for calculation of up-crossing rate of non-stationary Gaussian process. J Eng Mech 144(4):04018015.
95. Fujita M, Rackwitz R (1988) Updating first-and second-order reliability estimates by importance sampling. Doboku Gakkai Ronbunshu 392:53-59.

96. Marano GC, Rosso MM, Aloisio A, Cirrincione G (2024) Generative adversarial networks review in earthquake-related engineering fields. Bull Earthq Eng 22(7):3511-3562.
97. Karniadakis GE, Kevrekidis IG, Lu L, Perdikaris P, Wang S, Yang L (2021) Physics-informed machine learning. Nat Rev Phys 3(6):422-440.
98. Strang G (2019) Linear algebra and learning from data. Wellesley-Cambridge Press, Massachusetts.
99. Gardoni P (2017) Risk and reliability analysis. In: Gardoni P (ed) Risk and reliability analysis: theory and applications: in honor of Prof. Armen Der Kiureghian. Springer International Publishing, Cham, pp 3–24.
100. Ghanem RG, Spanos PD (2003) Stochastic finite elements: A spectral approach. Dover Publications, Mineola, New York.
101. Giovanis DG, Shields MD (2022) Imprecise subset simulation. Probab Eng Mech 69:103293.
102. Girsanov IV (1960) On transforming a certain class of stochastic processes by absolutely continuous substitution of measure. Theory Probab Appl 5:285-301.
103. Glasserman P, Heidelberger P, Shahabuddin P, Zajic T (1999) Multilevel splitting for estimating rare event probabilities. Oper Res 47(4):585-600.
104. Grigoriu M (1984) Crossing of non-Gaussian translation processes. J Eng Mech 110(4):610-620.
105. Grigoriu M (2002) Stochastic calculus: Applications in science and engineering. Birkhauser, Boston.
106. Grooteman F (2008) Adaptive radial-based importance sampling method for structural reliability. Struct Saf 30:533-542.
107. Grooteman F (2011) An adaptive directional importance sampling method for structural reliability. Probab Eng Mech 26:134-141.
108. Gupta S, Manohar CS (2006) Reliability analysis of randomly vibrating structures with parameter uncertainties. J Sound Vib 297:1000-1024.
109. Guimaraes H, Matos JC, Henriques AA (2018) An innovative adaptive sparse response surface method for structural reliability analysis. Struct Saf 73:12-28.
110. Guo H, Zhang J, Dong Y, Frangopol DM (2024) Probability-informed neural network-driven point-evolution kernel density estimation for time-dependent reliability analysis. Reliab Eng Syst Saf 249:110234.

111. Zhan H, Xiao NC, Ji Y (2022) An adaptive parallel learning dependent Kriging model for small failure probability problems. Reliab Eng Syst Saf 222:108403.

112. Harbitz A (1986) An efficient sampling method for probability of failure calculation. Struct Saf 3:109-115.

113. Hasofer AM, Lind NC (1974) Exact and invariant second moment code format. J Eng Mech Div 100:111-121.

114. Haukaas T, Der Kiureghian A (2006) Strategies for finding the design point in non-linear finite element reliability analysis. Probab Eng Mech 21:133-147.

115. Helal S, Elvira V (2026) Efficient rare event estimation for multimodal and high-dimensional system reliability via subset adaptive importance sampling. Reliab Eng Syst Saf 272:112473.

116. Hohenbichler M, Rackwitz R (1988) Improvement of second-order reliability estimates by importance sampling. J Eng Mech 114(12):2195-2199.

117. Hristov PO, DiazDelaO FA (2023) Subset simulation for probabilistic computer models. Appl Math Model 120:769-785.

118. Hsu WC, Ching J (2010) Evaluating small failure probabilities of multiple limit states by parallel subset simulation. Probab Eng Mech 25:291-304.

119. Hu Z, Du X (2018) Saddle point approximation reliability method for quadratic functions in normal variables. Struct Saf 71:24-32.

120. Goodfellow I, Bengio Y, Courville A (2016) Deep learning. MIT Press, Cambridge, Massachusetts.

121. Kaymaz I (2005) Application of kriging method to structural reliability problems. Struct Saf 27(2):133-151.

122. Kaymaz I, McMahon CA (2005) A response surface method based on weighted regression for structural reliability analysis. Probab Eng Mech 20(1):11-17.

123. Hurtado JE, Alvarez DA (2001) Neural-network-based reliability analysis: a comparative study. Comput Methods Appl Mech 191(1-2):113-132.

124. Hurtado JE, Alvarez DA (2010) An optimization method for learning statistical classifiers in structural reliability. Probab Eng Mech 25(1):26-34.

125. Hurtado JE (2001) Neural networks in stochastic mechanics. Arch Comput Methods Eng 8(3):303-342.

126. Hurtado JE (2004) Structural reliability: statistical learning perspectives. Springer-Verlag, Berlin.

127. Hurtado JE (2007) Filtered importance sampling with support vector margin: a powerful method for structural reliability analysis. Struct Saf 29(1):2-15.
128. Hurtado JE, Alvarez DA (2000) Reliability assessment of structural systems using neural networks. In: Proceedings of the European Congress on Computational Methods in Applied Sciences and Engineering, ECCOMAS. Barcelona, 11–14 September 2000.
129. Jafari-Asl J, Dong Y, Guo H (2025) Enhancing structural reliability through AI-Driven control variates and subset simulation. ASCE-ASME J Risk Uncertain Eng Syst Part A Civ Eng 11(4):04025056.
130. Li J, Xiu D (2010) Evaluation of failure probability via surrogate models. J Comput Phys 229(23):8966-8980.
131. Li J, Li J, Xiu D (2011) An efficient surrogate-based method for computing rare failure probability. J Comput Phys 230(24):8683-8697.
132. Bourinet JM (2016) Rare-event probability estimation with adaptive support vector regression surrogates. Reliab Eng Syst Saf 150:210-221.
133. Bourinet JM, Deheeger F, Lemaire M (2011) Assessing small failure probabilities by combined subset simulation and support vector machines. Struct Saf 33(6):343-353.
134. Wang J, Li C, Xu G, Li Y, Kareem A (2021) Efficient structural reliability analysis based on adaptive Bayesian support vector regression. Comput Methods Appl Mech 387:114172.
135. Wang J, Xu G, Mitoulis SA, Townsend JF, Li C, Kareem A (2026) Adaptive Bayesian support vector regression with advanced simulation and dimension reduction for efficient reliability analysis. Comput Methods Appl Mech 450:118606.
136. Wang J, Xu G, Li Y, Kareem A (2022) AKSE: A novel adaptive Kriging method combining sampling region scheme and error-based stopping criterion for structural reliability analysis. Reliab Eng Syst Saf 219:108214.
137. Wang J, Cao Z, Xu G, Yang J, Kareem A (2023) An adaptive Kriging method based on K-means clustering and sampling in n-ball for structural reliability analysis. Eng Comput 40(2):378-410.
138. Wang Z, Song J (2016) Cross-entropy-based adaptive importance sampling using von Mises-Fisher mixture for high dimensional reliability analysis. Struct Saf 59:42-52.

139. Yin J, Du X (2022) Active learning with generalized sliced inverse regression for high-dimensional reliability analysis. Struct Saf 94:102151.
140. Jafari-Asl J, Seghier ME, Ohadi S, Correia J, Barroso J (2022) Reliability analysis based improved directional simulation using Harris Hawks optimization algorithm for engineering systems. Eng Fail Anal 135:106148.
141. Jensen HA, Valdebenito MA (2007) Reliability analysis of linear dynamical systems using approximate representations of performance functions. Struct Saf 29:222-237.
142. Jia G, Tabandeh A, Gardoni P (2021) A density extrapolation approach to estimate failure probabilities. Struct Saf 93:102128.
143. Jia G, Taflanidis AA, Beck JL (2017) A new adaptive rejection sampling method using kernel density approximations and its application to subset simulation. ASCE-ASME J Risk Uncertain Eng Syst Part A Civ Eng 3(2):D4015001.
144. Chen K, Pan H, Zhang M, Li ZH (2025) Ground motion simulation via generative adversarial network. Appl Geophys 22(3):684-697.
145. Cheng K, Lu Z (2021) Adaptive Bayesian support vector regression model for structural reliability analysis. Reliab Eng Syst Saf 206:107286.
146. Cheng K, Papaioannou I, Straub D (2025) Enhanced sequential directional importance sampling for structural reliability analysis. Struct Saf 114:102574.
147. Murphy KP (2012) Machine learning: a probabilistic perspective. MIT Press, Cambridge, Massachusetts.
148. Murphy KP (2022) Probabilistic machine learning: an introduction. MIT Press, Cambridge, Massachusetts.
149. Papakonstantinou KG, Nikbakht H, Eshra E (2023) Hamiltonian MCMC methods for estimating rare events probabilities in high-dimensional problems. Probab Eng Mech 74:103485.
150. Eshra E, Papakonstantinou KG, Nikbakht H (2025) A direct importance sampling-based framework for rare event uncertainty quantification in non-Gaussian spaces. Reliab Eng Syst Saf 264:111200.
151. Eshra E, Papakonstantinou KG (2025) Gradient-free importance sampling scheme for efficient reliability estimation. J Eng Mech 151(12):04025076.
152. Murphy KP (2023) Probabilistic machine learning: advanced topics. MIT Press, Cambridge, Massachusetts.

153. Yuen KV (2010) Bayesian methods for structural dynamics and civil engineering. John Wiley, Chichester.

154. Kahn H, Harris TE (1951) Estimation of particle transmission by random sampling. Natl Bur Stand Appl Math Ser 12:27-30.

155. Kanjilal O, Manohar CS (2015) Markov chain splitting methods in structural reliability integral estimation. Probab Eng Mech 40:42-51.

156. Kanjilal O, Manohar CS (2017) Girsanov's transformation-based variance-reduced Monte Carlo simulation schemes for reliability estimation in nonlinear stochastic dynamics. J Comput Phys 341:278-294.

157. Kanjilal O, Manohar CS (2018) State-dependent Girsanov's controls in time variant reliability estimation in randomly excited dynamical systems. Struct Saf 72:30-40.

158. Kanjilal O, Manohar CS (2019) Estimation of time-variant system reliability of nonlinear randomly excited systems based on the Girsanov transformation with state-dependent controls. Nonlinear Dyn 95(2):1693-1711.

159. Kanjilal O, Papaioannou I, Straub D (2021) Cross entropy-based importance sampling for first-passage probability estimation of randomly excited linear structures with parameter uncertainty. Struct Saf 91:102090.

160. Kanjilal O, Papaioannou I, Straub D (2022) Series system reliability of uncertain linear structures under Gaussian excitation by cross entropy-based importance sampling. J Eng Mech 148(1):04021136.

161. Kanjilal O, Tatarevic A, Valdebenito M, Papaioannou I, Faes M, Straub D (2026) Revisiting the cross-entropy method for first-passage probability of Gaussian-process excited uncertain linear structural dynamic systems. Reliab Eng Syst Saf 272(1):112463.

162. Katafygiotis LS, Cheung SH (2007) Application of spherical subset simulation method and auxiliary domain method on a benchmark reliability study. Struct Saf 29:194-207.

163. Katafygiotis LS, Zuev KM (2008) Geometric insight into the challenges of solving high-dimensional reliability problems. Probab Eng Mech 23(2-3):208-218.

164. Keshtegar B, Meng Z (2017) A hybrid relaxed first-order reliability method for efficient structural reliability analysis. Struct Saf 66:84-93.

165. Kinnear HJ, DiazDelaO FA (2025) Niching subset simulation. Probab Eng Mech 79:103729.

166. Kinnear HJ, DiazDelaO FA (2026) Niching Importance Sampling for Multi-modal Rare-event Simulation. arXiv preprint arXiv:2604.06417.
167. Koo H, Der Kiureghian A, Fujimura K (2005) Design-point excitation for non-linear random vibrations. Probab Eng Mech 20:136-147.
168. Koutsourelakis PS (2004) Reliability of structures in higher dimensions, part II; Theoretical validation. Probab Eng Mech 19:419-423.
169. Koutsourelakis PS, Pradlwarter HJ, Schueller GI (2004) Reliability of structures in high dimensions, part I: algorithms and applications. Probab Eng Mech 19:409-417.
170. Kroese DP, Taimre T, Botev ZI (2011) Handbook of Monte Carlo methods. Wiley, New Jersey.
171. Kurtz N, Song J (2013) Cross-entropy-based adaptive importance sampling using Gaussian mixture. Struct Saf 42:35-44.
172. Lu L, Jin P, Pang G, Zhang Z, Karniadakis GE (2021) Learning nonlinear operators via DeepONet based on the universal approximation theorem of operators. Nat Mach Intell 3(3):218-229.
173. Lee I, Noh Y, Yoo D (2012) A novel second-order reliability method (SORM) using non-central or generalized chi-squared distributions. J Mech Des 134:100912.
174. Lei J, Ling C, Xie M (2025) Variational Bayesian Monte Carlo simulation for rare event assessment. Proc Inst Mech Eng Pt O J Risk Reliab 239(5):1206-1220.
175. Lemaire M, Chateauneuf A, Mitteau JC (2013) Structural reliability. Wiley-ISTE, London.
176. Li B, Xia W, Liao Z (2025) Relaxed subset simulation for reliability estimation. Reliab Eng Syst Saf 264:111302.
177. Li CQ, Firouzi A, Yang W (2016) Closed-form solution to first passage probability for non-stationary lognormal processes. J Eng Mech 142(12):04016103.
178. Li CQ, Melchers RE (1993) Outcrossings from convex polyhedrons for nonstationary Gaussian processes. J Eng Mech 119(11):2354-2361.
179. Li HS, Ma YZ, Cao Z (2015) A generalized Subset Simulation approach for estimating small failure probabilities of multiple stochastic responses. Comput Struct 153:239-251.
180. Li HS, Wang T, Yuan JY, Zhang H (2019) A sampling-based method for high-dimensional time-variant reliability analysis. Mech Syst Signal Process 126:505-520.
181. Li J, Chen JB (2009) Stochastic Dynamics of structures, John Wiley & Sons.

182. Liao Z, He X, Xia W (2024a) Enhancing subset simulation through Bayesian inference. Comput Methods Appl Mech Eng 432:117392.

183. Liao Z, Xia W, He X (2024b) An investigation into Markov chain Monte Carlo algorithms for Subset simulation: Emphasizing uncertainty analysis. Comput Struct 294:107268.

184. Lin YK, Cai GQ (1995) Probabilistic structural dynamics: advanced theory and applications. McGraw-Hill, Singapore.

185. Lu Z, Song S, Yue Z, Wang J (2008) Reliability sensitivity method by line sampling. Struct Saf 30(6):517-532.

186. Lyu MZ, Chen JB (2022) A unified formalism of the GE-GDEE for generic continuous responses and first-passage reliability analysis of multi-dimensional nonlinear systems subjected to non-white noise excitations. Struct Saf 98:102233.

187. Faber MH (ed) (2008) Risk assessment in engineering: principles, system representation and risk criteria. Joint Committee on Structural Safety, JCSS.

188. Macke M, Bucher CG (2003) Importance sampling for randomly excited dynamical systems. J Sound Vib 268(2):269-290.

189. Masoudifar M, Mahsuli M, Taciroglu E (2025) Deep learning-based stochastic ground motion modeling using generative adversarial and convolutional neural networks. Soil Dyn Earthq Eng 194:109306.

190. Matinfar M, Khaji N, Ahmadi G (2023) Deep convolutional generative adversarial networks for the generation of numerous artificial spectrum-compatible earthquake accelerograms using a limited number of ground motion records. Comput-Aided Civ Infrastruct Eng 38(2):225-240.

191. Papadrakakis M, Papadopoulos V, Lagaros ND (1996) Structural reliability analysis of elastic-plastic structures using neural networks and Monte Carlo simulation. Comput Methods Appl Mech 136(1-2):145-163.

192. Madsen HO, Krenk S, Lind NC (2006) Methods of structural safety. Dover Publications, New York.

193. Madsen HO, Tvedt L (1990) Methods for time-dependent reliability and sensitivity measures. J Eng Mech 116(10):2118-2135.

194. Manohar CS, Gupta S (2005) Modeling and evaluation of structural reliability: Current status and future directions. Research reviews in structural engineering, Golden Jubilee Publications of Department of Civil Engineering, Indian Institute of Science, Bangalore.

195. Mansour R, Olsson M (2014) A closed-form second-order reliability method using non-central chi-squared distributions. J Mech Des 136:101402.
196. Melchers RE (1989) Importance sampling in structural systems. Struct Saf 6(1):3-10.
197. Melnik-Melnikov PG, Dekhtyaruk ES (2000) Rare event estimation by "Russian Roulette and Splitting" simulation technique. Probab Eng Mech 15:125-129.
198. Meng Z, Li C, Pang Y, Li G, He W (2023) New bubble sampling method for reliability analysis. Struct Multidiscip Optim 66(8):180.
199. Miao F, Ghosn M (2011) Modified subset simulation method for reliability analysis of structural systems. Struct Saf 33:251-260.
200. Miao Q, Low YM (2025) Improved variance estimation for subset simulation by accounting for the correlation between Markov chains. Struct Saf 116:102606.
201. Milstein GN (1995) Numerical integration of stochastic differential equations. Springer, Netherlands.
202. Mostaghel N, Byrd RA (2000) Analytical description of multidegree bilinear hysteretic system. J Eng Mech 126(6):588-598.
203. Moustapha M, Marelli S, Sudret B (2022) Active learning for structural reliability: Survey, general framework and benchmark. Struct Saf 96:102174.
204. Lelievre N, Beaurepaire P, Mattrand C, Gayton N (2018) AK-MCSi: A Kriging-based method to deal with small failure probabilities and time-consuming models. Struct Saf 73:1-11.
205. Navaneeth N, Chakraborty S (2022) Koopman operator for time-dependent reliability analysis. Probab Eng Mech 70:103372.
206. Naderi L, Jia G (2026) Multi-fidelity Subset Simulation for rare event simulation. Reliab Eng Syst Saf 266:111739.
207. Nayek R, Manohar CS (2015) Girsanov transformation-based reliability modelling and testing of actively controlled structures. J Eng Mech 141(6):04014168.
208. Nelsen RB (2006) An introduction to copulas. Springer, New York.
209. Nie J, Ellingwood BR (2000) Directional methods for structural reliability analysis. Struct Saf 22:233-249.
210. Nie J, Ellingwood BR (2004a) A new directional simulation method for system reliability. Part I: application of deterministic point sets. Probab Eng Mech 19:425-436.
211. Nie J, Ellingwood BR (2004b) A new directional simulation method for system reliability. Part II: application of neural networks. Probab Eng Mech 19:437-447.

212. Nikolaidis E, Saleem S, Farizal, Zhang G, Mourelatos Z (2009) Probabilistic reanalysis using Monte Carlo simulation. SAE Int J Mater Manuf 1(1):22-35.
213. Oksendal B (2003) Stochastic differential equations: An introduction and applications. Springer, New York.
214. Olsen AI, Naess A (2006) Estimation of failure probabilities of linear dynamic systems by importance sampling. Sadhana 31(4):429-443.
215. Olsen AI, Naess A (2007) An importance sampling procedure for estimating failure probabilities of non-linear dynamic systems subjected to random noise. Int J Non-Linear Mech 42:848-863.
216. Ren P, Nakata R, Lacour M, Naiman I, Nakata N, Song J, Bi JZ, Malik OA, Morozov D, Azencot O, Erichson NB, Mahoney MW (2026) Learning earthquake ground motions via conditional generative modelling. Nat Commun 17(1):4021.
217. Papadimitriou C, Beck JL, Katafygiotis LS (1997) Asymptotic expansions for reliability and moments of uncertain systems. J Eng Mech 123(12):1219-1229.
218. Papaioannou I, Betz W, Zwirglmaier K, Straub D (2015) MCMC algorithms for subset simulation. Probab Eng Mech 41:89-103.
219. Dempster AP, Laird NM, Rubin DB (1977) Maximum likelihood from incomplete data via the EM algorithm. J R Stat Soc B Stat Methodol 39(1):1-22.
220. Geyer S, Papaioannou I, Straub D (2019) Cross entropy-based importance sampling using Gaussian densities revisited. Struct Saf 76:15-27.
221. Papaioannou I, Geyer S, Straub D (2019) Improved cross entropy-based importance sampling with a flexible mixture model. Reliab Eng Syst Saf 191:106564.
222. Papaioannou I, Straub D (2021) Combination line sampling for structural reliability analysis. Struct Saf 88:102025.
223. Pradlwarter HJ, Schueller GI (1997) On advanced Monte Carlo simulation procedures in stochastic structural dynamics. Int J Non-Linear Mech 32(4):735-744.
224. Pradlwarter HJ, Schueller GI (1999) Assessment of low probability events of dynamical systems by controlled Monte Carlo simulation. Probab Eng Mech 14:213-227.
225. Pradlwarter HJ, Schueller GI, Melnik-Melnikov PG (1994) Reliability of MDOF systems. Probab Eng Mech 9:235-243.
226. Lieu QX, Nguyen KT, Dang KD, Lee S, Kang J, Lee J (2022) An adaptive surrogate model to structural reliability analysis using deep neural network. Expert Syst Appl 189:116104.

227. Melchers RE, Beck AT (2018) Structural reliability and prediction. John Wiley, New Jersey.
228. Melchers RE (1990) Radial importance sampling for structural reliability. J Eng Mech 116(1):189-203.
229. Ghanem R, Owhadi H, Higdon D (eds) (2017) Handbook of Uncertainty Quantification. Springer, Cham, Switzerland.
230. Sutton RS, Barto AG (2018) Reinforcement learning. MIT Press, Cambridge, Massachusetts.
231. Rashki M (2018) Hybrid control variates-based simulation method for structural reliability analysis of some problems with low failure probability. Appl Math Model 60:220-234.
232. Rashki M (2021) SESC: A new subset simulation method for rare-events estimation. Mech Syst Signal Process 150:107139.
233. Rashki M, Faes MG, Wei P, Song J (2025) Asymptotic subset simulation: an efficient extrapolation tool for small probabilities approximation. Reliab Eng Syst Saf 260:111034.
234. Rubinstein RY, Kroese DP (2016) Simulation and the Monte Carlo method. John Wiley & Sons, New York.
235. Das S, Tesfamariam S (2024) Reliability assessment of stochastic dynamical systems using physics informed neural network based PDEM. Reliab Eng Syst Saf 243:109849.
236. Garg S, Gupta H, Chakraborty S (2022) Assessment of DeepONet for time dependent reliability analysis of dynamical systems subjected to stochastic loading. Eng Struct 270:114811.
237. Prince SJD (2023) Understanding deep learning. MIT Press, Cambridge, Massachusetts.
238. Huang SK, Chao WT, Lin YX (2024) Conditional generation of artificial earthquake waveforms based on adversarial networks. Soil Dyn Earthq Eng 180:108622.
239. Brunton SL, Kutz JN (2022) Data-driven science and engineering: Machine learning, dynamical systems, and control. Cambridge University Press, Cambridge.
240. Zhu SP, Keshtegar B (2025) Structural reliability analysis: analytical methods. Springer Nature, Switzerland.

241. Panda SS, Manohar CS (2008) Applications of meta-models in finite element-based reliability analysis of engineering structures. CMES-Comp Model Eng Sci 28:161-184.
242. Salari M, Safi M (2017) A new approach for finding the design point of nonlinear systems under random excitation. Struct Saf 69:47-56.
243. Santoso AM, Phoon KK, Quek ST (2011) Modified Metropolis-Hastings algorithm with reduced chain correlation for efficient subset simulation. Probab Eng Mech 26:331-341.
244. Schoenmakers JGM, Heemink AW, Ponnambalam K, Kloeden PE (2002) Variance reduction for Monte Carlo simulation for stochastic environment models. Appl Math Model 26:785-795.
245. Schueller GI, Pradlwarter HJ (2007) Benchmark study on reliability estimation in higher dimensions of structural systems – An overview. Struct Saf 29:167-182.
246. Schueller GI, Stix R (1987) A critical appraisal of methods to determine failure probabilities. Struct Saf 4(4):293-309.
247. Schueller GI, Pradlwarter HJ, Koutsourelakis PS (2004) A critical appraisal of reliability estimation procedures for high dimensions. Probab Eng Mech 19(4):463-474.
248. Sepúlveda JG, Glavind ST, Faber MH (2025) Enhancements and benchmarking of MCMC algorithms for subset simulation in structural reliability. Probab Eng Mech 81:103809.
249. Sharma A, Manohar CS (2023) Modified replica exchange-based MCMC algorithm for estimation of structural reliability based on particle splitting method. Probab Eng Mech 72:103448.
250. Shayanfar MA, Barkhordari MA, Barkhori M, Barkhori M (2018) An adaptive directional importance sampling method for structural reliability analysis. Struct Saf 70:14-20.
251. Shayanfar MA, Barkhordari MA, Barkhori M, Rakhshanimehr M (2017) An adaptive line sampling method for reliability analysis. Iran J Sci Technol Trans Civ Eng 41:275-282.
252. Shields MD, Giovanis DG, Sundar VS (2021) Subset simulation for problems with strongly non-Gaussian, highly anisotropic, and degenerate distributions. Comput Struct 245:106431.

253. Shields MD, Teferra K, Hapij A, Daddazio RP (2015) Refined stratified sampling for efficient Monte Carlo based uncertainty quantification. Reliab Eng Syst Saf 142:310-325.

254. Shinozuka M (1983) Basic analysis of structural safety. J Struct Eng 109(3):721-740.

255. Sonal SD, Ammanagi S, Kanjilal O, Manohar CS (2018) Experimental estimation of time variant system reliability of vibrating structures based on subset simulation with Markov chain splitting. Reliab Eng Syst Saf 178:55-68.

256. Song C, Kawai R (2023b) Adaptive radial importance sampling under directional stratification. Probab Eng Mech 72:103443.

257. Song C, Kawai R (2023a) Adaptive stratified sampling for structural reliability analysis. Struct Saf 101:102292.

258. Song S, Lu Z, Fu L (2007) Reliability sensitivity algorithm based on line sampling. Chin J Theor Appl Mech 23(4):564-570.

259. Song S, Lu Z, Qiao H (2009) Subset simulation for structural reliability sensitivity analysis. Reliab Eng Syst Saf 94(2):658-665.

260. Song S, Lu Z, Song Z (2011) Reliability sensitivity analysis involving correlated random variables by directional sampling. In: 2011 International Conference on Quality, Reliability, Risk, Maintenance, and Safety Engineering (ICQR2MSE), Xi’an, China, pp 845-850.

261. Song C, Kawai R (2023c) Monte Carlo and variance reduction methods for structural reliability analysis: A comprehensive review. Probab Eng Mech 73:103479.

262. Soong TT, Grigoriu M (1993) Random vibration of mechanical and structural systems. Prentice Hall, New Jersey.

263. Sundar VS, Ammanagi S, Manohar CS (2015) System reliability of randomly vibrating structures: Computational modeling and laboratory testing. J Sound Vib 351:189-205.

264. Sundar VS, Manohar CS (2014a) Estimation of time variant reliability of randomly parametered non-linear vibrating systems. Struct Saf 47:59-66.

265. Sundar VS, Manohar CS (2014b) Random vibration testing with controlled samples. Struct Control Health Monit 21:1269-1283.

266. Tripura T, Thakur A, Chakraborty S (2024) Multi-fidelity wavelet neural operator surrogate model for time-independent and time-dependent reliability analysis. Probab Eng Mech 77:103672.

267. Tabandeh A, Jia G, Gardoni P (2022) A review and assessment of importance sampling methods for reliability analysis. Struct Saf 97:102216.
268. Tanaka H (1997) Application of an importance sampling method to time-dependent system reliability analyses using the Girsanov transformation. In: Proceedings of the Seventh International Conference on Structural Safety and Reliability, Kyoto, Japan.
269. Teixeira R, Nogal M, O'Connor A (2021) Adaptive approaches in metamodel-based reliability analysis: A review. Struct Saf 89:102019.
270. Thaler D, Dhulipala SL, Bamer F, Markert B, Shields MD (2024) Reliability analysis of complex systems using subset simulations with Hamiltonian Neural Networks. Struct Saf 109:102475.
271. Thedy J, Liao KW (2021) Multisphere-based importance sampling for structural reliability. Struct Saf 91:102099.
272. Au SK, Lam HF, Ng CT (2007a) Reliability analysis of single-degree-of-freedom elastoplastic systems. I: Critical excitations. J Eng Mech 133(10):1072-1080.
273. Au SK, Lam HF, Ng CT (2007b) Reliability analysis of single-degree-of-freedom elastoplastic systems. II: Suboptimal excitations. J Eng Mech 133(10):1081-1085.
274. Vrouwenvelder T, Beck AT, Proske D, Faber M, Köhler J, Schubert M, Straub D, Teichgräber M (2025) Interpretation of probability in Structural Safety–A philosophical conundrum. Struct Saf 113:102473.
275. Triantafyllou SP, Koumousis VK (2014) Hysteretic finite elements for the nonlinear static and dynamic analysis of structures. J Eng Mech 140(6):04014025.
276. Tripura T, Chakraborty S (2023) Wavelet neural operator for solving parametric partial differential equations in computational mechanics problems. Comput Methods Appl Mech Eng 404:115783.
277. Wolpert DH, Macready WG (1995) No free lunch theorems for search. Technical Report SFI-TR-95-02-010, Santa Fe Institute.
278. Wolpert DH, Macready WG (1997) No free lunch theorems for optimization. IEEE Trans Evolut Comput 1(1):67-82.
279. Papaioannou I, Papadimitriou C, Straub D (2016) Sequential importance sampling for structural reliability analysis. Struct Saf 62:66-75.
280. Ullmann E, Papaioannou I (2015) Multilevel estimation of rare events. SIAM/ASA J Uncertainty Quantif 3(1):922-953.

281. Valdebenito MA, de Angelis M, Patelli E (2021a) Line sampling simulation: recent advancements and applications. In: Reliability-based analysis and design of structures and infrastructure. CRC Press, Boca Raton, pp 215–226.
282. Valdebenito MA, Jensen HA, Hernández HB, Mehrez L (2018) Sensitivity estimation of failure probability applying line sampling. Reliab Eng Syst Saf 171:99-111.
283. Valdebenito MA, Jensen HA, Labarca AA (2014) Estimation of first excursion probabilities for uncertain stochastic linear systems subjected to Gaussian load. Comput Struct 138:36-48.
284. Valdebenito MA, Misraji MA, Faes MG (2024) Multidomain line sampling for reliability and sensitivity analysis in stochastic linear dynamics. In: International Conference on Modeling, Simulation and Optimization. Springer Nature Singapore, Singapore, pp 740-750.
285. Valdebenito MA, Wei P, Song J, Beer M, Broggi M (2021b) Failure probability estimation of a class of series systems by multidomain line sampling. Reliab Eng Syst Saf 213:107673.
286. Rashki M, Faes MG (2023) No-free-lunch theorems for reliability analysis. ASCE-ASME J Risk Uncertain Eng Syst Part A Civ Eng 9(3):04023019.
287. Vořechovský M (2022) Reliability analysis of discrete-state performance functions via adaptive sequential sampling with detection of failure surfaces. Comput Methods Appl Mech Eng 401:115606.
288. Fauriat W, Gayton N (2014) AK-SYS: an adaptation of the AK-MCS method for system reliability. Reliab Eng Syst Saf 123:137-144.
289. Zhang W, Zhao M, Du X, Gao Z, Ni P (2023) Probabilistic machine learning approach for structural reliability analysis. Probab Eng Mech 74:103502.
290. Wang H, Li L, Liu J, Yuan X (2025) An efficient method for solving system failure probability functions based on subset simulation and probability reanalysis techniques. Reliab Eng Syst Saf 262:111248.
291. Wang Z, Broccardo M, Der Kiureghian A (2016) An algorithm for finding a sequence of design points in reliability analysis. Struct Saf 58:52-59.
292. Wang Z, Broccardo M, Song J (2019) Hamiltonian Monte Carlo methods for subset simulation in reliability analysis. Struct Saf 76:51-67.

293. Wang Z, Mourelatos ZP, Li J, Baseski I, Singh A (2014) Time-dependent reliability of dynamic systems using subset simulation with splitting over a series of correlated time intervals. J Mech Des 136(6):061008.

294. Wang Z, Song J (2018) Hyper-spherical extrapolation method (HEM) for general high dimensional reliability problems. Struct Saf 72:65-73.

295. Wei N, Lu Z, Hu Y (2023a) An eccentric radial-based importance sampling method for reliability analysis. Expert Syst Appl 219:119687.

296. Wei N, Lu Z, Hu Y (2023b) Stochastic collocation enhanced line sampling method for reliability analysis. Reliab Eng Syst Saf 240:109552.

297. Wen YK (1976) Method for random vibration of hysteretic systems. J Eng Mech Div 102(2):249-263.

298. Wen YK, Chen HC (1987) On fast integration for time variant structural reliability. Probab Eng Mech 2(3):156-162.

299. Huang X, Chen J, Zhu H (2016) Assessing small failure probabilities by AK–SS: an active learning method combining Kriging and Subset Simulation. Struct Saf 59:86-95.

300. Xia W, Liao Z (2025) Enhanced generalized subset simulation with multiple importance sampling for reliability estimation. Comput Struct 313:107741.

301. Xiao S, Reuschen S, Köse G, Oladyshkin S, Nowak W (2019) Estimation of small failure probabilities based on thermodynamic integration and parallel tempering. Mech Syst Signal Process 133:106248.

302. Xu Z, Saleh JH (2021) Machine learning for reliability engineering and safety applications: Review of current status and future opportunities. Reliab Eng Syst Saf 211:107530.

303. Gong Y, Cheung SH (2025) A physics-informed global-local neural operator for parametric stochastic differential equations. Eng Appl Artif Intell 159:111523.

304. Jiang Y, Luo J, Liao G, Zhao Y, Zhang J (2015) An efficient method for generation of uniform support vector and its application in structural failure function fitting. Struct Saf 54:1-9.

305. Jiang Y, Zhao L, Beer M, Patelli E, Broggi M, Luo J, He JY, Zhang J (2017) Multiple response surfaces method with advanced classification of samples for structural failure function fitting. Struct Saf 64:87-97.

306. Lee KY, Li B, Chiaromonte F (2013) A general theory for nonlinear sufficient dimension reduction: formulation and estimation. Ann Stat 41:221–249.

307. Matsumoto Y, Yaoyama T, Lee S, Hida T, Itoi T (2023) Fundamental study on probabilistic generative modeling of earthquake ground motion time histories using generative adversarial networks. Jpn Archit Rev 6(1):e12392.

308. Ma YZ, Zhu YC, Li HS, Nan H, Zhao ZZ, Jin XX (2022) Adaptive Kriging-based failure probability estimation for multiple responses. Reliab Eng Syst Saf 228:108771.

309. Yuan X, Zheng W, Zhao C, Valdebenito MA, Faes MG, Dong Y (2024) Line sampling for time-variant failure probability estimation using an adaptive combination approach. Reliab Eng Syst Saf 243:109885.

310. Hu Z, Mahadevan S (2016) A single-loop kriging surrogate modelling for time-dependent reliability analysis. J Mech Des 138(6):061406.

311. Hu Z, Du X (2015) Mixed efficient global optimization for time-dependent reliability analysis. J Mech Des 137(5):051401.

312. Liu Z, Lu Z, Ling C, Feng K, Hu Y (2022) An improved AK-MCS for reliability analysis by an efficient and simple reduction strategy of candidate sample pool. Structures 35:373-387.

313. Lv Z, Lu Z, Wang P (2015) A new learning function for Kriging and its applications to solve reliability problems in engineering. Comput Math Appl 70(5):1182-1197.

314. Xiang Z, Bao Y, Tang Z, Li H (2020) Deep reinforcement learning-based sampling method for structural reliability assessment. Reliab Eng Syst Saf 199:106901.

315. Zayed A, Garbatov Y, Guedes Soares C (2013) Time variant reliability assessment of ship structures with fast integration techniques. Probab Eng Mech 32:93-102.

316. Zhang X (2026) Line importance sampling for reliability analysis with complex failure domain. Reliab Eng Syst Saf 271:112296.

317. Zhang X, Lu Z, Cheng K (2022) Cross-entropy-based directional importance sampling with von Mises-Fisher mixture model for reliability analysis. Reliab Eng Syst Saf 220:108306.

318. Zhang X, Lu Z, Cheng K, Wang Y (2020a) A novel reliability sensitivity analysis method based on directional sampling and Monte Carlo simulation. Proc Inst Mech Eng Pt O J Risk Reliab 234(4):622-635.

319. Zhang X, Zhenzhou L, Wanying Y, Kaixuan F, Wang Y (2020b) Line sampling-based local and global reliability sensitivity analysis. Struct Multidiscip Optim 61(1):267-281.

320. Zuev KM, Beck JL, Au SK, Katafygiotis LS (2012) Bayesian post-processor and other enhancements of subset simulation for estimating failure probabilities in high dimensions. Comput Struct 92-93:283-296.

321. Zuev KM, Katafygiotis LS (2011) Modified Metropolis-Hastings algorithm with delayed rejection. Probab Eng Mech 26:405-412.

322. Zuniga MM, Garnier J, Remy E, de Rocquigny E (2011) Adaptive directional stratification for controlled estimation of the probability of a rare event. Reliab Eng Syst Saf 96(12):1691-1712.